\documentclass[twocolumn]{aastex61}
\accepted{07 14, 2018}%{\today}
\shorttitle{Intra-cluster GCs in the Virgo cluster core}
\shortauthors{Longobardi et al.}
\begin{document}

\title{The Next Generation Virgo Cluster Survey (NGVS) XXXI. The kinematics of intra-cluster globular clusters in the core of the Virgo cluster}

\correspondingauthor{Alessia Longobardi, Eric Peng}
\email{alongobardi@pku.du.cn, peng@pku.edu.cn}

\author[0000-0001-5569-6584]{ALESSIA LONGOBARDI}
\affil{Kavli Institute for Astronomy and Astrophysics, Peking University, Beijing 100871, China; alongobardi@pku.du.cn; peng@pku.edu.cn}

\author[0000-0002-2073-2781]{ERIC W.\ PENG}
\affiliation{Department of Astronomy, Peking University, Beijing 100871, China}
\affiliation{Kavli Institute for Astronomy and Astrophysics, Peking University, Beijing 100871, China; alongobardi@pku.du.cn; peng@pku.edu.cn}

\author[0000-0003-1184-8114]{PATRICK C\^{O}T\'{E}}
\affiliation{NRC Herzberg Astronomy and Astrophysics, National Research Council, 5071 West Saanich Road, Victoria, BC V9E 2E7, Canada}

\author[0000-0002-7089-8616]{J. CHRISTOPHER MIHOS}
\affiliation{Department of Astronomy, Case Western Reserve University, Cleveland, OH 41106, USA}

\author[0000-0002-8224-1128]{LAURA FERRARESE}
\affiliation{NRC Herzberg Astronomy and Astrophysics, National Research Council, 5071 West Saanich Road, Victoria, BC V9E 2E7, Canada}

\author[0000-0003-0350-7061]{THOMAS H. PUZIA}
\affiliation{Institute of Astrophysics, Pontificia Universidad Cat\'{o}lica de Chile, Av. Vicu\~{n}a Mackenna 4860, 7820436 Macul, Santiago, Chile}

\author[0000-0002-7214-8296]{ARIANE LAN\c{C}ON}
\affiliation{Observatoire astronomique de Strasbourg, Universit\'{e} de Strasbourg, CNRS, UMR 7550, 11 rue de l'Universit\'{e}, F-67000 Strasbourg, France}

\author[0000-0003-1632-2541]{HONG-XIN ZHANG}
\affiliation{CAS Key Laboratory for Research in Galaxies and Cosmology, Department of Astronomy, University of Science and Technology of China, Hefei 230026, China}
\affiliation{Institute of Astrophysics, Pontificia Universidad Cat\'{o}lica de Chile, Av. Vicu\~{n}a Mackenna 4860, 7820436 Macul, Santiago, Chile}

\author{ROBERTO P. MU\~{N}OZ}
\affiliation{Institute of Astrophysics, Pontificia Universidad Cat\'{o}lica de Chile, Av. Vicu\~{n}a Mackenna 4860, 7820436 Macul, Santiago, Chile}

\author[0000-0002-5213-3548]{JOHN P. BLAKESLEE}
\affiliation{NRC Herzberg Astronomy and Astrophysics, 5071 West Saanich Road, Victoria, BC V9E 2E7, Canada}
\affiliation{Gemini Observatory, Casilla 603, La Serena 1700000, Chile}

\author[0000-0001-8867-4234]{PURAGRA GUHATHAKURTA}
\affiliation{UCO/Lick Observatory, University of California, Santa Cruz, 1156 High Street, Santa Cruz, CA 95064, USA}

\author[0000-0001-9427-3373]{PATRICK R. DURRELL}
\affiliation{Department of Physics and Astronomy, Youngstown State University, Youngstown, OH 44555, USA}

\author[0000-0003-4945-0056]{R\'{U}BEN S\'{A}NCHEZ-JANSSEN}
\affiliation{UK Astronomy Technology Centre, Royal Observatory Edinburgh, Blackford Hill, Edinburgh, EH9 3HJ, UK}

\author{ELISA TOLOBA}
\affiliation{Department of Physics, University of the Pacific, 3601 Pacific Avenue, Stockton, CA 95211, USA}

\author[0000-0002-5389-3944]{ANDR\'{E}S JORD\'{A}N}
\affiliation{Institute of Astrophysics, Pontificia Universidad Cat\'{o}lica de Chile, Av. Vicu\~{n}a Mackenna 4860, 7820436 Macul, Santiago, Chile}
\affiliation{Millennium Institute of Astrophysics, Av.\ Vicu\~na Mackenna 4860, 7820436 Macul, Santiago, Chile}

\author{SUSANA EYHERAMENDY}
\affiliation{Millennium Institute of Astrophysics, Av.\ Vicu\~na Mackenna 4860, 7820436 Macul, Santiago, Chile}
\affiliation{Department of Statistics, Faculty of Mathematics Pontificia Universidad Cat\'{o}lica de Chile, Av. Vicu\~na Mackenna 4860, 7820436 Macul, Santiago, Chile}

\author{JEAN-CHARLES CUILLANDRE}
\affiliation{CEA/IRFU/SAp, Laboratoire AIM Paris-Saclay, CNRS/INSU, Universit\'{e} Paris Diderot, Observatoire de Paris, PSL Research University, F-91191 Gif-sur-Yvette Cedex, France}

\author{STEPHEN D. J. GWYN}
\affiliation{NRC Herzberg Astronomy and Astrophysics, 5071 West Saanich Road, Victoria, BC V9E 2E7, Canada}

\author{ALESSANDRO BOSELLI}
\affiliation{Aix Marseille Universit\'{e}, CNRS, LAM (Laboratoire d'Astrophysique de Marseille) UMR 7326, F-13388 Marseille, France}

\author[0000-0003-3343-6284]{PIERRE-ALAIN DUC}
\affiliation{Observatoire astronomique de Strasbourg, Universit\'{e} de Strasbourg, CNRS, UMR 7550, 11 rue de l'Universit\'{e}, F-67000 Strasbourg, France}

\author[0000-0002-4718-3428]{CHENGZE LIU}
\affiliation{Department of Astronomy, Shanghai Key Laboratory for Particle Physics and Cosmology, Shanghai Jiao Tong University, Shanghai 200240, China}

\author[0000-0002-5897-7813]{KARLA ALAMO-MART\'{I}NEZ}
\affiliation{Institute of Astrophysics, Pontificia Universidad Cat\'{o}lica de Chile, Av. Vicu\~{n}a Mackenna 4860, 7820436 Macul, Santiago, Chile}
%\author{}
%\affiliation{}
\author[0000-0002-1218-3276]{MATHIEU POWALKA}
\affiliation{Observatoire astronomique de Strasbourg, Universit\'{e} de Strasbourg, CNRS, UMR 7550, 11 rue de l'Universit\'{e}, F-67000 Strasbourg, France}
\author{SUNGSOON LIM}
\affiliation{Department of Astronomy, Peking University, Beijing 100871, China}

%\author{the NGVS team}
%\collaboration{(AAS Journals Data Scientists collaboration)}

%\author{Butler Burton}
%\affiliation{National Radio Astronomy Observatory}
%\affiliation{AAS Journals Associate Editor-in-Chief}
%\nocollaboration

%\altaffiliation{Creator of AASTeX v6.1}
%\collaboration{(LaTeX collaboration)}

%\affiliation{AAS Director of Publishing}
%\affiliation{American Astronomical Society \\
%2000 Florida Ave., NW, Suite 300 \\
%Washington, DC 20009-1231, USA}

%\author{Jeff Lewandowski}
%\affiliation{IOP Senior Publisher for the AAS Journals}
%\affiliation{IOP Publishing, Washington, DC 20005}



\begin{abstract}
Intra-cluster (IC) populations are expected to be a natural result of the hierarchical assembly of clusters, yet their low space densities make them difficult to detect and study. We present the first definitive kinematic detection of an IC population of globular clusters (GCs) in the Virgo cluster, around the central galaxy, M87. 
This study focuses on the Virgo core for which the combination of NGVS photometry and follow-up spectroscopy allows us to reject foreground star contamination and explore GC kinematics over the full Virgo dynamical range. The GC kinematics changes gradually with galactocentric distance, decreasing in mean velocity and increasing in velocity dispersion, eventually becoming indistinguishable from the kinematics of Virgo dwarf galaxies at $\mathrm{R>320\, kpc}$. By kinematically tagging M87 halo and intra-cluster GCs we find that 1) the M87 halo has a smaller fraction ($52\pm3\%$) of blue clusters with respect to the IC counterpart ($77\pm10\%$), 2) the $(g'-r')_{0}$ vs $(i'-z')_{0}$ color-color diagrams reveal a galaxy population that is redder than the IC population that may be due to a different composition in chemical abundance and progenitor mass, and 3) the ICGC distribution is shallower and more extended than the M87 GCs, yet still centrally concentrated. The ICGC specific frequency, $S_{N,\mathrm{ICL}}=10.2\pm4.8$, is consistent with what is observed for the population of quenched, low-mass galaxies within 1~Mpc from the cluster's center. The IC population at Virgo's center is thus consistent with being an accreted component from low-mass galaxies tidally stripped or disrupted through interactions, with a total mass of $\mathrm{M_{ICL,tot}=10.8\pm0.1\times10^{11}M_{\odot}}$.
\end{abstract}




\section{Introduction}

Mergers are believed to play a dominant role in the assembly of massive galaxies \citep{delucia07}. A two-phase formation scenario \citep{naab09} predicts that in a first dissipative phase (high redshift, $z \le 2$) stars form quickly and build the compact innermost regions of massive galaxies \citep[with stars rich in $\alpha$-elements;][]{thomas05}; later, minor mergers dominate the mass assembly in the outermost regions \citep{oser10} and the accretion of metal-poor stars from smaller stellar systems results in the variation of the galaxy stellar properties. In dense environments, accretion is even more dramatic, such that close to the dynamical center of clusters, central galaxies are expected to have accreted a majority of their stars \citep{cooper15}.

In galaxy clusters, a fraction of their baryonic content is represented by the intra-cluster light (ICL, or ``diffuse light''), a component that is gravitationally unbound to cluster galaxies, but bound to the cluster potential \citep[e.g.,][]{murante04,rudick06,dolag10,puchwein10,contini14}. Galaxy interactions, as well as tidal interactions between galaxies and the cluster potential, are both believed to play an important role in the production of the ICL. Hence, the ICL is thought to be intimately linked to the dynamical history of the cluster, so that the ICL's observable features
contain information about the evolutionary processes that took place in these dense environments. 

Despite its dynamical definition, and low surface brightness (SB) \citep[its peak in SB correspond to $\sim$ 1\% of the brightness of the night sky,][]{vilchez99}, the ICL is usually identified on the basis of its photometric properties. It is either identified as any optical light below a fixed SB limit \citep{feldmeier04,mihos05,zibetti05,mihos17}, or as the less concentrated light profile that overlaps onto the one of the galaxy's halo \citep{gonzales05,seigar07}. These studies have shown that the detected ICL is often a discernibly separate entity from the host galaxies, with well defined transitions in the surface brightness profile, axis ratio, and position angle, and whose evolution is tied to the cluster as whole rather than to the central galaxy. In theoretical studies, in which the ICL is dynamically identified in terms of either binding energy \citep[stars that are not bound to identified galaxies, including the central one;][]{murante04} or velocity distribution \citep[broader for the ICL than for the galaxy halo;][]{dolag10,contini14}, the authors find two distinct stellar populations in terms of kinematics, spatial distribution and physical properties like age and metallicity, suggesting that the identification of the system galaxy+ICL as a single entity is ill-defined. However, in several observational and theoretical analysis \citep[e.g.,][]{gonzales07,cooper15}, no separation is made between galaxy halos and ICL, and the two components are treated as a continuum. 

In nearby clusters, individual tracers can be used to study the diffuse cluster light around galaxies. They can trace its spatial distribution, age, metallicity, and, when spectroscopy is available, they allow us to gather information on the kinematics of this diffuse component, leading to a less ambiguous definition of galaxy halos and ICL \citep{longobardi15a}. Individual stellar tracers include supernovae \citep{galyam03,dilday10,sand11,barbary12}, red-giant and super-giant stars \citep{ferguson98,durrell02,oyama13}, planetary nebulae
\citep[PNe;][]{arnaboldi96,feldmeier04,arnaboldi04,gerhard05,aguerri05,ventimiglia11,longobardi13,longobardi15a,hartke17,hartke18}, and globular clusters \citep[GCs;][]{mclaughlin94,jordan04,williams07,lee10,peng11,strader11,romanowsky12,durrell14,alamo17,ko17}.

GCs are compact groups of stars that are found to inhabit all types of galaxies more luminous than $\sim 3 \times 10^{6}\, L_{\odot}$ \citep[e.g.,][]{georgiev10}, with the most luminous systems characterized by higher GC specific frequencies \citep[e.g.,][]{harris81,brodie06,peng08,georgiev10}. Several GC studies have shown that the properties of these systems are correlated with those of their host galaxies \citep{harris91,brodie91,peng06,peng08,harris13}, suggesting a link between their formation and the evolution of the galaxy itself. It is now well established that the GC systems, within the majority of large galaxies, possess two or more sub-populations of clusters characterized by very different chemical composition \citep[e.g.,][]{gebhardt99,puzia99,forbes01,kundu01,larsen01,puzia05a,puzia05b,peng06,leaman13,tonini13,harris17}.
The most popular division uses optical colors to identify a red (metal-rich) and a blue (metal-poor) population of GCs, although the existence of an additional, intermediate, population of GCs has been claimed \citep{peng09,strader11,agnello14}. To these different classes correspond different observed properties, so that red and blue GCs are measured to have different kinematics and spatial distribution \citep{zepf00,cote01,schroder02,cote03,perrett03,peng04,schuberth10,strader11,coccato13,agnello14,zhang15}. Also, the relative fraction of blue-to-red GCs increases with galactocentric radius, probably tracing the hierarchical processes that built up the galaxies stellar halos \citep[e.g.,][]{cote98,cote00,cote02,tonini13,lim17}. It is important to emphasize, however, that the interpretation of a single color as a measure of metallicity is a simplification. Such an interpretation presents the uncertainty driven by the non-linearity in the relation \citep{peng06,cantiello07,richtler13}, and wrongly assumes that the same stellar population characteristics found in the Local Group apply to dense environments \citep{blakeslee12,powalka17}.

As consequence of their brightness and spatial extension, GCs are powerful tracers of the galaxies' outer halos, also bridging the transition region between the galaxies and the ICL that surrounds them. Single intra-group/cluster GCs have been detected in A1185 \citep{jordan03,west11}, Fornax \citep{firth08,shubert08}, Virgo \citep{williams07,firth08,lee10,ko17}, Coma \citep{peng11}, and A1689 \citep{alamo17}. However, the identification of a large sample of tracers has been elusive, mainly due to a significant level of contamination (see Sect.~\ref{obj_classification}).

The Virgo cluster, the nearest large-scale structure in the Universe, and its central galaxy M87 have long been the targets of GC studies with the aim of tracing their formation and evolution \citep[e.g.,][]{baum55,mould87,jordan02,cote04,ferrarese12,brodie14}. Virgo is characterized by both spatial and kinematic substructures \citep{binggeli87}. Moreover, the evidence that many galaxies are presently falling towards the cluster core \citep{tully84,conselice01,boselli08,boselli14}, as well as the presence of a complex network of extended tidal features \citep{mihos05,mihos17}, suggest that Virgo core is not completely in equilibrium. Close to its dynamical center lies M87 \citep{binggeli87,nulsen95,mei07}. It is considered a cD galaxy \citep[e.g.,][]{carter78} with an extended stellar envelope that reaches out to projected galactocentric radii of $\sim150$ kpc \citep{ferrarese06,kormendy09,janowiecki10}. Evidence of the galaxy experiencing a gas-rich major-merger event can be found in its central regions, hosting a kinematically distinct core (KDC) in its inner 5 kpc \citep{emsellem14}. Moreover H$\alpha$ studies have shown the presence of prominent filaments of ionized gas extending out to 10~kpc from the galaxy's nucleus \citep{arp67,ford79,sparks93,gavazzi00}. Evidence of accretion are traced by age and metallicity gradients in the galaxy's inner region \citep{liu05,montes14}, together with blue color gradients towards the outer regions \citep{rudick10,mihos17}. Furthermore, at large radii, kinematic signatures of accretion events are found in the orbital distribution of GCs \citep{agnello14,oldham16}, in the kinematics of UCDs \citep{zhang15}, as well as in the presence of kinematic substructures in the velocity phase-space of GCs \citep{romanowsky12} and PNs \citep{longobardi15b}. A large accretion event \citep[\textit{the crown} of M87,][]{weil97,longobardi15b} was found to contribute 60\% of the total light at its highest density point \citep[60--90 kpc NW of the galaxy;][]{longobardi15b}, showing that M87 is still assembling in a substantial way. 

This work follows other recent Next Generation Virgo Cluster Survey analyses with the aim of studying the Virgo core \citep{zhu14,grossauer15,liu15a,zhang15,sanchez16,ferrarese16,roediger17}, and the correlations between the GC properties and the evolution of the environment that surround them \citep[e.g.,][]{powalka16a,powalka16b,powalka17}.

We make use of the deep and extended photometric information provided by the Next Generation Virgo Cluster surveys \citep[NGVS/NGVS-IR][]{ferrarese12,munoz14} in the central 2x2 deg$^{2}$ of  Virgo, to study the transition region between the central galaxy and the IC space. The aim is to identify and separate the GC populations associated with these components and study their properties separately. The paper is structured as follows: in Section~\ref{sec:data} we present our data, the different sources of contamination, hence our working sample of GCs. In Section~\ref{kinematic_separation} we study the GC line-of-sight velocity distribution (LOSVD) and kinematically separate the M87 halo component from the Virgo ICL. The properties of both M87 halo and Virgo intra-cluster GCs are analyzed in Section~\ref{subsec:populations} in terms of colors and spatial distributions.
We discuss our results in Section~\ref{sec:discussion} and finally report our conclusions in Section~\ref{sec:conclusion}. 

Throughout the paper we assume a distance for Virgo and M87 of 16.5 Mpc \citep{mei07,blakeslee09}, implying a physical scale of $80\ {\rm pc\ arcsec}^{-1}$.

\section{Observations \& Data} 
\label{sec:data}
This work examines the GC population in the core region of the Virgo cluster through a synergy between optical/near-infrared photometry from the Next Generation Virgo Cluster Survey \citep[NGVS,][]{ferrarese12}/ Next Generation Virgo Cluster Survey-Infrared \citep[NGVS-IR,][]{munoz14}, and spectroscopy, the latter coming from a compilation of different campaigns. In what follows we give a brief description of these surveys and refer the reader to the references therein for additional details.

\subsection{NGVS photometry}
The NGVS is an optical imaging survey of the Virgo cluster carried out with MegaCam on the CFHT \citep{ferrarese12}. It covers a total area of 104 deg$^{2}$, covering the two main sub-clusters  (Virgo A centered on M87, and Virgo B centered on M49) out to their virial radii. \citet{ferrarese12} adopted several methods of background subtraction and image combination to produce stacked NGVS images. Here we work with the stacks built using the MegaPipe global background subtraction, combined with the artificial skepticism algorithm \citep{stetson89}. This provides high accuracy photometry for sources of small spatial extent, making the NGVS a deep photometric survey in the $u^*\, g'\, r'\, i'\, z'$ bands, that for point sources reaches a depth  of $g'=25.7$ mag \footnote{Deep imaging in the $u^*$ and $r'$ bands do not cover the entire NGVS footprint, with the $r'$ band that is limited to the central 2 by 2 square degree around M87. Magnitudes are on the CFHT MegaPrime photometric system.}. In addition, images in all filters have a FWHM that is lower than  1\farcs0, with the best seeing conditions obtained for the $i'$ band with an upper limit of $0\farcs6$. All final images have the same astrometric reference frame, tied to the positions of stars in the Sloan Digital Sky Survey, and the same spatial resolution with a physical scale of 0\farcs186 px$^{-1}$.

For the Virgo core, a 3.62 deg$^{2}$ region centered on M87, a deep $K_{s}$ band imaging survey (NGVS-IR)  was  also  carried out with WIRCam on the CFHT \citep{munoz14}. Any raw images with a seeing worse than 0\farcs7 were rejected before stacking. This made it possible to produce stacked images with the same pixel scale as the MegaCam stacks, although the original WIRCam pixel scale is 0\farcs3 px$^{-1}$. The Swarp software \citep{bertin02} coupled with a Lanczos-2 interpolation method was used to stack the sky-subtracted images. Thus, the  NGVS-IR  has a limiting magnitude of $K_{s}\sim24.4$~AB~mag for point sources, and an image quality better than $0\farcs7$.  Photometry of all objects on the final, stacked NGVS/NGVS-IR images was conducted using Source Extractor \citep{bertin96}. The aperture corrections (due to small variations in the stellar point-spread function across the field) were derived using samples of bright, unsaturated stars in each field for each filter \citep[more details can be found in][]{durrell14,liu15a,munoz14}.

\subsection{Spectroscopy}
We have compiled a catalog containing all the known redshifts within the NGVS footprint. It combines data from the literature \citep[e.g.,][]{binggeli85,hanes01,strader11}, public archives such as SDSS \citep[DR12,][]{alam15} and NED\footnote{The NASA/IPAC Extragalactic Database (NED) is operated by the Jet Propulsion Laboratory, California Institute of Technology, under contract with the National Aeronautics and Space Administration.}, as well as data from our own spectroscopic surveys dedicated to the detection of compact objects in Virgo. In this work we only focus on the inner $2^{\circ}\times 2^{\circ}$ around M87, and in what follows we will give a brief description of the surveys that provided data for this analysis.

\paragraph{{\bf Spectroscopic data in the Virgo core}}
\begin{itemize}
\item[] \textit{The Literature sample.} In addition to our own surveys (described below), the central region of the Virgo cluster has been the target of several spectroscopic campaigns. \citet{strader11} (S11) combined the pre-2011 published data together with their new observations to compile a sample of radial velocities covering M87 within a 40\arcmin\ radius. As a compilation of different surveys \citep[e.g.,][]{hanes01,cote01}, data from the S11 catalog come from observations carried out with different instruments, hence characterized by different resolutions and spectral coverage. The authors extensively describe the different observational strategies and the compilation procedure that led to the creation of the S11 catalog, hence we refer the reader to their work for a detailed description of the surveys. Here we note that 
we use the radial velocities provided in their catalog, but in some cases have modified the object classification based on our high-quality photometric information (see Sec.~\ref{back_contamination}). 

\item[] \textit{The MMT campaign.} In 2009 and 2010 we used the Hectospec multifiber spectrograph on the 6.5m MMT telescope to survey the compact stellar systems in the central $2^{\circ}\times 2^{\circ}$ area around M87 \citep{zhang15}. Data covered the spectral region within 3650\AA--9200\AA\, with a resolution of $R=1000$. By exposing for $\sim$2 hours, the survey depth is $g'\sim 22.5$ mag and velocities have a typical uncertainty of 
$\sigma_{V_{LOS}}\mathrm{= 30\, km\, s^{-1}}$.

\item[] \textit{The AAT campaign.} Compact objects in Virgo were further followed up spectroscopically in March/April 2012 when we surveyed the central region of the Virgo cluster (Virgo A sub-cluster), using the 2dF AAOmega multi-fiber spectrograph on the AAT. Spectra were obtained in the wavelength range 3700\AA--8000\AA\ and with a spectral resolution of $R=1300$ \citep{zhang15}. The survey consisted of nine 2dF pointings with a typical exposure time of 1.5 hours, covering a total sky area of $\sim30\, \mathrm{deg}^2$. At the limiting magnitude of $g'\sim 20.5$ mag and with such an observational set-up the typical uncertainty on the velocity estimates is $\sigma_{V_{LOS}}\mathrm{= 30\, km\, s^{-1}}$. 

\item[] \textit{The Keck campaign.} In April 2013 additional data were collected using the DEIMOS spectrograph located at the Keck II 10 m telescope. The instrumental configuration provided a wavelength coverage of 4800\AA--9500\AA\ with a spectral resolution of 2.8\AA\, (FWHM). This campaign provided velocity measurements with a mean precision of $\sigma_{V_{LOS}}\mathrm{= 10\, km\, s^{-1}}$ and down to $g'\sim 24.7$ mag \citep{toloba16}. 24.7 must be the faintest object, but certainly not representative of the GC population. 

\item[] \textit{The Magellan campaign.} Finally, in March 2016 we used the IMACS multi-slit spectrograph on the Magellan Baade telescope to obtain spectra for compact objects in Virgo with $g' < 22$ mag. With a field of view of $27\farcm5 \times 27\farcm5$, the observations surveyed two regions, one centered on M87 and the second one offset by 30\arcmin\ towards the NW along the major axis of the galaxy. Data were obtained in the wavelength range 3900\AA\, - 9000\AA\, with a spectral resolution of 6.5\AA. With a total integration time of 3.5 hrs this survey provided data with a mean velocity uncertainty  $\sigma_{V_{LOS}} \mathrm{= 30\, km\, s^{-1}}$ \citep{zhang18}. 
\end{itemize}

As we show below the NGVS photometry provides vital information needed to select the GC sample that we use to analyze the kinematics of the GC system and identify the LOSVDs of the different dynamical components. Hence our working sample is selected from the matched objects between the NGVS photometric and spectroscopic catalogs in the pilot region, resulting in a total sample of $\mathrm{N_{obj}=2809}$ sources. We further decide to analyze objects down to $g' \le 24.0$~mag and $\sigma_{V_{LOS}}\mathrm{ \le 50\, km\, s^{-1}}$. This leads to a total number of sources $\mathrm{N_{obj}=2551}$, with  median photometric uncertainties $\sigma_\mathrm{mag}=[0.01, 0.005, 0.004, 0.005, 0.008, 0.02]$ in the  $u^*\, g'\, r'\, i'\, z'\, \mathrm{and}\, K_{s}$ bands, respectively. Of these objects, 897 have high photometric probability to be a GC (see Sec.~\ref{back_contamination}). Among these, 592 were gathered by S11, while 232, 9, 29, and 18 radial velocities were observed for the first time in the MMT, AAT, Keck, and Magellan campaign, respectively (the remaining velocities are from public archives). We emphasize that from repeated measurements in different campaigns the velocity estimates are in agreement within the spectroscopic uncertainty threshold of $\sim \sigma_{V_{LOS}} = 50\, \mathrm{kms^{-1}}$. The two spectroscopic catalogs with the highest number of GCs in common (67 common objects from the S11 and MMT samples) only present two velocity estimates that deviate $< \sim 1.5\, \sigma_{V_{LOS}}$.

\subsection{Objects classification}
\label{obj_classification}
The selection of a bona-fide sample of GCs is a crucial first step for our kinematic analysis. The two main types of contaminants in our sample are i) background galaxies and ii) foreground Milky Way stars. A third, more subtle, type of potential contaminants are iii) Virgo UCD galaxies. In what follows, we examine such a contamination and its contribution to the sample of GCs.

\subsubsection{Background galaxies/Foreground stars contamination}
\label{back_contamination}

\begin{figure}
\includegraphics[width = 9 cm]{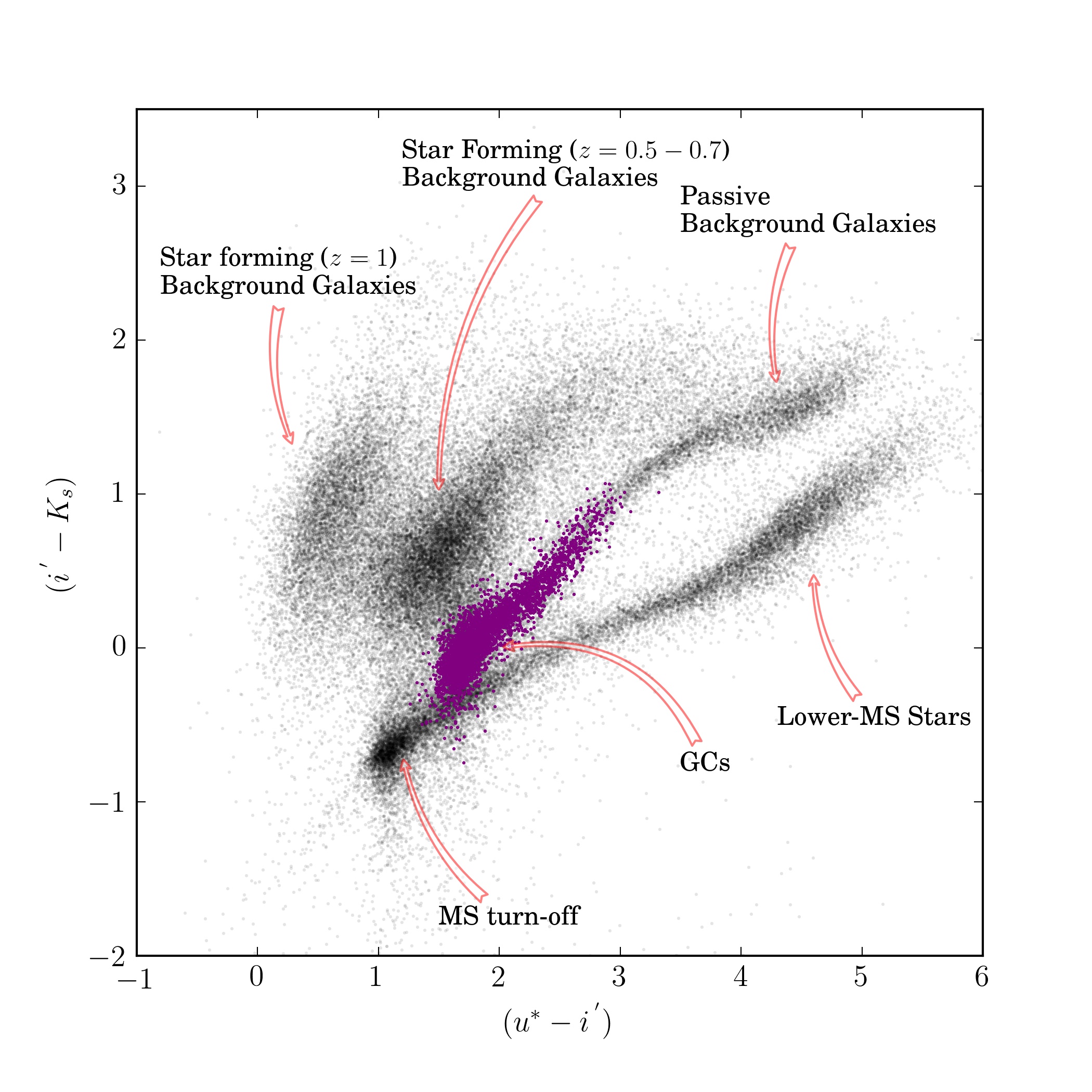}
\caption{$u^*i'K_{s}$ color-color diagram for the NGVS photometric sample in the pilot region with $g' \le 24.0$ AB mag. In this plane background galaxies, GCs, and foreground MW stars separate into different regions \citep{munoz14}. Magenta dots identify objects with high photometric probability, $\mathrm{p_{gc}} > 0.5$, to be GCs (see text for more details).} 
\label{fig:CC_all}
\end{figure}

\citet{munoz14} showed that background galaxies, GCs, and foreground stars define different regions in the $u^*i'K_{s}$ color-color diagram. In Fig.~\ref{fig:CC_all}, this relation is shown for the photometric sample in the NGVS pilot region. From redder to bluer $(i'-K_{s})$ colors, we can see the different contribution from background galaxies with various star forming histories at redshifts up to $z \simeq 1$, GCs (which merge into the redshift sequence of passive galaxies at the red end), and foreground main sequence stars. 
However, a successful compilation of a GC sample implies accounting for the shared contribution between different populations in the transition regions of these three sequences. To do so, we have used the full photometric information gathered by the NGVS/NGVS-IR surveys, and adapted the "extreme deconvolution" (XD) algorithm from \citet{bovy11} to classify our data according to multi-dimensional color and concentration information. The details of this classification procedure will be presented in a future work. Here we give a brief description of the procedure and validate the results in the next Section, with the spectroscopic information we have.

In the pilot region the XD approach uses multi-dimensional Gaussians to model the density of objects in the 5-dimensional space: $(u^*-g')_{0}$, $(g'-i')_{0}$, $(i'-z')_{0}$,  $(i'-K_{s})_{0}$, and a measure of concentration $iC=i'_{4}-i'_{8}$, where  $i'_{4}$ and $i'_{8}$ are the point source aperture-corrected $i$-band magnitudes measured in fixed apertures of diameter $r=4$ pixels (0\farcs72) and $r=8$ pixels (1\farcs44), respectively. The $i$-band seeing in the NGVS images has $FWHM<0\farcs6$ with a median seeing of $0\farcs54$. For point-like objects $iC=0$, and extended objects have $iC>0$. The Gaussians in the model are convolved with the observational uncertainties of the data. Therefore, each data point is assumed to be drawn from the model convolved with its own set of measurement uncertainties, and at the end of the XD procedure we obtain for each object the probabilities, $\mathrm{p_{galx}},\mathrm{p_{gc}},\, \mathrm{and}\, \mathrm{p_{star}}$, to be either a galaxy, a GC or a star, respectively. Magenta dots in Fig.~\ref{fig:CC_all} are objects with high photometric probability, $\mathrm{p_{gc}} > 0.5$, to be GCs.

\subsubsection{Spectroscopic validation of the photometric selection}
\label{spec_validation}

\begin{figure*}[!ht]
\includegraphics[width = 18cm]{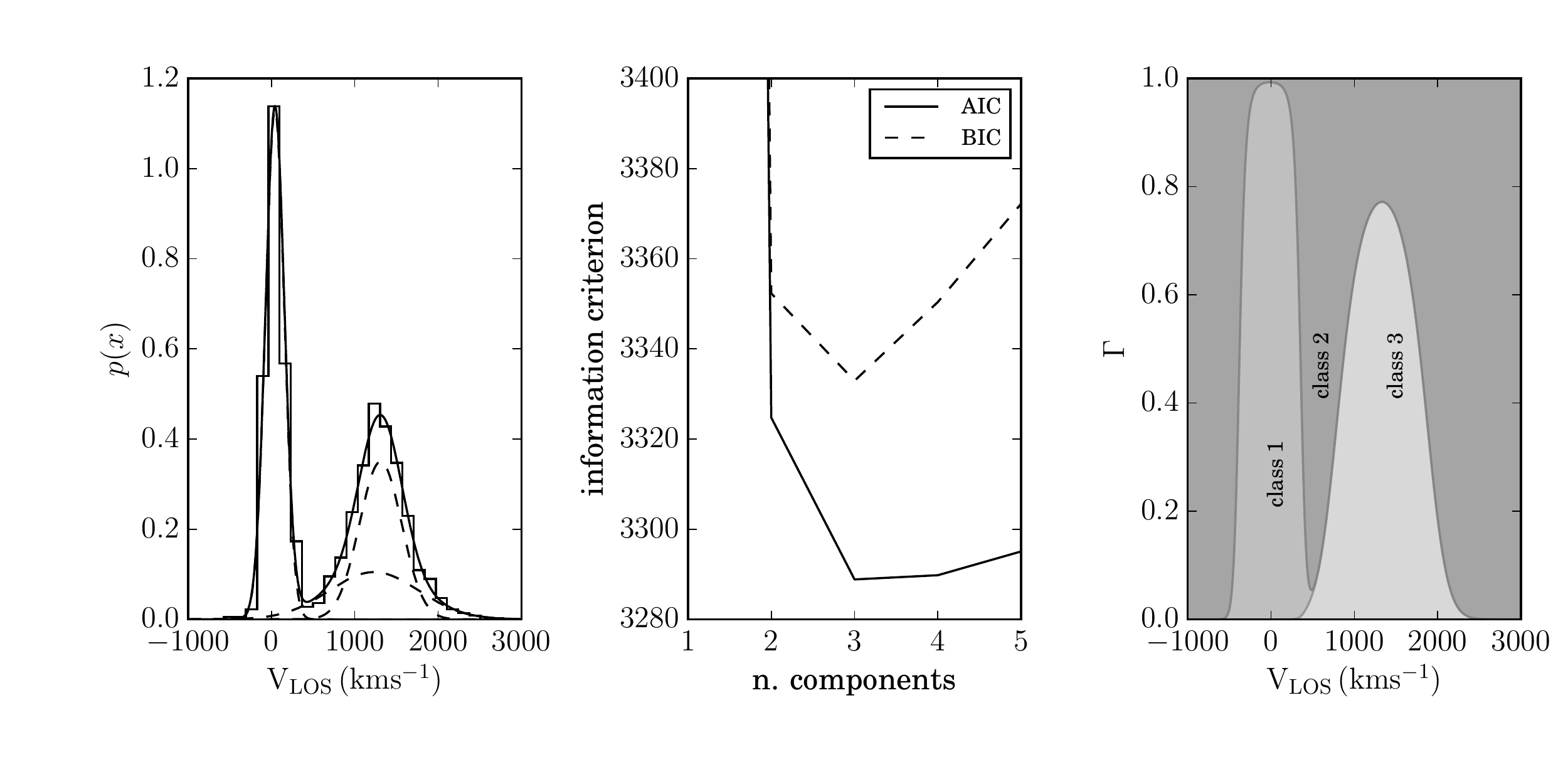}
\caption{{\bf Left Panel}: LOSVD of the spectroscopically confirmed objects in the NGVS pilot region that define our sample. The GMM best-fit model describes the total LOSVD as a mixture of three Gaussians (black continues line) centered on the star, Virgo ICL, and M87 halo distributions, respectively (black dashed lines). {\bf Central Panel}: BIC and AIC scores for the Gaussian mixtures as function of the number of components. Both the information criteria prefer three mixtures to describe the data as in the left panel. {\bf Right Panel}: Posterior probabilities that an object is drawn from one of the three components, i.e. the  stars (light gray, class 1), the Virgo IC (gray, class 2), and the M87 galaxy (dark gray, class 3) distributions, as function of its LOSV. For a given velocity value, the vertical extent of each region is proportional to that probability.}
\label{fig:GMM_total}
\end{figure*}

\paragraph{{ \bf Background galaxies}}
There is a well-defined gap between the Virgo members and background galaxies at radial velocities $\mathrm{V_{galx}=3000\, km\, s^{-1}}$ \citep{binggeli87,conselice01,kim14}. Therefore, the background component is identified by selecting objects with $\mathrm{V_{LOS} > 3000\, km\, s^{-1}}$. We find that 721 sources in our sample satisfy this criterion, representing $\sim 28\%$ of our working sample. The XD photometric classification identifies these contaminants successfully: $> 99\%$ of the background objects are assigned high photometric probability to be background galaxies, with mean value $\langle \mathrm{p_{galx}}\rangle = 0.98$, and only 1\% would result in having high photometric probability to be GCs.

\paragraph{{ \bf Foreground stars}}

The velocity distribution of the Galactic stellar halo is measured to peak at a heliocentric radial velocity of $\sim\!40\ \mathrm{km\, s^{-1}}$ with a velocity dispersion of $\sim\! 120\, \mathrm{km\, s^{-1}}$ \citep[e.g.,][]{battaglia05}. Therefore, the MW star LOSVD overlaps with the LOSVD of objects in Virgo, the latter measured to cover a velocity range $ \mathrm{-500\, km\, s^{-1} \le V_{LOS} \le 3000\, km\, s^{-1}}$ \citep{binggeli87,conselice01}. The contamination by foreground stars is visible as a strong and narrow Gaussian that peaks at around 40 km s$^{-1}$ in Fig.~\ref{fig:GMM_total} (left-hand panel), where we plot the LOSVD of our working sample with $\mathrm{V_{LOS}\le 3000\, km\, s^{-1}}$.

To isolate this kinematic component, we run a Gaussian Mixture Model (GMM), i.e., a probabilistic model that divides the observed LOSVD into a linear combination of $K$ independent Gaussian probability density functions. At the end of the procedure, the posterior probability, $\Gamma$, for a data value to belong to either of the K Gaussian components are returned (Fig.~\ref{fig:GMM_total} right-hand panel, and see Sec.\ref{kinematic_separation} for details on the GMM). Tests with the Bayesian Information Criterion (BIC) and Akaike information criterion (AIC) show that the GMM best-fit model is a mixture of three Gaussians (Fig.~\ref{fig:GMM_total} central panel), identifying the Galactic stellar contamination as component with $\mathrm{V_{star}=41\pm4\, km\, s^{-1}}$ and $\mathrm{\sigma_{star}=114\pm3\, km\, s^{-1}}$. In Sec.~\ref{kinematic_separation} we will show that the remaining two Gaussians are tracing the Virgo cluster and the M87 halo distributions, however as for the moment our goal is to validate the photometric classification, we only isolate the objects with velocities that fall within the stars' LOSVD, and consider as one single class the remaining objects.

\begin{figure}
\centering
\includegraphics[trim=0.cm 0.cm 0.cm 0cm,clip,width =8cm]{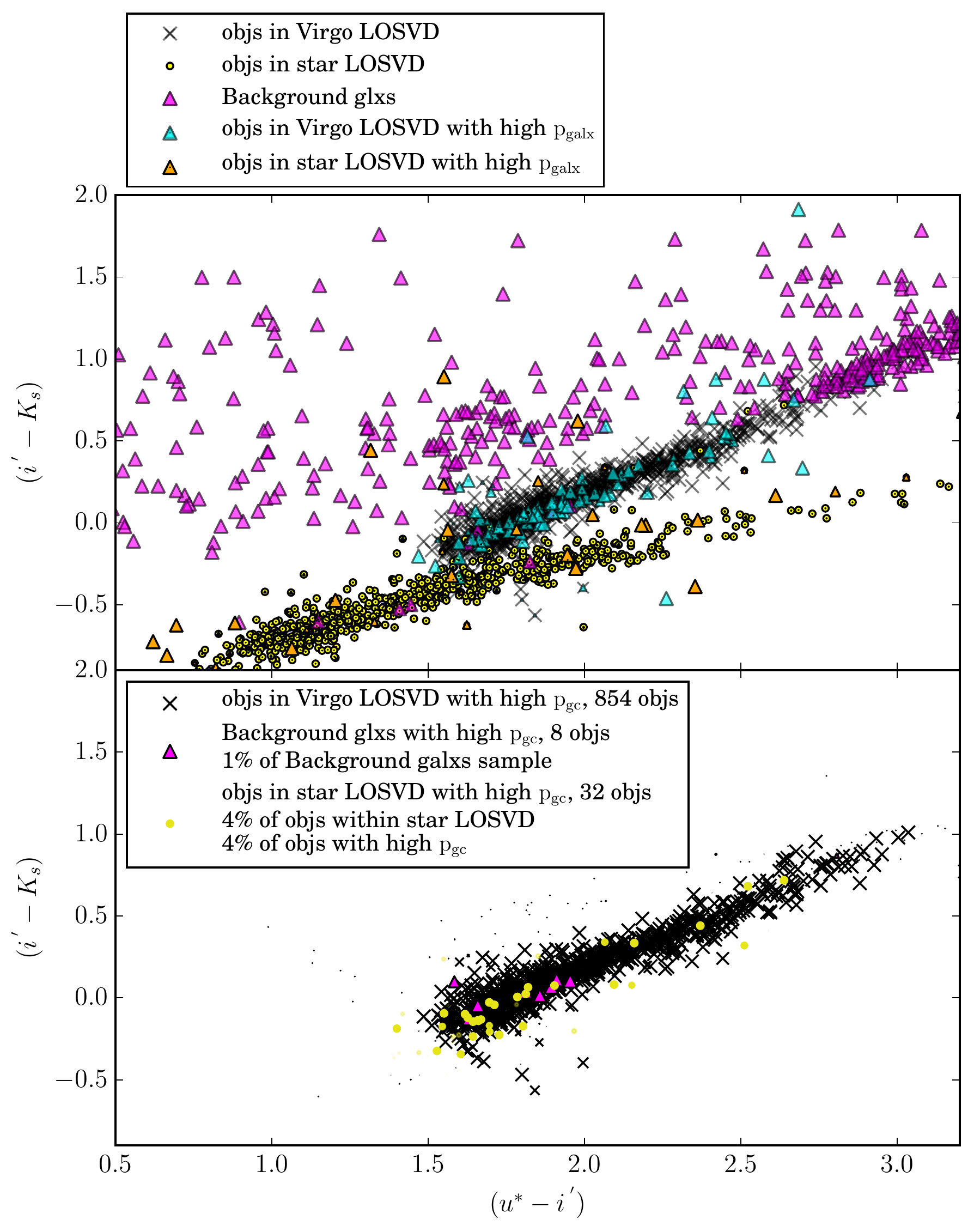}
\caption{{\bf Top Panel:} $u^*i'K_{s}$ diagram color coded based on the spectroscopic information: yellow dots are objects with higher posterior probabilities to belong to the star LOSVD (class 1), while black crosses are more likely to be drawn from the M87 galaxy/ICL distributions (class 2/3). Magenta triangles represent background galaxies. Cyan and orange triangles are objects with velocities consistent with being part of class 1 (orange) or class 2/3 (cyan), however they have photometric probability, to be background galaxies. {\bf Bottom Panel:} Same as Top panel with the objects size scaled by its photometric probability to be a GC. The photometric information allows us to retrieve the fraction of objects with colors consistent with being GCs but with velocities within the stars LOSVD (yellow dots). Only 1\% of background galaxies (magenta triangles) is misclassified as GCs following the photometric classification. }
\label{fig:color_Star_GC}
\end{figure}

In Fig.~\ref{fig:color_Star_GC} the $(u^*-i')$ vs $(i'-K_{s})$ diagram encodes the spectroscopic information in the color: objects with higher GMM posterior probabilities to belong to the star LOSVD are shown in yellow and orange. From this plot we can see that the majority of these objects are consistent with being foreground stars on the basis of their $u^*i'K_{s}$ properties (yellow circles). However, a fraction of objects with $u^*i'K_{s}$ colors consistent with being GCs have velocities that fall within the stars LOSVD. Their total number is 32, i.e., $\sim$4\% of the sample of star candidates based on their kinematics. In the bottom panel of Fig.~\ref{fig:color_Star_GC}, where the color-color diagram is re-plotted by weighting the contribution of each data point by the XD probability, $\mathrm{p_{gc}}$, to be a GC, we show that we can successfully retrieve this fraction using the photometric classification previously described. 
Going back to the top panel of Fig.~\ref{fig:color_Star_GC}, we also see that there are few misclassified objects ($\sim$ 9\% of our working sample with $\mathrm{V_{LOS}\le 3000\, km\, s^{-1}}$) with high photometric probability, $\mathrm{p_{galx}}$ , to be background galaxies (cyan and orange triangles). These objects have been visually inspected and reintroduced in the bona-fide sample of GCs when their velocity is consistent with the Virgo IC/M87 halo components (cyan triangles, 85 objects). For those whose velocities fall within the star LOSVD (orange triangles, 77 objects), we only reintroduce those that are in the locus of the color-color diagram where the majority of GCs sits (4 objects). All the reintroduced objects have been assigned $\mathrm{p_{gc}}=1.0$. To summarize, we have shown that the photometric information gathered by the NGVS is enough to properly separate foreground stars and Virgo compact objects, allowing us to avoid hard cuts in the velocity distribution for $\mathrm{V_{LOS} < 500\, km\, s^{-1}}$.

\subsubsection{Virgo UCD contamination}
In the last 20 years evidence has emerged for the existence of a family of compact stellar systems, called ``ultra-compact dwarfs" (UCDs). Their sizes and luminosities are intermediate between compact elliptical galaxies and the most massive GCs \citep{hilker99,drinkwater00,hasegan05,misgeld11}, and their colors and stellar populations are similar to GCs and nuclear star clusters \citep{taylor10,spengler17}. Recently, \citet{zhang15} studied the properties of a sample of UCD galaxies within $\sim1$~deg of M87, and compared it to the properties of the red and blue GC populations. They found that they do not share the same kinematics as the GC system, however no distinction was made between galaxy halo and intra-cluster GCs, leading to the question as to whether UCDs may trace the IC component instead, hence sharing similar properties as the ICGCs. 
The evidence that the Virgo UCDs never reach the cluster dispersion, nor do they trace the Virgo systemic velocity, is enough to argue that these systems are not to be associated with the ICL. However, the argument gets stronger if we consider that at large radii the UCDs behave more closely to the red GCs (in terms of density profiles and rotational axis). As we will show in this work (see Sec.~\ref{kinematic_separation}), the presence of an IC component particularly affects the GC properties at large radii and it is dominated by the blue GC population (see Sec.~\ref{subsec:populations}). Thus, if the UCDs did trace the ICL they would rather show similar density distributions and kinematics as the blue GC population at large radii, that we do not observe. 

We then use the information gathered by \citet{zhang15,liu15a} to exclude the UCDs from our sample of bona-fide GCs. These authors identified UCDs based on their photometric properties, such as size and colors. By cross matching the catalogs from \citet{zhang15} and \citet{liu15a} with our sample of compact objects we identify 85 UCD candidates that we remove from our sample of bona-fide GCs. The selection of these objects is reliable, only missing $\sim6\%$ of genuine UCDs for $g' < 21.5$ \citep{zhang15,liu15a}. However, under the assumption that UCDs are the surviving nuclei of tidally stripped galaxies, \citet{ferrarese16} argued that we expect a contribution in the $21.5 < g' \le 24.0$ mag range that amounts to $\sim 100$ objects. Given our spectroscopic incompleteness (on average we only observe 10\% of the photometric candidates, see Sec.~\ref{density_p}), this implies a residual contamination by UCDs of $\sim 10$ objects in our spectroscopic sample. Such a fraction is negligible with respect to the final sample of GCs, and does not affect any of the scientific results that we present in this study.

\subsubsection{GCs bound to other Virgo members}
\label{GC_Virgo_memb}
In the previous sections we have subtracted the contribution of stars, background galaxies and Virgo UCDs from our working sample. However, as our goal is to study the GC population of the M87 halo and the Virgo IC component we also have to flag GCs that are likely to be bound to other Virgo members. To do this we considered GCs to belong to another galaxy if 1) The GC velocity relative to the galaxy velocity, $\mathrm{V_{g}}$, is within $\mathrm{3 \times \sigma_{g}}$, where  $\sigma_{g}$ is the velocity dispersion of the galaxy (assuming a Gaussian LOSVD, 99.7\% of the distribution falls within this limit), and 2) their distance from the galaxy's center is within 10 effective radii,  $\mathrm{R_e}$, of the galaxy. This generous radial criterion was chosen to account for possible differences between the galaxy light and the GC spatial distribution.  Fits to the surface brightness and surface density profiles of galaxies' light and GC systems show that the latter are, on average, more extended than that of the host galaxies' \citep[e.g.,][]{puzia04,kartha14,forbes17,hudson18}. For less massive, dwarf galaxies the scaling relation can be reduced to $\mathrm{R_{e,GC}} = 1.5 \times \mathrm{R_{e,gal}}$ \citep[e.g.,][]{peng16}.

Within a 2$^{\circ}$ radius around M87 there are 197 Virgo galaxies with known velocity, 129 of which are early-type dwarf galaxies, as compiled by the GOLDMine project \citep[but see also \citet{toloba14},][]{gavazzi03}. The GOLDMine catalog also provided us the velocity dispersion for a fraction of galaxies, while the effective radius information was retrieved from \citet{mcdonald11}. For those galaxies with no measured velocity dispersion and/or effective radius, mostly dwarfs, we have given a fixed value of $\mathrm{\sigma_{g} = 50\, km\, s^{-1}}$ and $\mathrm{R_{e} = 30\arcsec}$. At the end of this procedure we have identified 56 GCs as belonging to other Virgo galaxies. Finally, we are aware that there will be a fraction of galaxies in the surveyed area with no velocity information, hence that is not considered in this analysis. However, we note that additional GCs bound to other galaxies would show a correlation in position and velocity that we checked for. This test showed no evidence of any possible spatial/kinematic correlation, thus we can state that if there is a contribution from GCs belonging to other nearby galaxies in our bona-fide sample of objects, it is negligible.

\begin{figure}[!t]
\centering
\includegraphics[trim=0.2cm 0.cm 0.cm 0cm,clip,width = 8.5cm]{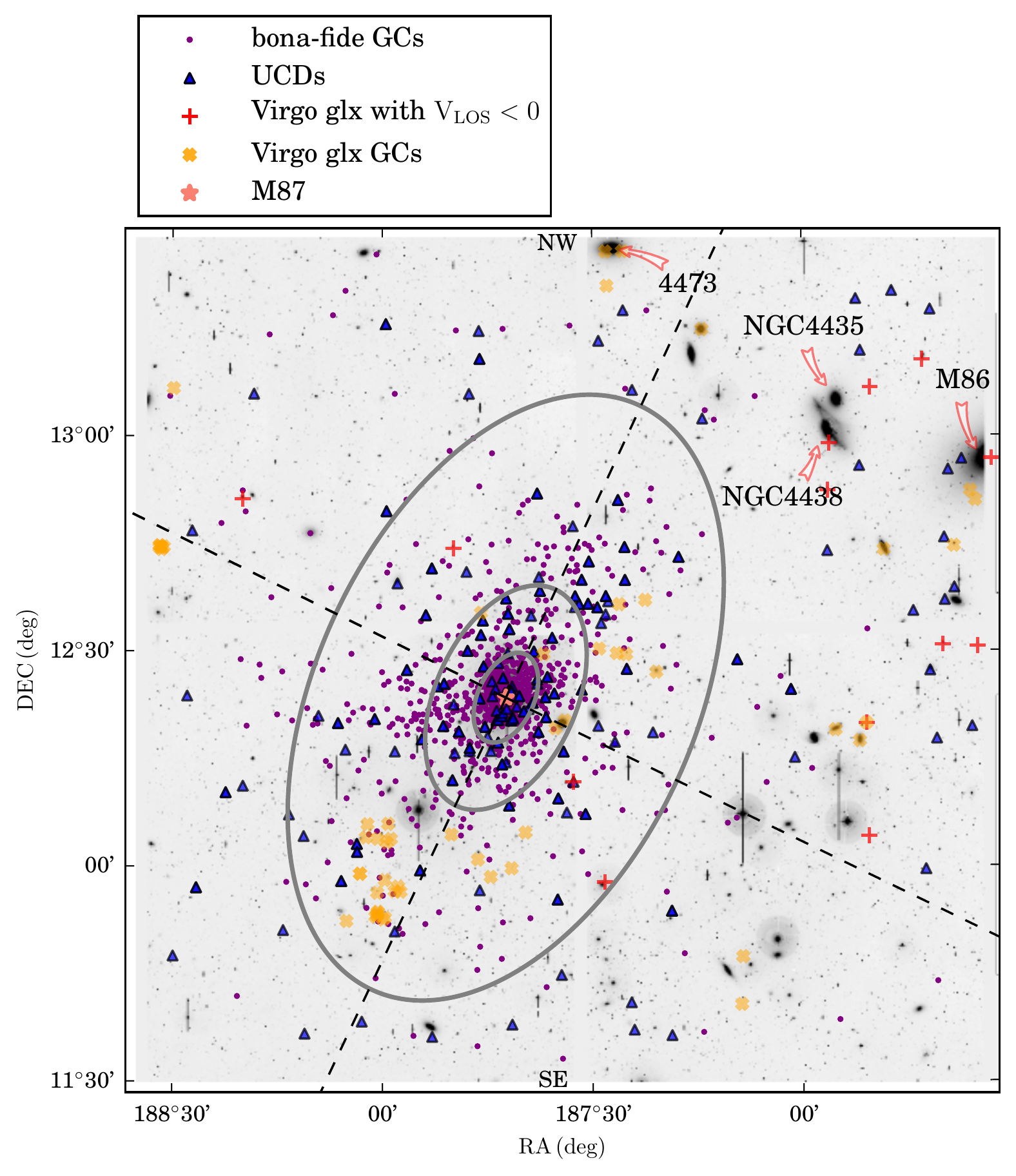}
\caption{Spatial distribution of GCs in Virgo with spectroscopic information (purple dots) superimposed on to the g-band NGVS image of the Virgo core. Blue triangles show the position on the sky of UCDs, while orange crosses show GCs that we associated to other Virgo galaxies. Aside M87 (salmon star), we highlight the position of other four main galaxies as given in the plot. Red crosses identify Virgo galaxies with negative $\mathrm{V_{LOS}}$. The gray ellipses, oriented as the M87's isophotes (major and minor-axes inclination as depicted by the dotted lines), divide the sky in four regions. North is up, East is to the left.}
\label{fig:GC_spatial_dist}
\end{figure}

To summarize, in this section we have analyzed and removed the contribution from contaminants in our working sample to define our final sample of bona-fide GCs.
We have found that:
\begin{enumerate}
\item Background galaxies (i.e. no Virgo members with $\mathrm{V_{LOS} > 3000\, km\, s^{-1}}$) represent the $\sim 28\%$ of our working sample;

\item Once the background contamination is subtracted, foreground MW stars contribute $\sim$ 46\% of the data, leaving 975 sources with high photometric probability, $\mathrm{p_{gc}}$ to be a compact Virgo object;

\item Among the compact objects, 85, i.e. 9\%, are identified as UCDs;

\item Finally, 56 GCs (6\% of the total GC sample) are consistent with belonging to other Virgo members.
\end{enumerate}

This selection led to a final sample of 837 confirmed GCs that we will analyze in the next Sections. Their spatial distribution is shown in Fig.~\ref{fig:GC_spatial_dist} (purple dots) together with the position on the sky of UCDs (blue triangles) and GCs associated to other Virgo members (orange crosses; the centers of other Virgo galaxies are signaled by red crosses).

\section{The M87 galaxy halo and Virgo ICL as traced by globular clusters}
\label{kinematic_separation}

When studying the light at the center of a cluster reaching out $\sim 300\, \mathrm{kpc}$ ($1^{\circ}$) in radius from the central galaxy we are tracing the transition region between the central galaxy halo and the IC component.  \citet{longobardi15a} analyzed a 0.5 $\mathrm{deg^{2}}$ area around M87 using PN kinematics to dynamically identify and separate the galaxy halo and the ICL. They showed that the two components overlap at all radii, but with different PN properties, hence consistent with having different parent stellar populations.

In what follows, we will investigate whether GCs trace a similar galaxy halo--IC interface. If there are, indeed, ICGCs, in what systems did they form and do they show photometric properties that differ from the M87 halo GCs?

\subsection{Projected GC velocity phase-space}
\label{sec:GC_PSpace}
\begin{figure*}
\centering
\includegraphics[trim=0.cm 0.cm 0.cm 0cm,clip,width = 15cm]{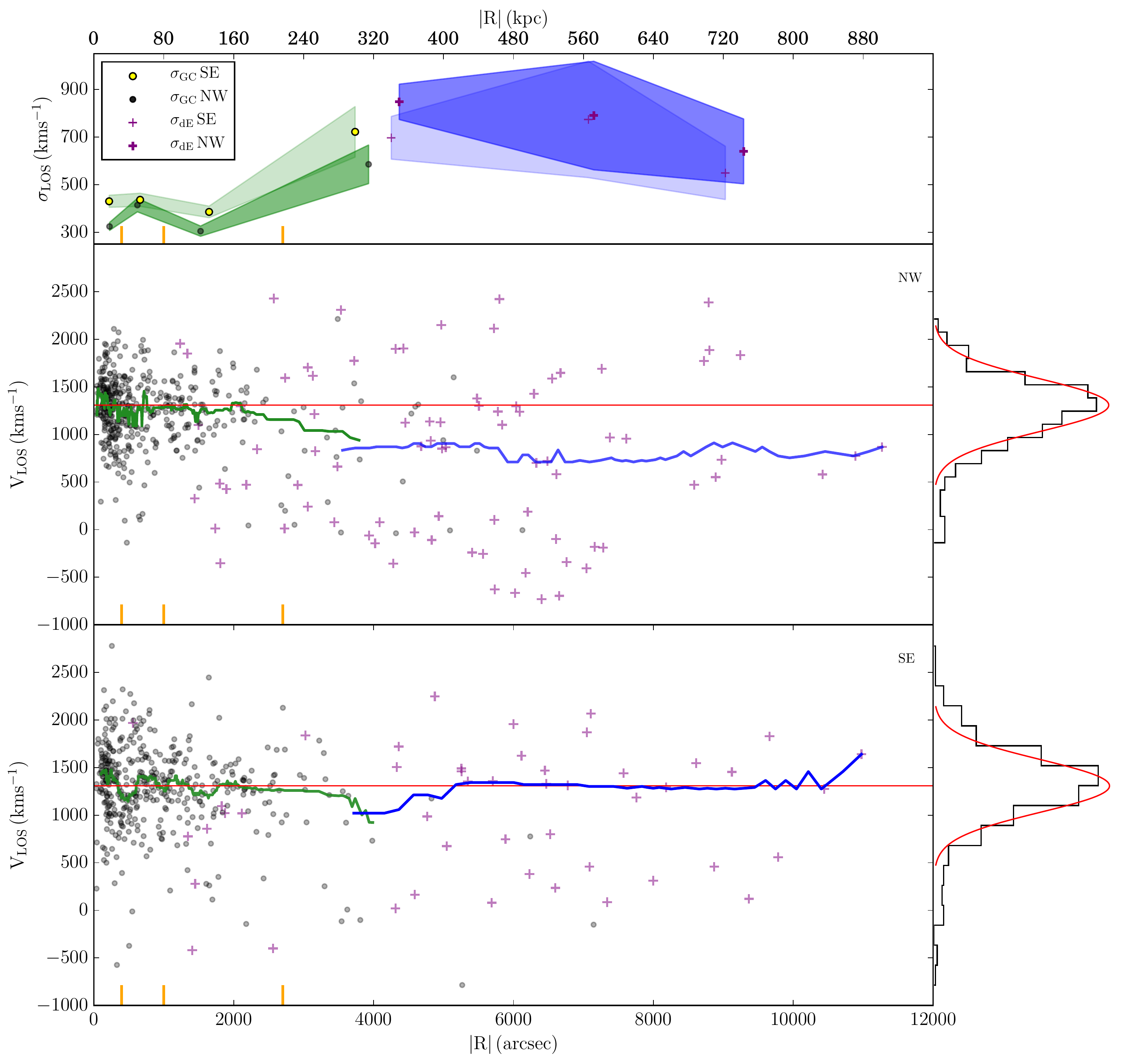}
\caption{{\bf Top Panel:} LOS velocity dispersion profile for the GCs in the NGVS pilot region (dots with the green shaded area delimiting the 1 $\sigma$ region) and for the dwarf early-type galaxies within a $2^{\circ}$ radius from the cluster's center (purple crosses with blue shaded area delimiting the 1 $\sigma$ region). The GC velocity dispersion profile is observed to increase on both the NW (dark dots and dark shaded area) and SE (yellow dots and light shaded area) sides of M87 reaching the velocity dispersion of the dwarfs in Virgo. {\bf Central Panel:} Projected velocity phase-space that compares the velocity distributions of GCs (black dots) and Virgo dwarf early-type galaxies (purple crosses) in the NW region. The continuous lines trace the running average computed from the total GC sample (green) and the dwarf population (blue). The histogram of the velocities, plotted on the side, shows a LOSVD with strong and asymmetric tails compared to a Gaussian distribution centered on the M87 systemic velocity with a velocity dispersion of $\sim 300\, \mathrm{kms^{-1}}$, and normalized to the total number of GCs (red curve).  {\bf Bottom Panel:} Same as central panel, however the relation is shown for the SE region with respect to M87. On both sides the GC mean velocity decreases as function of the distances, deviating from the M87 systemic velocity ($\mathrm{V_{sys}=1308\, km\, s^{-1}}$, red line) and merging into the mean velocity of the Virgo dwarfs at $\mathrm{|R|\sim320\, kpc}$. The vertical orange lines separate the phase-space in four elliptical bins, as given in the text.}
\label{fig:GC_pspace}
\end{figure*}

In Fig.~\ref{fig:GC_pspace} we show the velocity--position phase space diagram (central and bottom panels) for our GC sample. The GC LOS velocities are plotted against the major axis distance, the latter computed as $\mathrm{R=[(x/(1-e))^{2}+y^{2}}]^{1/2}$ for a system aligned along the y-axis, with position angle $\mathrm{P.A.}=155.0^{\circ}$, and ellipticity $\mathrm{e}=0.4$ \citep{ferrarese06,janowiecki10}. The relations for 
positive and negative $\mathrm{R}$ are plotted separately to trace the North-West (NW) (central panel) and South-East (SE) (bottom panel) halves with respect to the center of M87, respectively (see Fig.\ref{fig:GC_spatial_dist} for a representation of the spatial configuration).

This plot shows a system dominated by random motion, with the GC kinematics concentrated around the M87 systemic velocity $\mathrm{V_{sys}=1308\, km\, s^{-1}}$, together with a scattered distribution of high- and low-velocity objects on both sides of the galaxy. As shown in the histograms of the velocities (side panels), the wings (or tails) of the GC LOSVDs are strong and quite asymmetric with respect to a Gaussian distribution (red curve) centered on the M87 systemic velocity and with a velocity dispersion of $\sigma \sim 300\, \mathrm{kms^{-1}}$ \footnote{Here we consider as velocity dispersion representative of M87 the one traced by the red GC population over a range 0\farcm5-30.0\arcmin \citep{zhang15}}. 
Such behavior contrasts with what is observed for non-rotating early-type galaxies, whose LOSVD are symmetric (i.e. $h_{3} \sim 0$), hence well described by a single Gaussian to within few percent \citep{gerhard93,bender94}. Orbital anisotropy would create a symmetric variation around the mean velocity (i.e. $h_{4} \ne 0$), and is unlikely to be measured at all radii (see Fig.~\ref{fig:GMM_bin} for a plot of the GC LOSVDs as function of the distance from the M87's center). On the other hand, strong tails are seen in simulations, observed to characterize the LOSVD of galaxies in dense environments, and consistent with the presence of an intra-cluster/group component overlapping with the galaxy halo \citep{dolag10,longobardi15a,barbosa18,veale17}. In fact, the relative importance of these ``tails'' increases as function of the major axis distance, with the GCs gradually deviating from the central galaxy's kinematics and merging into the cluster potential. 
This is shown in the same plot by comparing the GC kinematics with the kinematics of the population of early-type dwarfs in Virgo with measured radial velocities \citep[129 objects,][]{gavazzi03}. We study the \textit{running average} of the LOS velocities of GCs (continuous green lines in the central and bottom panels) and early-type dwarfs (continuous blue lines),  i.e., we plot the running mean in bins of $n$ objects, with $n=30$ for the GCs, $n=35$ for the NW dwarfs, and $n=25$ for the SE dwarfs. By comparing the different running average profiles it is clear that the GC mean velocity decreases going towards larger distances, deviating more than 2.5 $\sigma$ at large distances and on both sides from the M87 systemic velocity (red line) and merging into the mean velocity of the Virgo dwarfs at $\mathrm{|R|\sim 320\, kpc}$. At these distances the GC velocity dispersion reaches its maximum, $\mathrm{\sigma_{LOS}=590\pm 80\, km\, s^{-1}}$, merging with the cluster potential (Fig.~\ref{fig:GC_pspace}, top panel). The comparison between the NW and SE fields around M87 shows that the NW has a more dE galaxies with negative velocities (red crosses in Fig.~\ref{fig:GC_spatial_dist}). It is likely that a substantial fraction of these objects comes from the accretion of material that is falling towards the cluster's center dragged along by the well known infalling M86 subgroup, and it is contributing to what we see today as an IC component.

\subsection{Kinematic separation between M87 halo and ICL}
The comparison with the Virgo dE/dS0 galaxies showed that the GC system around M87 is tracing the transition toward the Virgo cluster potential. Hence, we now further analyze the GC LOSVD, and we do it i) separately for the NW and SE halves, and ii) in four different elliptical bins, or stripes in phase-space (gray ellipses in Fig.~\ref{fig:GC_spatial_dist}, or orange lines in Fig.~\ref{fig:GC_pspace}), covering the entire GC range. The four regions in space are defined as $\mathrm{|R| \leq 32\, kpc}$, $\mathrm{32 < |R| \leq 80\, kpc}$, $\mathrm{80 < |R| \leq 216\, kpc}$, and $\mathrm{|R| > 216\, kpc}$, with bin size increasing towards the outer regions to ensure a statistically significant sample of tracers even at large radii, where the GC number density is the lowest. Thus, our choice of radial binning represents the best compromise between similar radial extension and statistical noise. We emphasize that the stability of the results was tested against different choices of radial limits: within the uncertainties the results are consistent with each other. In our analysis, we assume that both the M87 halo and Virgo IC LOSVDs are Gaussians. As argued in the previous Section, this is a reliable approximation for the galaxy's stellar velocity distribution, and, from hydrodynamical simulations \citep{dolag10,cui14}, Gaussian shapes also describe well the IC component. We are aware, however, that a Gaussian may be a less good model for the Virgo IC LOSVD if the Virgo cluster is not fully virialized. 

\begin{figure*}[!htbp]
\small
\centering
\includegraphics[width = 8.5cm]{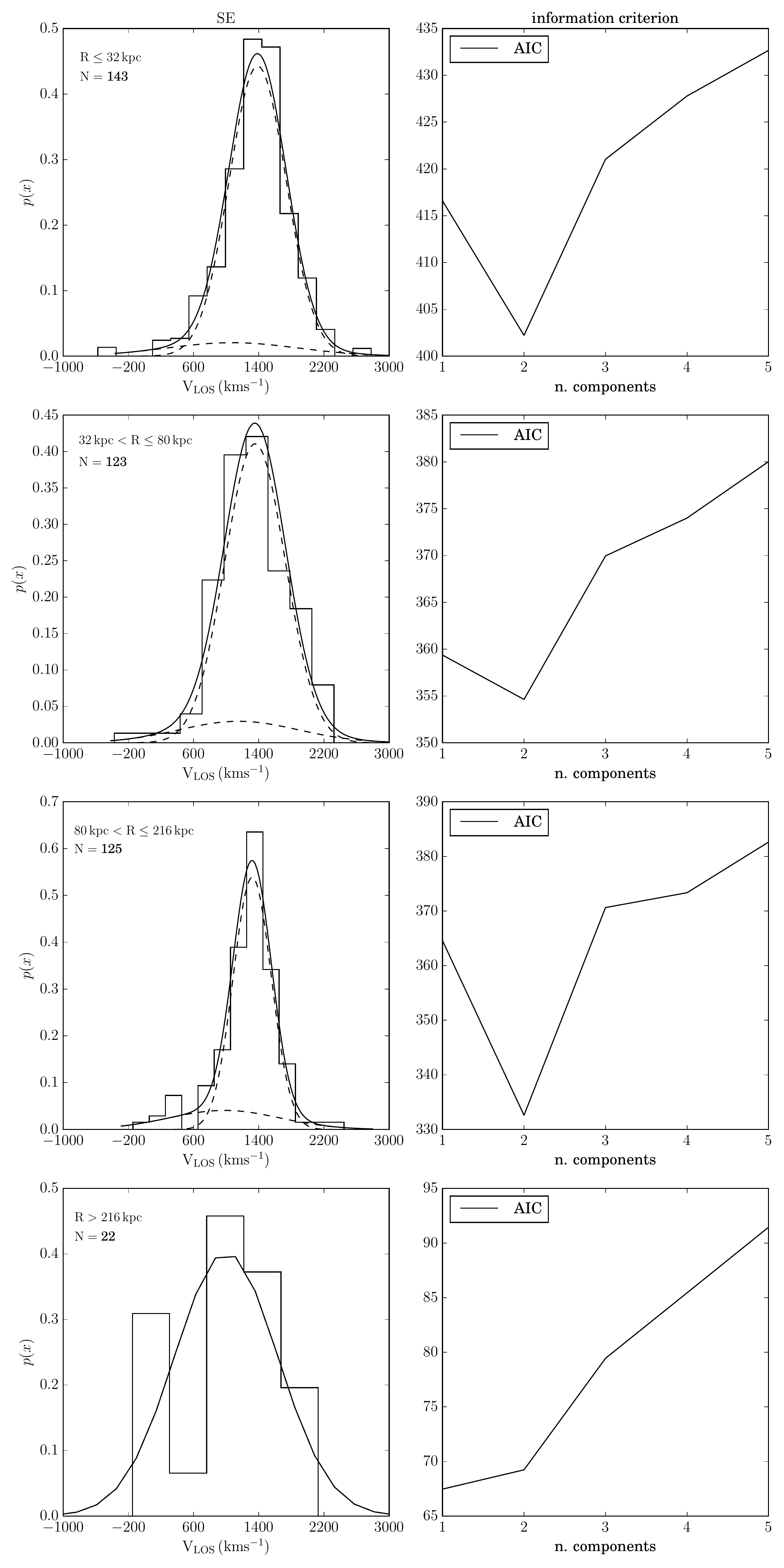}
\includegraphics[width = 8.5cm]{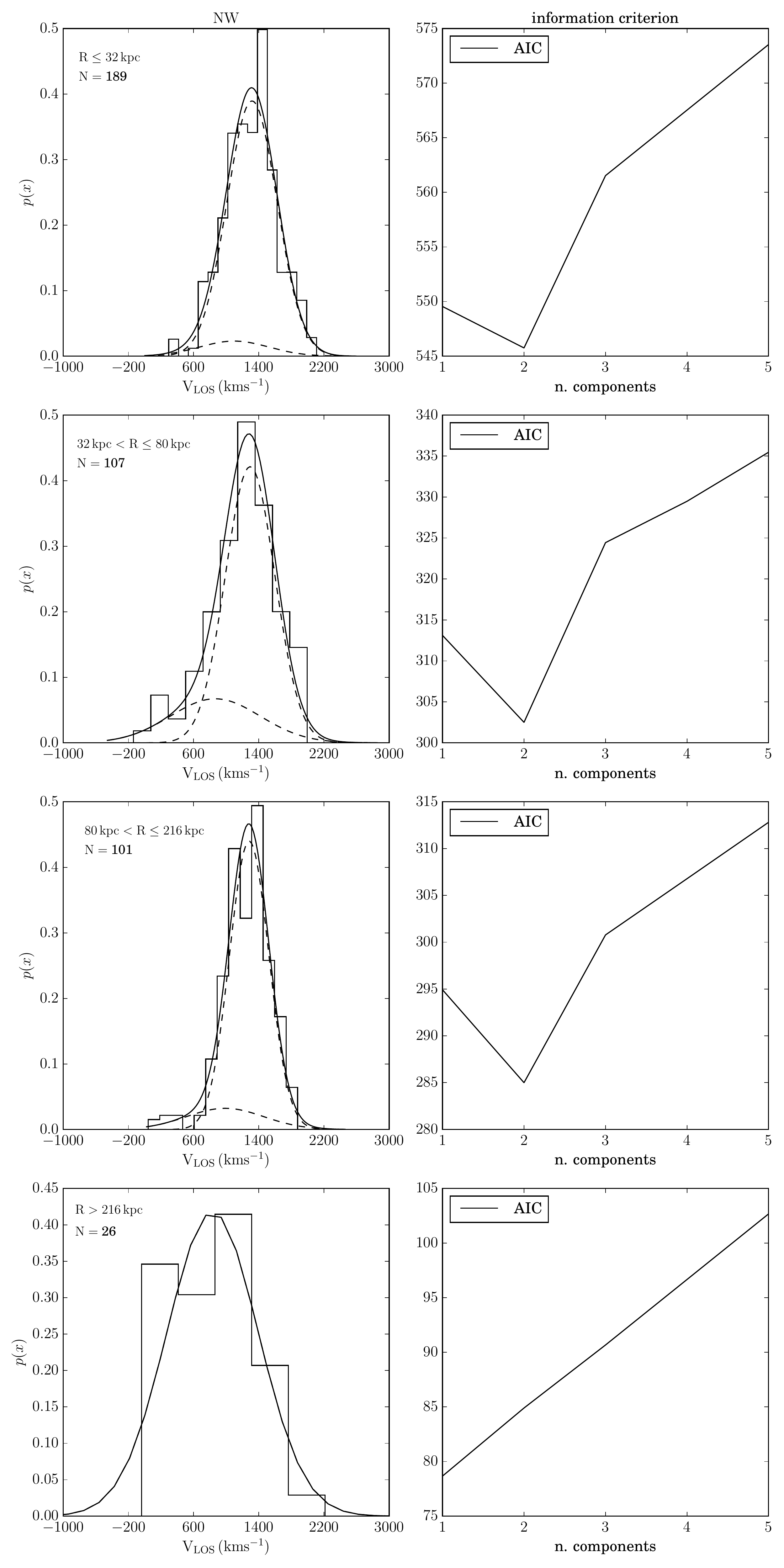}
\caption{GMM analysis of the GC LOSVD carried out separately for the GCs located to the SE (Left Panels) and NW (Right Panels) of M87. From top to bottom the sample is divided in four elliptical annuli as given in Fig.~\ref{fig:GC_pspace}, with $\mathrm{N}$ representing the total number of GCs used in the GMM fitting.  A AIC analysis (2nd and 4th panels) prefers a two components Gaussian mixture out to  $\mathrm{R = 216\, kpc}$, implying the contribution from the M87 galaxy halo and Virgo ICL to the total LOSVD at all distances out to this radius. For $\mathrm{R > 216\, kpc}$ the best fit mixture is a model with one component tracing the Virgo cluster potential.}
\label{fig:GMM_bin}
\end{figure*}

\begin{table*}[!htbp]
\renewcommand{\thetable}{\arabic{table}}
\centering
\small
\caption{Fitting parameters for the GMM best fit models for the M87 halo and Virgo ICL component and separately for the SE and NW regions} \label{tab:GMM_out}
%\resizebox{1.\textwidth}{!}{
\label{GMM_parameter}
\centering
\begin{tabular}{cccccc}
\tablewidth{0pt}
\hline
\hline
 & $\rm{SE_{M87, halo}}$ & $\rm{NW_{M87,halo}}$ & & $\rm{SE_{ICL}}$ & $\rm{NW_{ICL}}$\\
\hline
 & & & $\mathrm{|R| \leq 32\, kpc}$ & & \\
\hline
$\mu$& $1385\pm30\, \rm{km\, s^{-1}}$ &$1317\pm22\, \rm{km\, s^{-1}}$ &  & $1074\pm403\, \rm{km\, s^{-1}}$ & $1098\pm300\, \rm{km\, s^{-1}}$ \\
$\sigma$& $350\pm21\, \rm{km\, s^{-1}}$ &$307\pm16\, \rm{km\, s^{-1}}$ & & $806\pm285\, \rm{km\, s^{-1}}$ & $424\pm212\, \rm{km\, s^{-1}}$ \\
 $\mathrm{W}$& $0.90\pm0.07$ & $0.90\pm0.07$& & $0.10\pm0.04$ & $0.10\pm0.07$ \\
\hline
 & & & $\mathrm{32 < |R| \leq 80\, kpc}$ & & \\
\hline
$\mu$&$1355\pm34\, \rm{km\, s^{-1}}$ &$1293\pm31\, \rm{km\, s^{-1}}$ & & $1153\pm421\, \rm{km\, s^{-1}}$ &$865\pm159\, \rm{km\, s^{-1}}$  \\
$\sigma$& $371\pm24\, \rm{km\, s^{-1}}$ &$304\pm22\, \rm{km\, s^{-1}}$ & & $729\pm287\, \rm{km\, s^{-1}}$ & $551\pm112\, \rm{km\, s^{-1}}$ \\
 $\mathrm{W}$&$0.88\pm0.08$ & $0.77\pm0.7$& & $0.12\pm0.07$ &$0.22\pm0.06$  \\
\hline
 & & & $\mathrm{80 < |R| \leq 216\, kpc}$ & & \\
\hline
$\mu$& $1319\pm22\, \rm{km\, s^{-1}}$ &$1282\pm24\, \rm{km\, s^{-1}}$ & & $977\pm182\, \rm{km\, s^{-1}}$ & $986\pm237\, \rm{km\, s^{-1}}$ \\
$\sigma$&  $234\pm16\, \rm{km\, s^{-1}}$ &$241\pm17\, \rm{km\, s^{-1}}$ & & $680\pm129\, \rm{km\, s^{-1}}$ &$474\pm167\, \rm{km\, s^{-1}}$\\
 $\mathrm{W}$& $0.82\pm0.08$ &$0.87\pm0.09$ & & $0.18\pm0.05$ &$0.12\pm0.06$  \\
\hline
 & & & $\mathrm{|R| > 216\, kpc}$ & & \\
\hline
$\mu$& -- &-- & &  $999\pm135\, \rm{km\, s^{-1}}$ &  $833\pm109\, \rm{km\, s^{-1}}$\\
$\sigma$& -- &-- & & $634\pm95\, \rm{km\, s^{-1}}$ & $559\pm77\, \rm{km\, s^{-1}}$ \\
 $\mathrm{W}$&-- &-- & & $1.0\pm0.2$ & $1.0\pm0.2$\\

\hline

\end{tabular}
\end{table*}
We then run a GMM analysis for each of these samples to isolate individual kinematic structures and trace the coexistence of the M87 halo/Virgo IC components. The GMM is a probabilistic model which assumes that a distribution of points, in our case the objects' velocities, can be described as a linear combination of $K$ independent Gaussian probability density functions (PDFs). The best set of Gaussians is selected by implementing the expectation-maximization (EM) algorithm, an iterative process that continuously updates the PDF parameters until convergence in the likelihood is reached. The process ends with the assignment of posterior probabilities, $\Gamma$, that each data point belongs to each of the $K$ Gaussians. This analysis is based on the public scientific Python package \textit{scikit-learn} \citep{pedregosa11}. However, our estimated Gaussian mixture distribution starts from a weighted sample of data, the weights, $\mathrm{p_{gc}}$, being the normalized probabilities to be a GC given its photometric properties (see Sec.~\ref{back_contamination}). We then modified the python GMM routine accordingly, and the details on this will be given in Longobardi et al. (2018, in preparation). Here we only point out that the EM for mixture models consists of two steps. The first step, or E step, consists of calculating the expectation of the component assignments, $\Gamma_{i,k}$  for each data point given the model  parameters. Thus, at the $m^{th}$ iteration of the EM process the posterior probabilities of the $i^{th}$ particle to belong to the $k^{th}$ Gaussian component become:
\begin{displaymath}
\Gamma_{i,k}^{m}=\frac{p_{k}(x_{i}\, |\, \mu_{k},\sigma_{k})P_{k}^{m}}{ \sum_{k=1}^{K}{{p_{k}(x_{i}\, |\, \mu_{k},\sigma_{k})P_{k}^{m}}}
}\times \mathrm{p_{gc,i}}
\end{displaymath}
where the Gaussian density function (PDF) is written as $p(x_{i}) = \sum_{k=1}^{K}{{p_{k}(x_{i}\, |\, \mu_{k},\sigma_{k})P_{k}}}$, with $p_{k}(x\, |\, \mu_{k},\sigma_{k})$ being the individual mixture component centered on $\mu_{k}$, with a dispersion $\sigma_{k}$, and mixture weight $P_{k}$. The second step, or M step, consists of maximizing the expectations parameters, i.e., in updating the values $\mu_{k}$, $\sigma_{k}$, and $P_{k}$, as weighted parameters, the weights being $\Gamma_{i,k}$. In this way the EM process always takes into account the starting weight of each particle $\mathrm{p}_{\mathrm{gc},i}$. Tests with the AIC show that the mixture that best describes the total GC LOSVD is a two component model in the first three elliptical bins, while in the outermost bin the GMM identifies only one component.  A comparison of the performances of AIC and BIC scores was tested against mock sets of data that mimic different LOSVDs and sample sizes. AIC scores were found to be more reliable, hence we only report these scores. Moreover, the stability of the fitted parameters were tested with 1000 GMM runs for different mock data sets and initialization values. 

The results of our GMM analysis, the histograms of the data together with the GMM best-fit model, and relative AIC scores are shown in Fig.~\ref{fig:GMM_bin}, while we summarize the fitting Gaussian parameters in Table~\ref{tab:GMM_out}.

\begin{figure*}
\centering
\includegraphics[trim=0.cm 0.cm 0.cm 0cm,clip,width = 15cm]{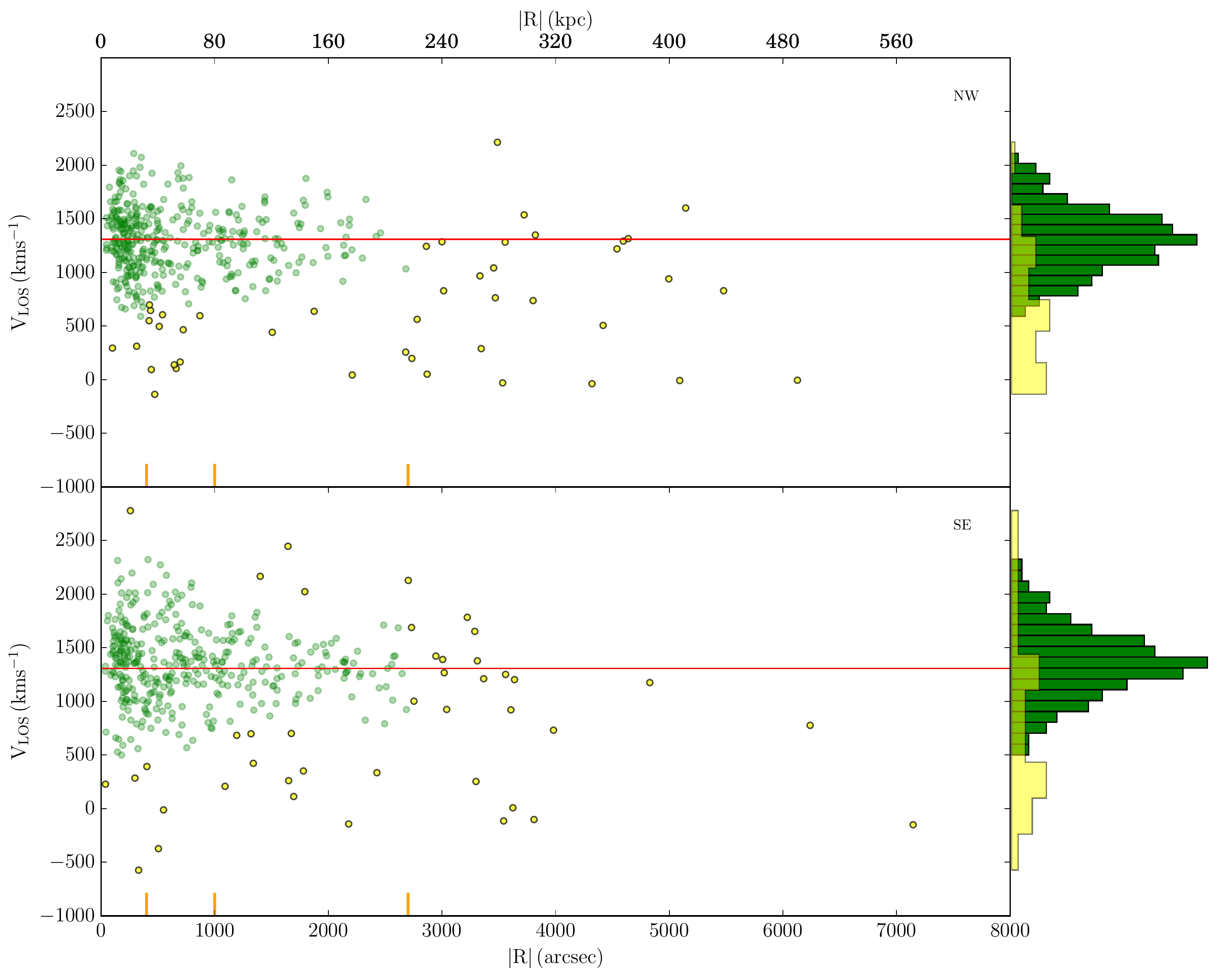}
\caption{{\bf Top Panel:} Projected velocity phase-space that compares the velocity distributions of the M87 halo GCs (green dots) and Virgo ICGCs (yellow dots) in the NW region. The histograms of the velocities, plotted on the side, compare the GC LOSVDs associated to the two dynamical components (green is M87 halo and yellow is ICL).  {\bf Bottom Panel:} Same as top panel, however the relation is shown for the SE region with respect to M87. The M87 systemic velocity ($\mathrm{V_{sys}=1308\, km\, s^{-1}}$) is plotted as red line. The vertical orange lines separate the phase-space in four elliptical bins, as given in the text.}
\label{fig:GC_pspace_separation}
\end{figure*}

The GMM analysis shows that out to 216 kpc there is the coexistence of a hotter (broader) and colder (narrower) component, while outside this radius the
narrower Gaussian is not detectable, and the GC kinematics are well described by one single, broad Gaussian. 
Within the uncertainties the narrow Gaussians describe the galaxy's kinematics with a mean, systemic velocity consistent with $\mathrm{V_{LOS} = 1308\, km\, s^{-1}}$, and with a velocity dispersion that decreases as function of the distance, reaching a value of $\mathrm{\sigma_{LOS} \sim 230\, km\, s^{-1}}$ in the outermost bin.
The broad Gaussians, instead, are consistent with the kinematics traced by the early-type dwarf population in the core of Virgo, the latter measured to peak at $\mathrm{V_{dE+dS0} = 1139\pm 67\, km\, s^{-1}}$, with a velocity dispersion of $\mathrm{\sigma_{dE+dS0} = 649\, km\, s^{-1}}$ \citep{binggeli93}. The mean velocities and velocity dispersions of the ICGC component are consistent not only with those of the dwarf galaxies, but also with those of the more massive systems in the Virgo core \citep{binggeli93}. However, i) very broad, asymmetric wings, as the one we show in Fig.~\ref{fig:GC_pspace} and Fig.~\ref{fig:GMM_bin}, are only observed for the LOSVD of the dwarf spheroidal galaxies \citep[see][]{binggeli93,conselice01}, and ii) in the next Sections we will show that the photometric properties of the detected ICGCs are similar to what is measured for the dE system in Virgo, hence our comparison with the lower mass Virgo galaxies. The broad GC LOSVD component points toward a different kinematic behavior when measured separately in the two halves: SE of M87 the velocity dispersion is larger than in the NW region. However, more kinematic tracers will be needed to reduce the uncertainties associated with these results (see Table~\ref{tab:GMM_out}). For clarity, the NW and SE GC velocity phase-spaces, together with the LOSVDs of the M87 halo and Virgo intra-cluster GCs are shown in Fig.~\ref{fig:GC_pspace_separation}.

Different  LOSVDs  for  the  galaxy halo  and  the ICL  are  predicted  by
cosmological  analysis  of  structure  formation. At the center of simulated clusters the total LOSVD splits in two Gaussians characterized by very different velocity dispersions \citep{dolag10,cui14}. One component is gravitationally bound to the
galaxy and more spatially concentrated; the other is more diffuse
and its high velocity dispersion reflects the satellites' orbital distribution in the cluster gravitational potential. 
Using this framework, we can identify the narrow and the broad Gaussians traced by the GC kinematics with the M87 halo and Virgo ICL, the latter contributing $\sim$13\% of the GC sample within 216 kpc, beyond which the M87 halo component is no longer detected.

\begin{figure}
\centering
\includegraphics[trim=0.cm 0.cm 0.cm 0cm,clip,width = 8cm]{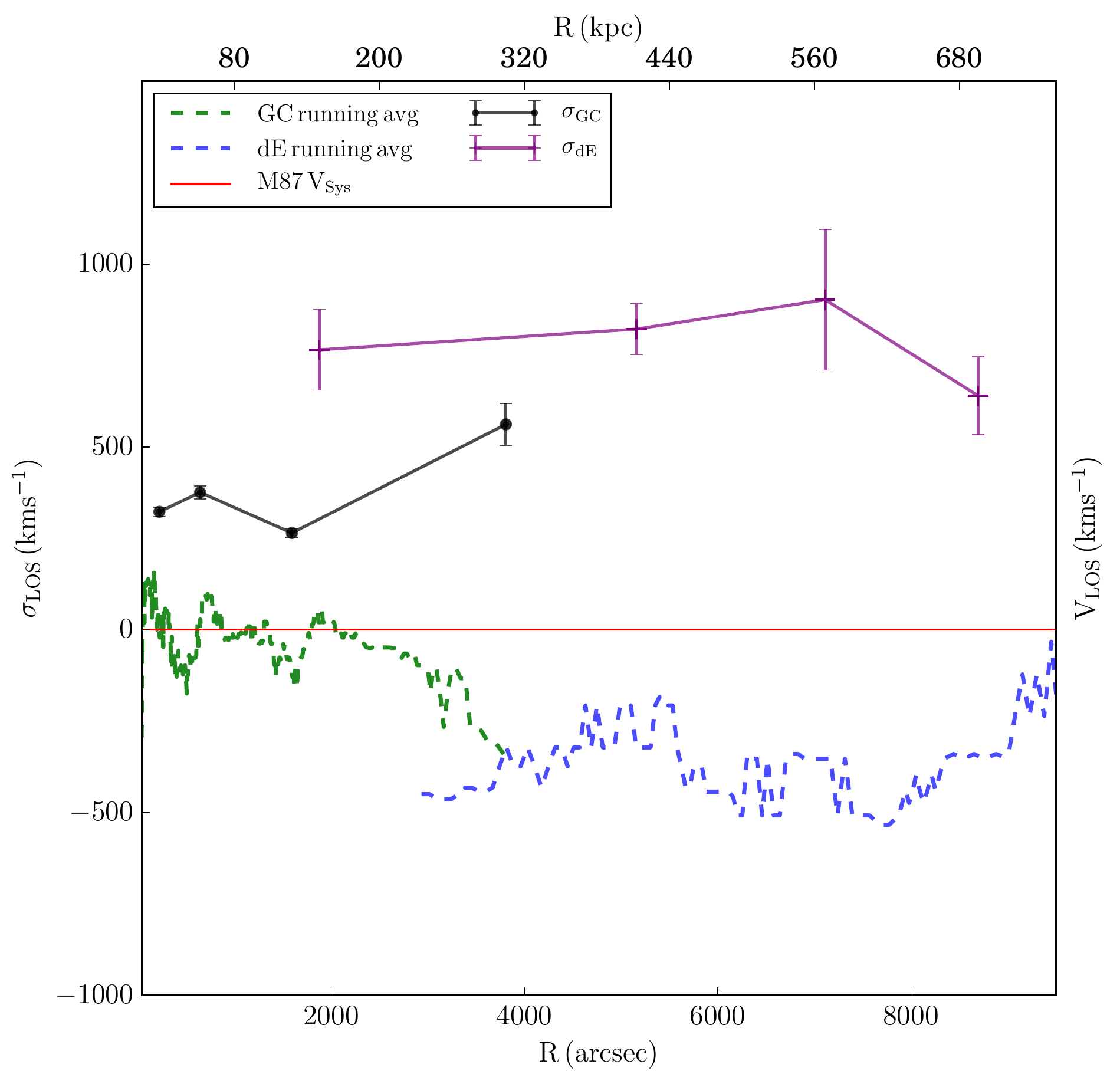}
\caption{Comparison between the LOS kinematics of the GCs in the NGVS pilot region  and of the dwarf early-type galaxies within a $2^{\circ}$ radius from the cluster's center. At a projected distance $\mathrm{R = 128\, kpc} $ the GC velocity dispersion (black dots) reaches its minimum and the mean velocity (green line) starts deviating from the systemic velocity of M87 (red line). Beyond this radius, the GC kinematics traces the IC potential: the GC velocity dispersion rises till reaching the velocity dispersion values of the Virgo dE (purple crosses) at $\mathrm{R\sim320\, kpc}$; at these distances the GC mean velocity merges into the mean velocity of the Virgo dwarfs (blue line). The mean velocity values are centered on $\mathrm{V_{Sys}=1308\, kms^{-1}}$. The error bars
show the uncertainties from Poissonian statistics.}  
\label{fig:GC_sigma_V}
\end{figure}

In Fig.~\ref{fig:GC_sigma_V} we plot again the comparison between the kinematics of the GC system and the Virgo dE population, however, this time, no separation between NW and SE fields is made. Interestingly, the kinematic behavior of the GC system exhibits a change in dynamics at a projected distance $\mathrm{R = 128\, kpc} $ where the velocity dispersion reaches its minimum (black dots), and the mean velocity (green line) starts deviating from the systemic velocity of M87 (red line). Thus, interpreting such a distance as the $\textit{truncation radius}$, $\mathrm{R_{T}}$, of the galaxy, would imply a total dark matter mass for M87, 
$\mathrm{M_{M87} = 4.6 \times 10^{13} M_{\odot}}$. This is a rough estimate, that uses the relation:

\begin{equation}
\rm{R_{T} = R_{Virgo} \times \left(\frac{M_{M87}}{3\times M_{Virgo}}\right)^{1/3}},
\end{equation}
were $\mathrm{R_{T} = 127.5\, kpc \times ((1-e)^{1/2})}$ is the average distance within which systems are gravitationally bound to the galaxy, $\mathrm{R_{Virgo}}$ is the distance between M87 and the center of the cluster \citep[we have assumed $\mathrm{R_{Virgo} = 1^{\circ}}$;][]{binggeli93}, and $\mathrm{M_{Virgo} = 4\times 10^{14} M_{\odot}}$ is the total Virgo dark matter mass \citep{mclaughlin99}.
Moreover, we can also speculate that the different velocity dispersions measured for the IC component, higher in the south than in the north, are tracing the un-relaxed nature of the Virgo cluster. Young galaxy clusters still in the process of assembly are expected to show signs of dynamically non-relaxed substructures.

\section{M87 halo and Virgo intra-cluster GC populations} 
\label{subsec:populations}
The statistical analysis we have carried out allows us to assign membership to either the M87 halo or the Virgo ICL. In this Section we further analyze the properties of the M87 galaxy and intra-cluster GCs that have spectroscopic information to investigate whether the different kinematic components also correspond to two different populations in terms of photometric properties and spatial distributions.

\subsection{Photometric properties}
\begin{figure*}[!htbp]
\begin{center}
\includegraphics[trim=0.cm 0.cm 0.cm 0.cm,clip, width=11cm]{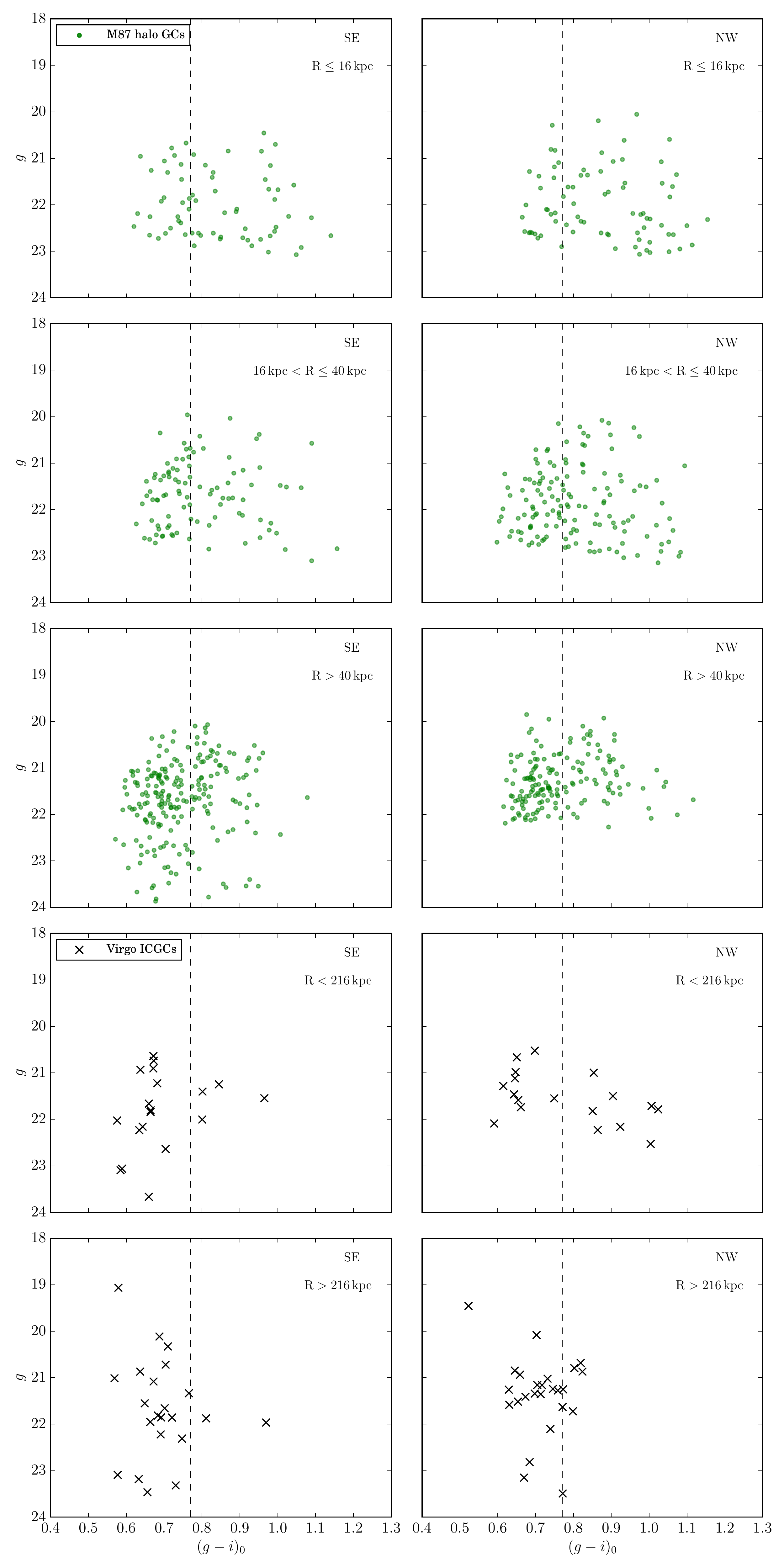}
\caption{ $(g'-i')_{0}$ vs $g'$ color-magnitude diagram for the M87 galaxy (green dots)
and the Virgo intra-cluster (black crosses) GCs, at increasing distances from the galaxy's center (see legend) and for the SE and NW halves around M87. As a consequence of the smaller number of objects, the ICGC color-magnitude diagrams are compared in two different elliptical bins, with $\mathrm{R = 216\, kpc}$ being the boarder. Both dynamical components have a blue and a red population of GCs. In the M87 halo the number of blue GCs increases as function of the distance: in the outermost bin it equals to $\sim 70\%$. }  
\end{center}
\label{fig:GC_RB}
\end{figure*}

Both the M87 halo and the Virgo intra-cluster GCs are characterized by a large spread in $(g'-i')_{0}$ colors, indicating that both dynamical components have two distinct, red (metal-rich) and blue (metal-poor), sub-populations of GCs. By fitting a double-Gaussian profile to the GC color distributions, we measure that the red and blue GC components cross at $(g'-i')_{0} = 0.77$~mag. The relative fraction of blue-to-red GCs differs, however: $77\%\pm10\%$ of the Virgo ICGC population is blue, while only $52\%\pm3\%$ of the GCs bound to the M87 halo are metal-poor\footnote{Such a red-to-blue cluster ratios are consistent with those we would obtain integrating the extrapolated number density profiles for the red/blue populations related to the two dynamical components (see Sec.\ref{density_p})}. In Fig.~\ref{fig:GC_RB}, we show the $(g'-i')_{0}$ vs.\  $g'_0$ color-magnitude diagram for both the M87 halo (green dots) and the Virgo intra-cluster GCs (black crosses), separately for the SE and NW fields, and as function of the distance from the galaxy's center. The M87 halo sample is compared in three different elliptical bins, while, given the lower number of tracers, the ICGCs are analyzed within two elliptical bins with $\mathrm{R = 216\, kpc}$ representing the boundary. Despite the incompleteness that characterizes our sample of spectroscopically confirmed GCs (higher for $\mathrm{R > 216\, kpc}$ and with the NW side affected the most), we see that the blue GC component of M87 is higher at larger distances, increasing from 40\% close to the center, to 70\% in the outermost bin. This is not true for the IC component that, independently from the distance, is dominated by the blue GCs.  

\begin{figure*}[!t]
\includegraphics[width = 10.5cm]{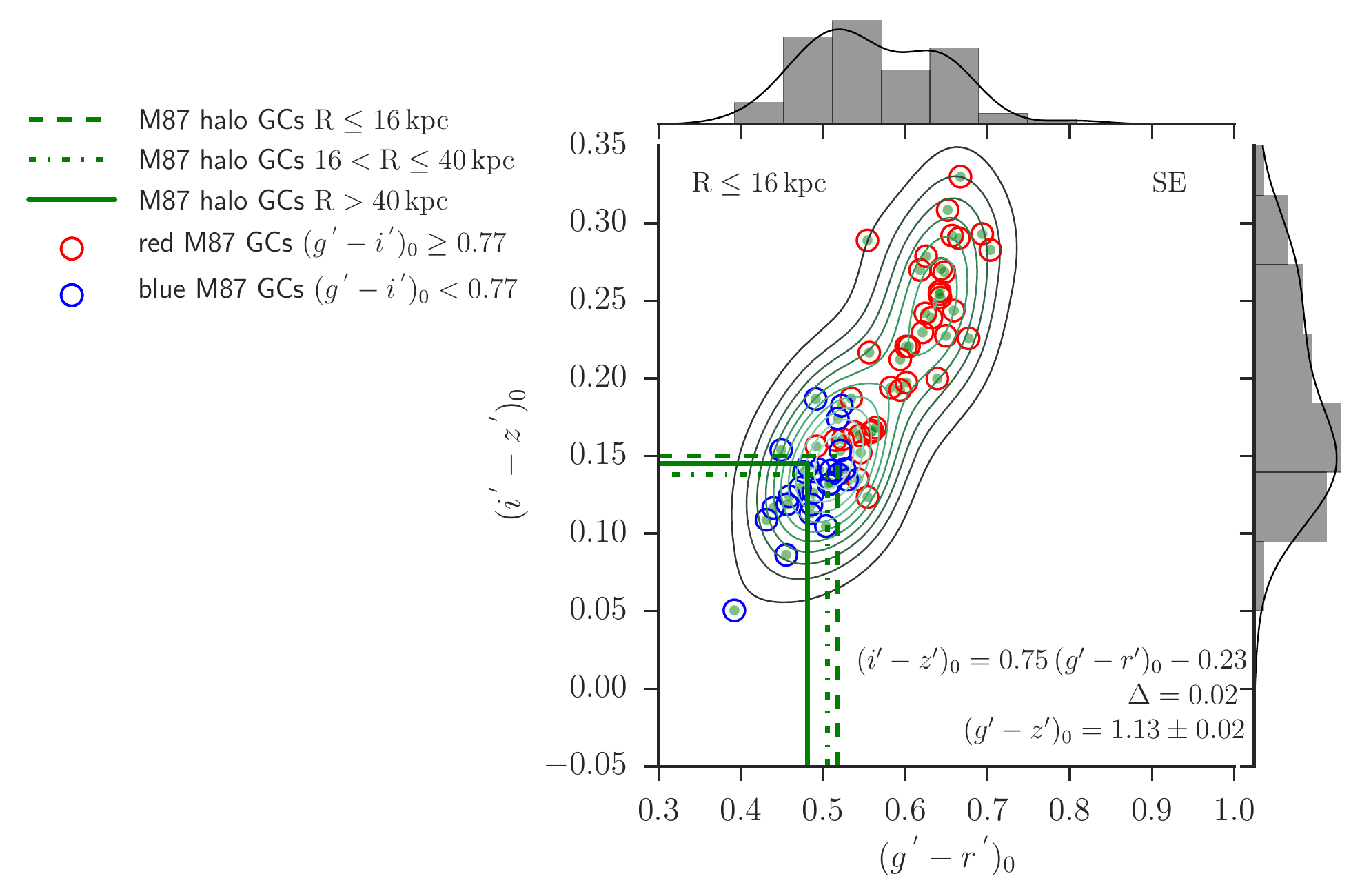}
\includegraphics[width = 6.9cm]{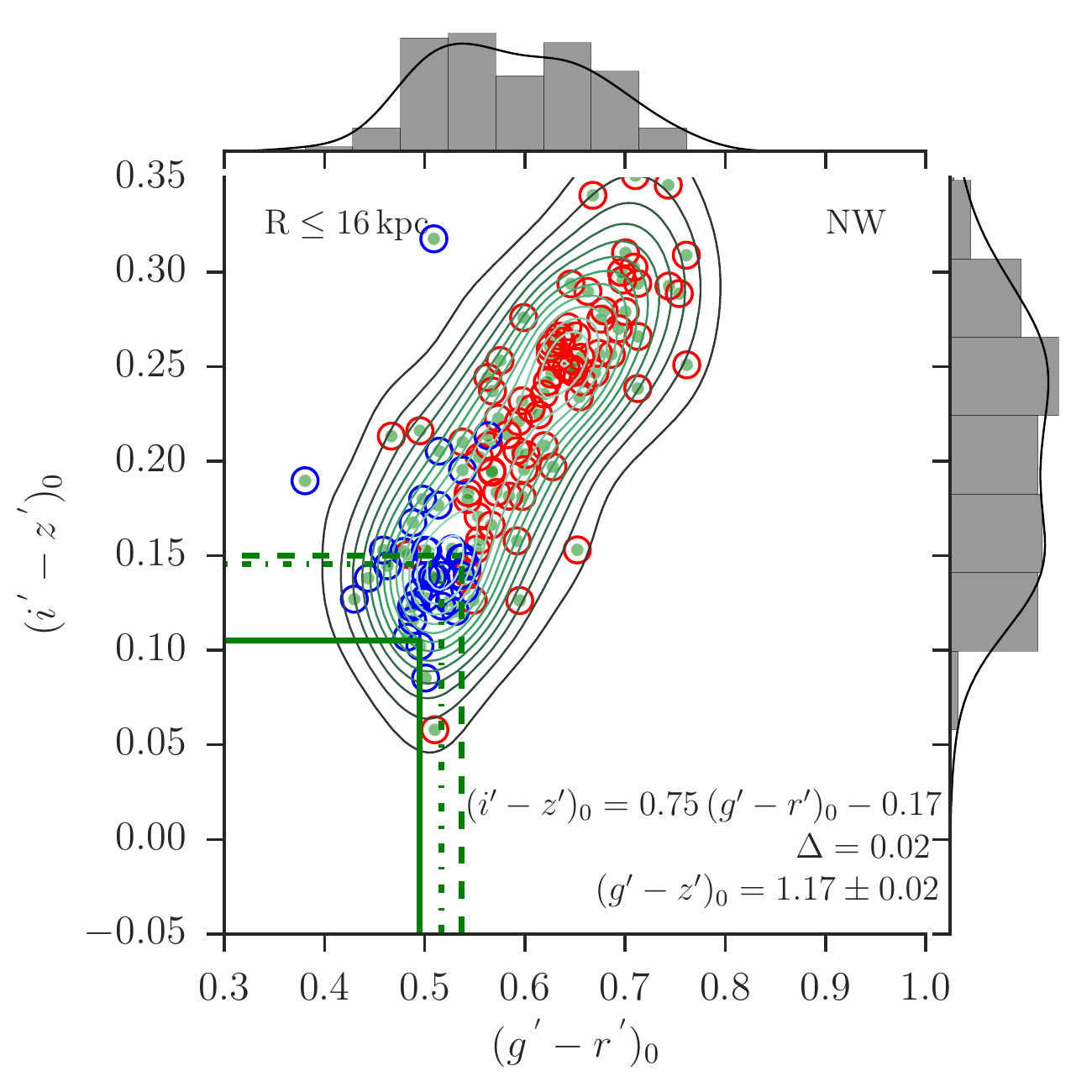}\\

\hspace{+3.5cm}
\includegraphics[width = 6.9cm]{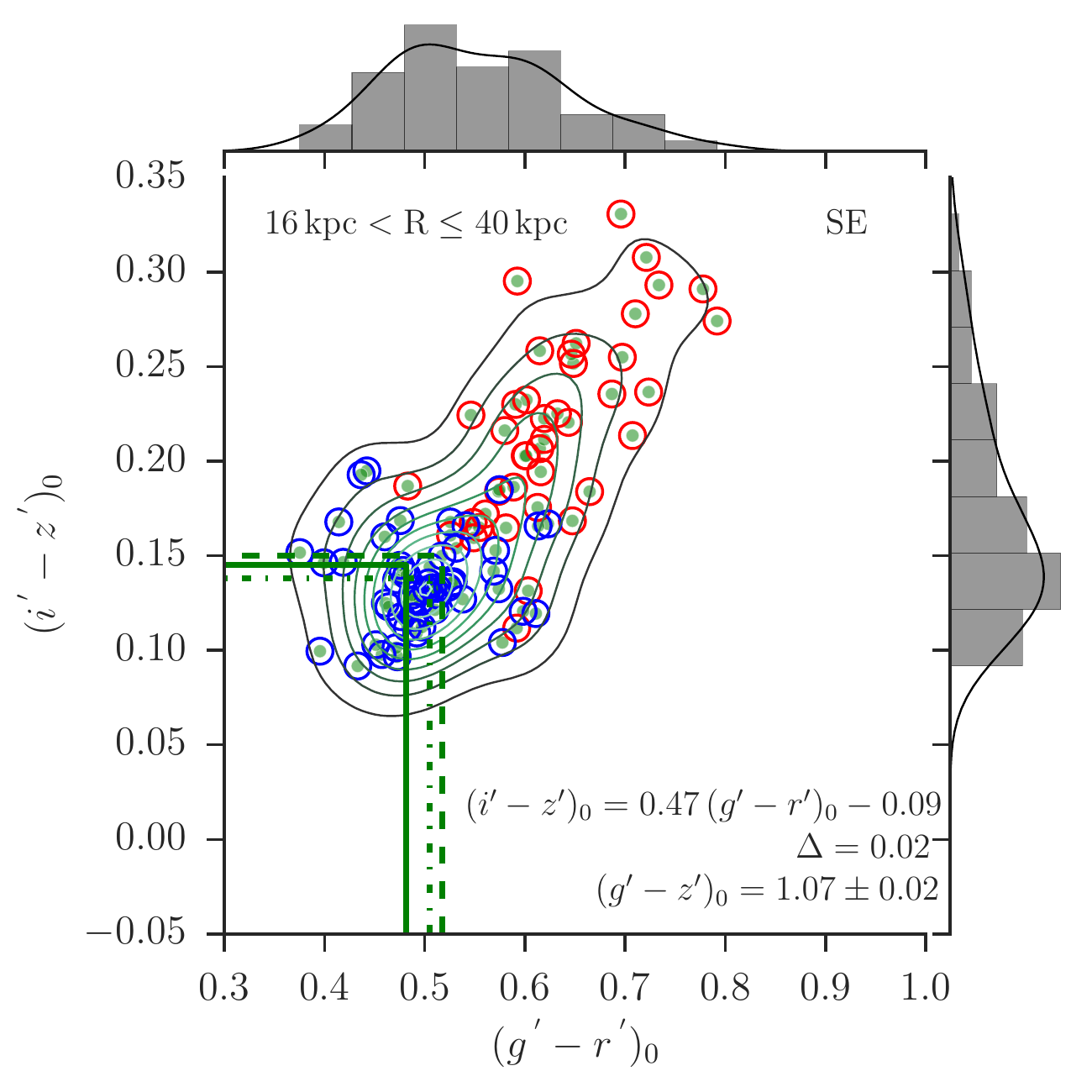}
\includegraphics[width = 6.9cm]{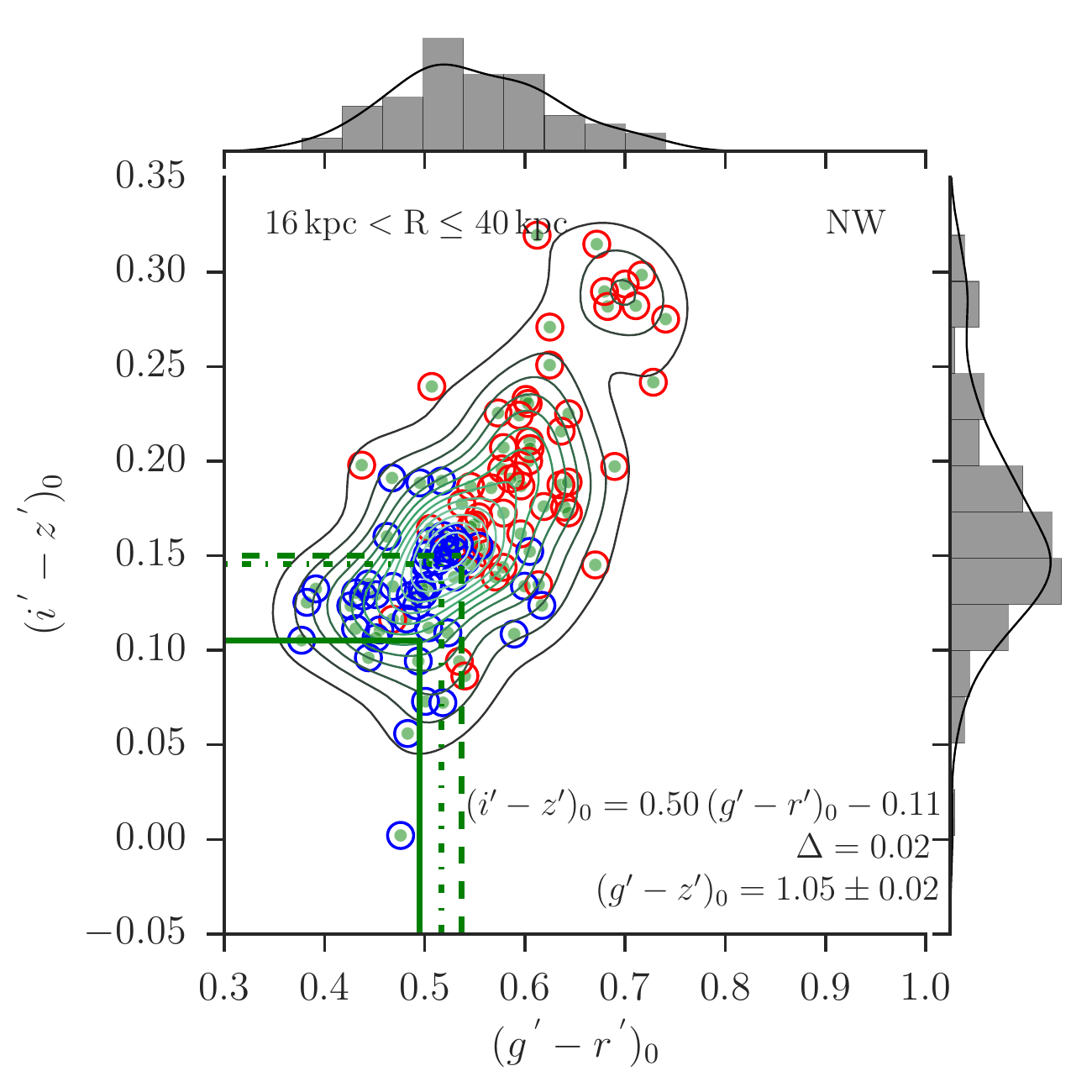}\\

\hspace{+3.5cm}
\includegraphics[width = 6.9cm]{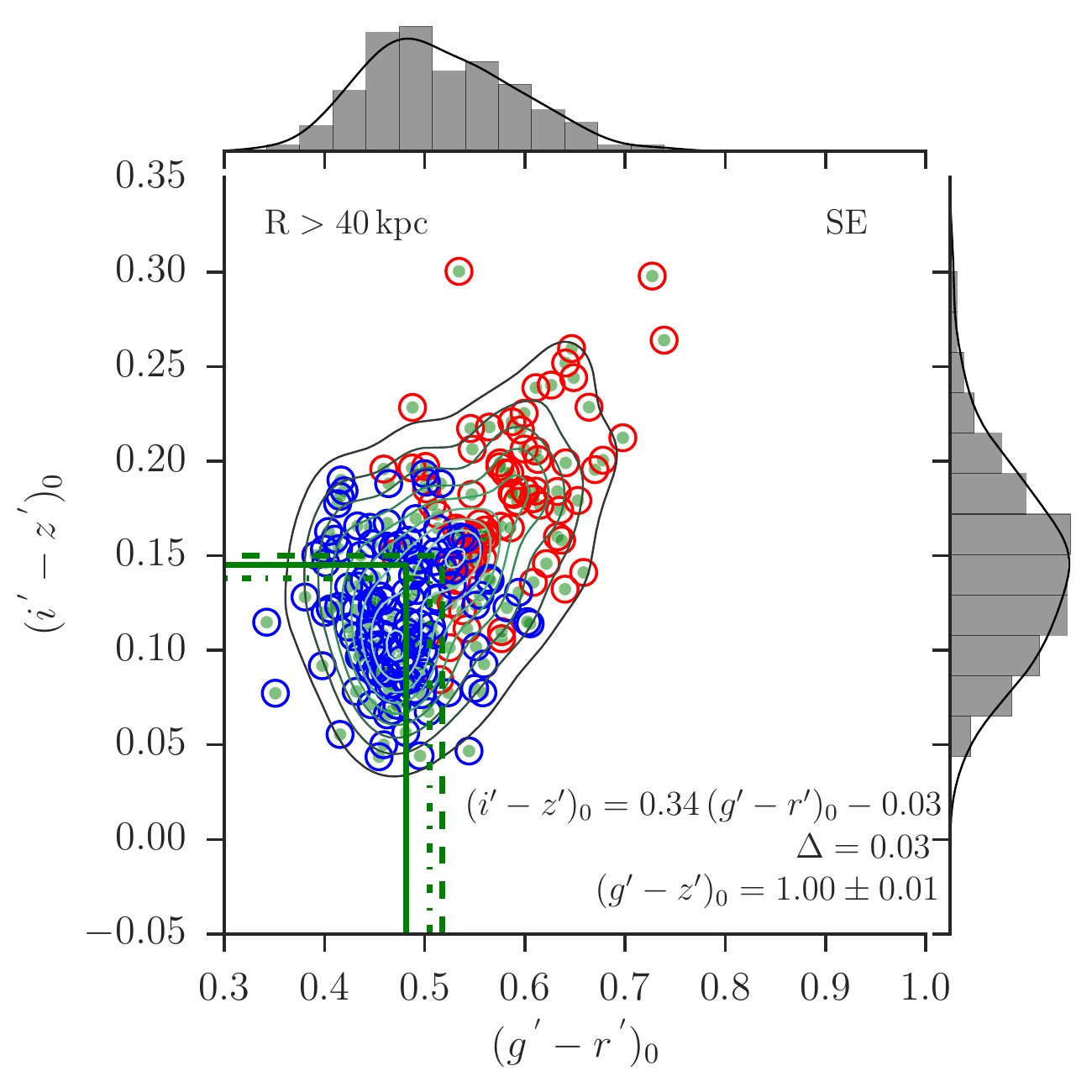}
\includegraphics[width = 6.9cm]{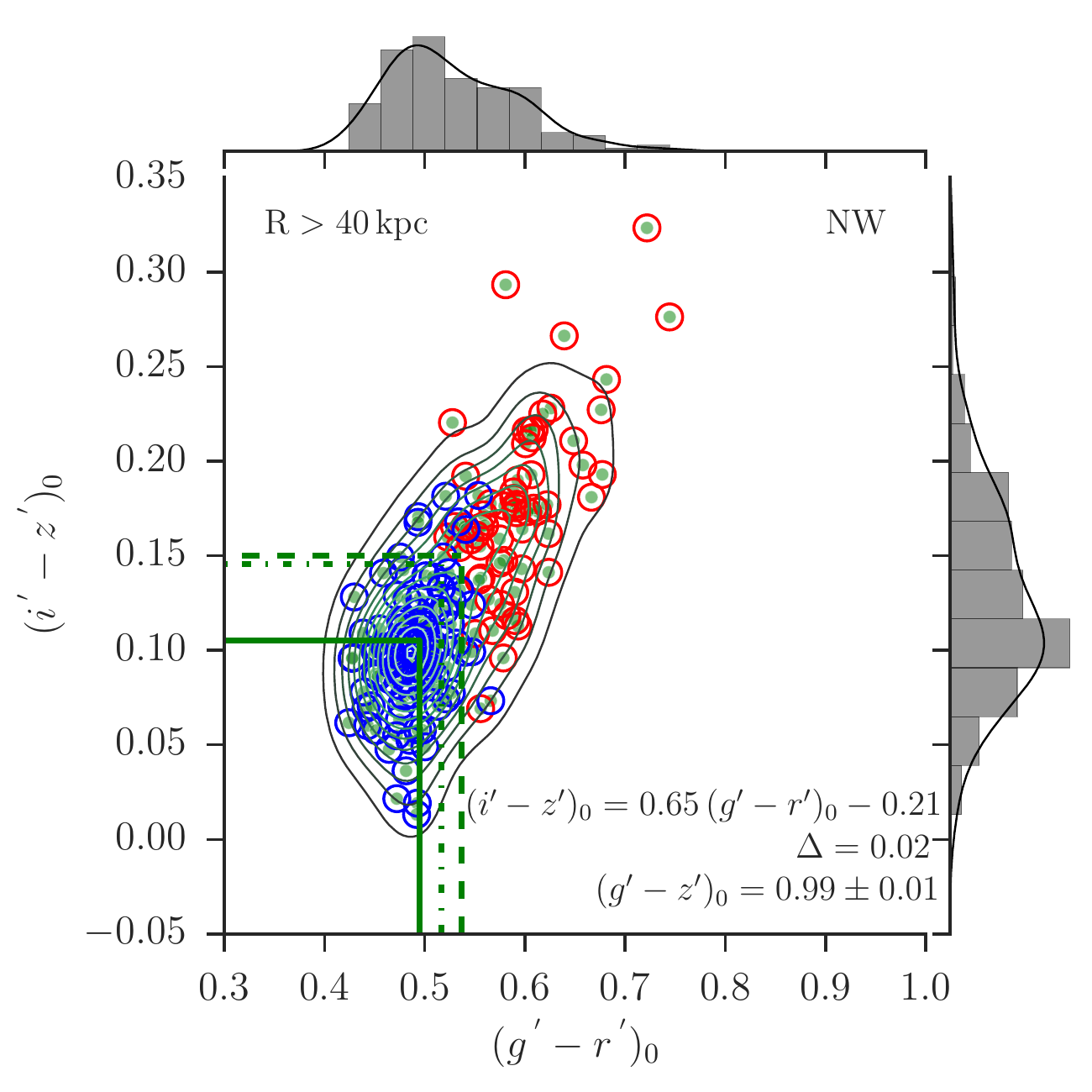}
\caption{$(g'-r')_{0}\, \mathrm{vs.}\, (i'-z')_{0}$ color-color diagrams and corresponding kde density contours for the GC populations associated to the M87 halo. {\bf Left Panel:} The relation is shown for the GC population that lies in the SE side with respect to M87, and from top to bottom for GC sub-samples at increasing distances from the center of the galaxy. Independent of galactocentric distance, the high-density peaks are close in value as shown by the green lines.  {\bf Right Panel:} Same as in the left panel but for the sample of GCs in the NW half of M87.  In this region the GCs show bluer $(i'-z')_{0}$ colors. In this diagram we flag the position of the blue and red class of GCs as identified based on their $(g'-i')_{0}$ values: green dots in red circles refer to the metal-rich population, while green dots in blue circles represent metal-poor GCs. The linear maximum-likelihood fits to the color-color relations, the $\Delta$ values, as well as the mean $(g'-z')_{0}$ colors for each GC sub-sample are given in the legend. The uncertainty on the magnitudes is less than 0.06 mag. }
\label{fig:CC_narrow}
\end{figure*}
\begin{figure*}[!t]
\centering
\includegraphics[width = 10.5cm]{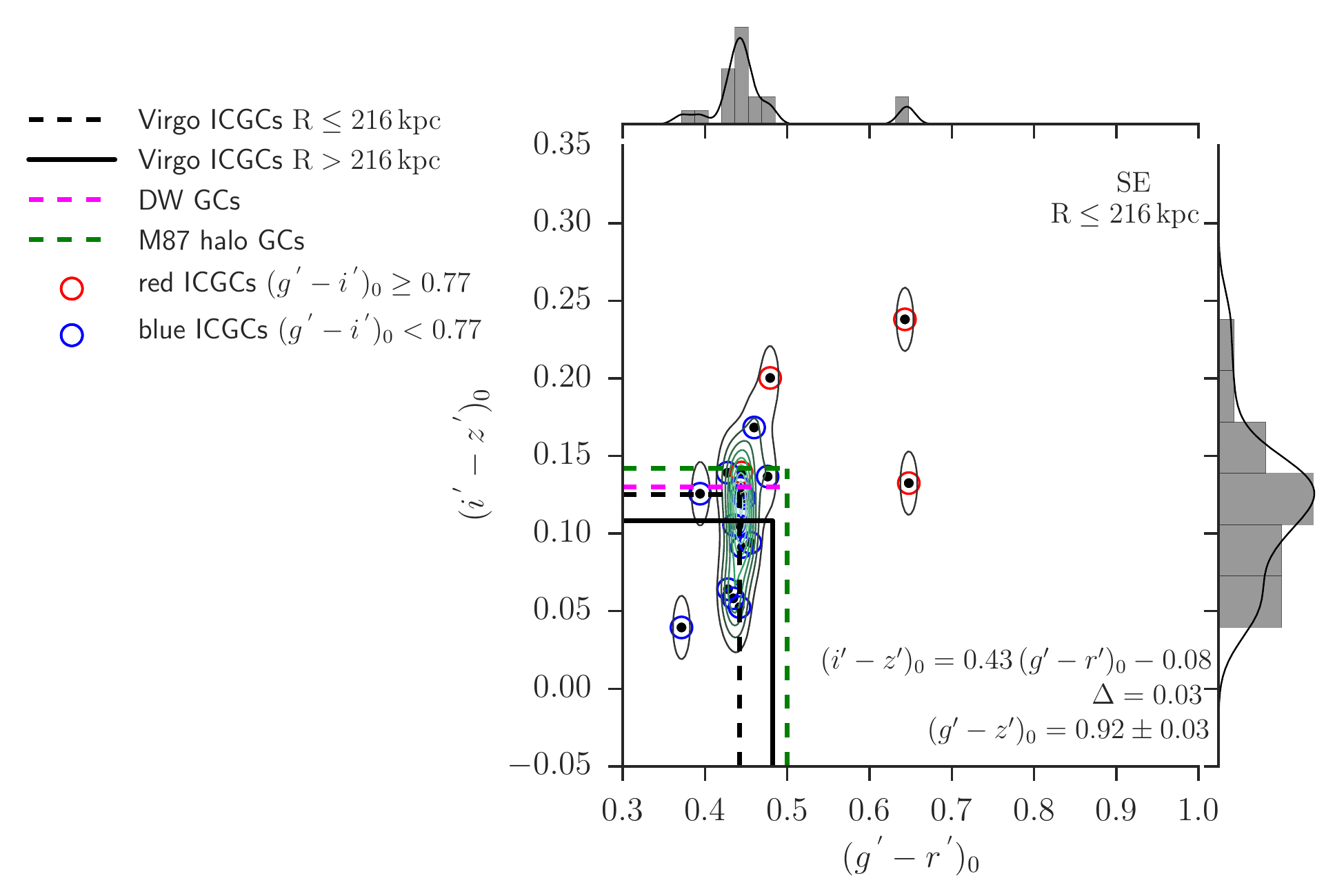}
\includegraphics[width = 6.9cm]{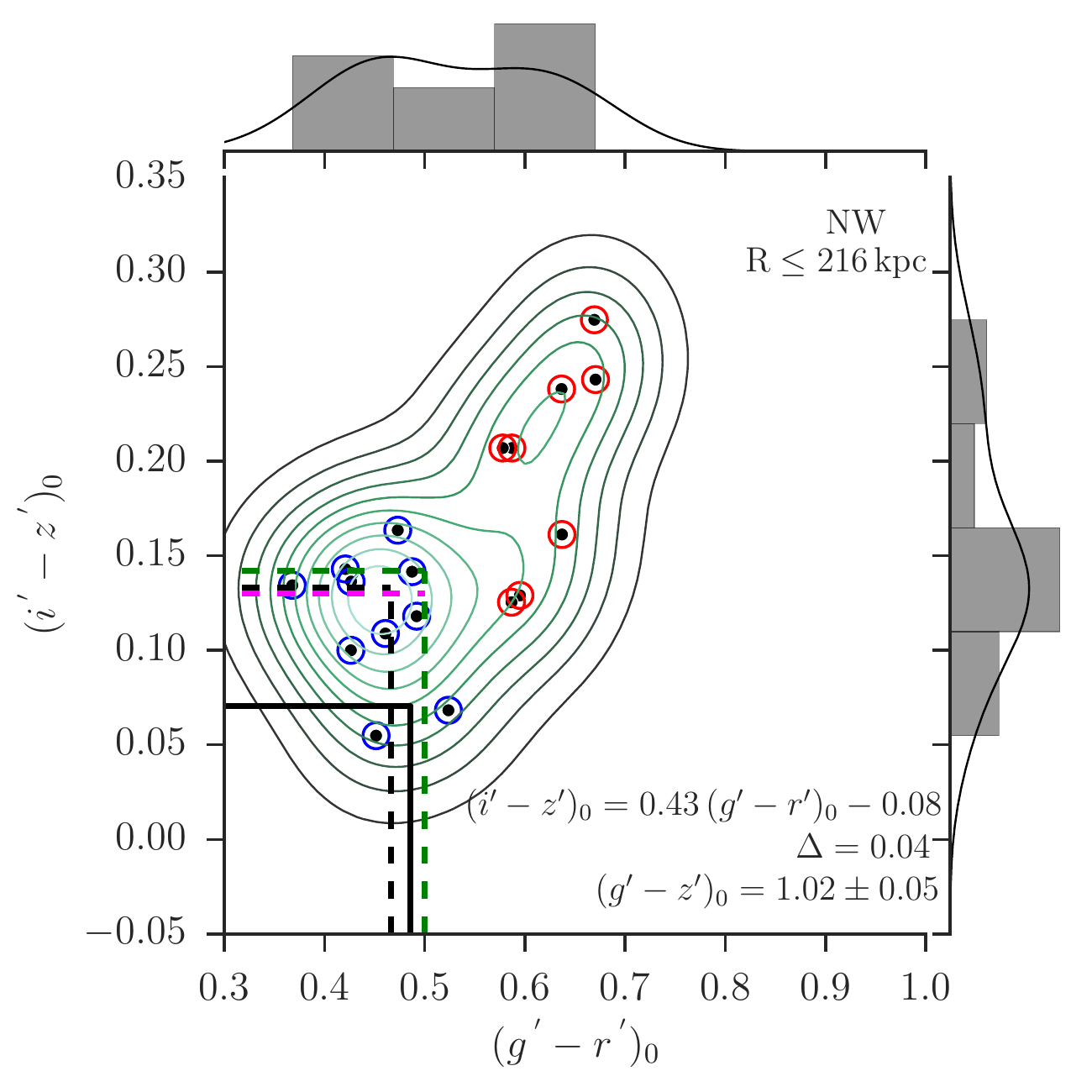}\\
\hspace{+3.5cm}
\includegraphics[width = 6.9cm]{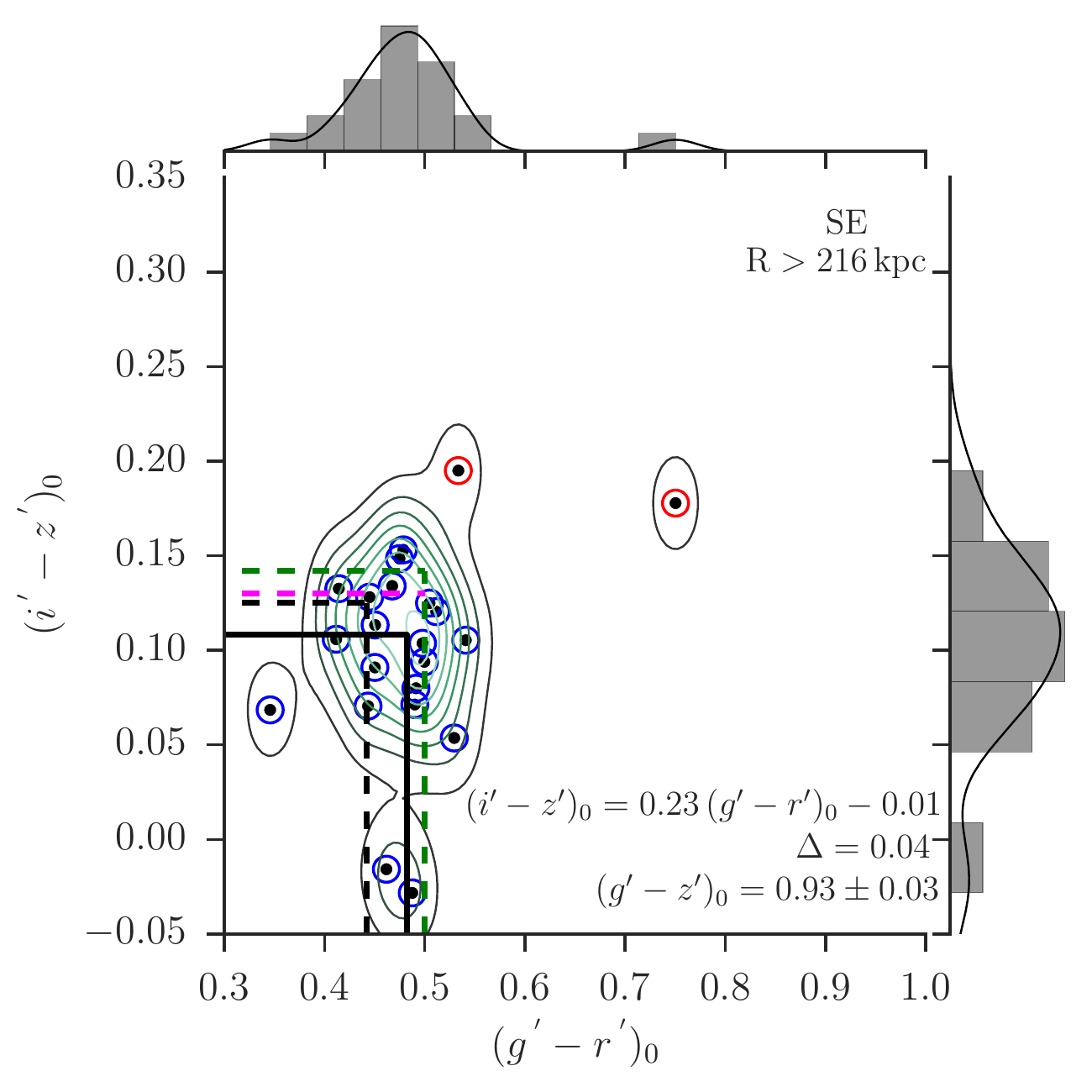}
\includegraphics[width = 6.9cm]{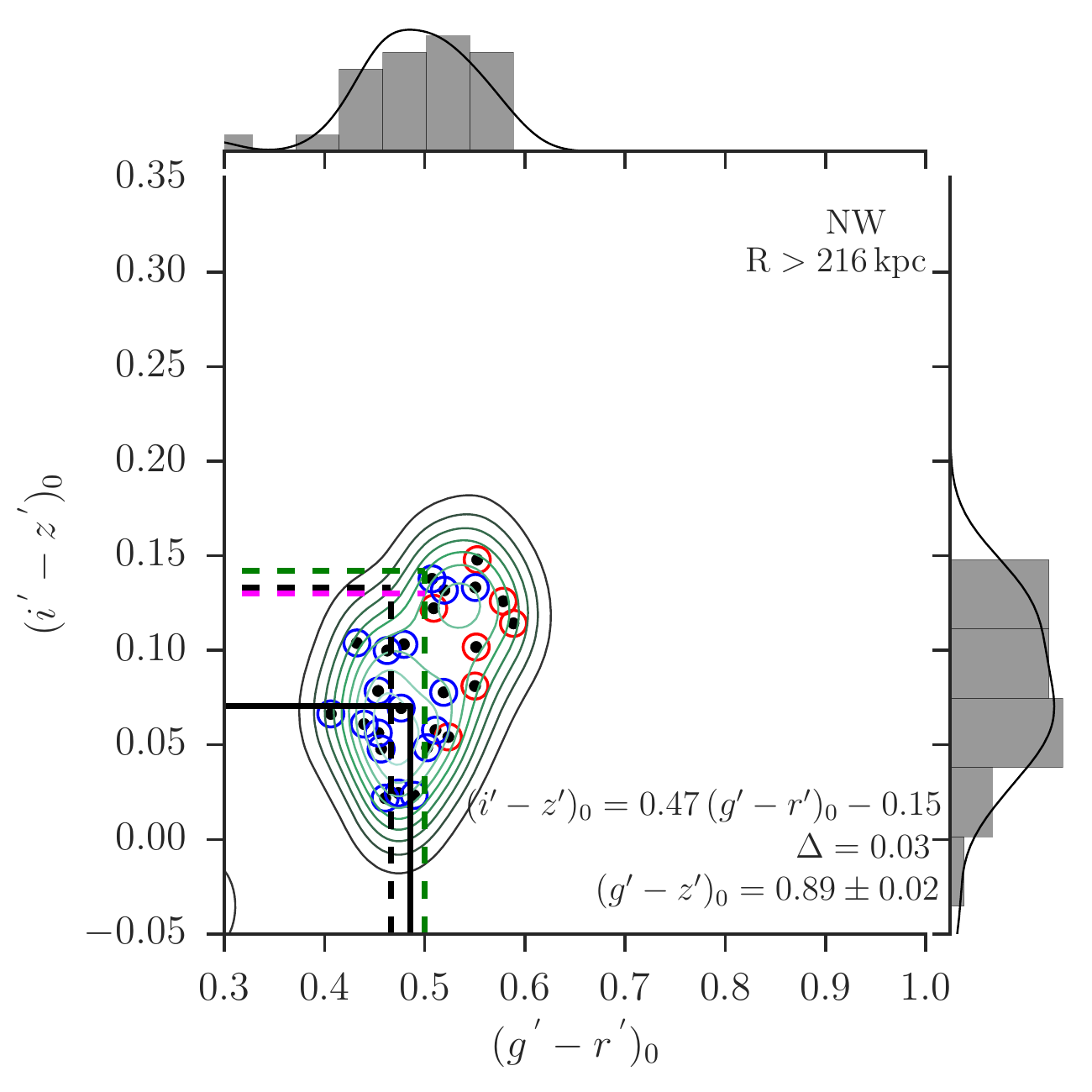}\\
\caption{Same as in Fig.~\ref{fig:CC_narrow} but for the Virgo ICGCs. The ICGCs are characterized by bluer colors than the galaxy GCs (the green dashed line shows a representative peak at  $[(g'-r')_{0},(i'-z')_{0}] \sim [0.51, 0.14]$), with the bluest population found at $\mathrm{R > 216}$ kpc. The magenta dashed lines show the color-color high-density peak for a sample of GCs associated to Virgo early-type dwarfs. The linear maximum-likelihood fits to the color-color relations, the $\Delta$ values, as well as the mean $(g'-z')_{0}$ colors for each GC sub-sample are given in the legend. The uncertainty on the magnitudes is less than 0.06 mag. }
\label{fig:CC_broad}
\end{figure*}

We also present the $(g'-r')_{0}\, \mathrm{vs.}\, (i'-z')_{0}$ color-color diagrams for the M87 halo (Fig.~\ref{fig:CC_narrow}) and the Virgo intra-cluster GCs (Fig.~\ref{fig:CC_broad}). This analysis was motivated by the recent finding of a correlation between GC color-color relations and environment, and interpreted as driven by different stellar abundance ratios \citep[][altough interpretations related to different underlying ages are also given, e.g. \citet{usher15,powalka18} ]{powalka16b,powalka18}. In this study, the working sample was selected through a combination of the $u^*i'K_{s}$ color properties and the compactness of the objects, similar to what we have described in Sec.~\ref{back_contamination}. However, they only studied GCs whose magnitudes were measured with an uncertainty $\sigma_{\mathrm{mag}} < 0.06$ mag in all the bands, and also offset the $z'$ magnitudes to account for known deviations of the SDSS-DR10 magnitudes from the AB magnitude system \citep{powalka16a,powalka16b}. Hence, we consider the same constraints and restrict our analysis to GCs with the same precision on the magnitude values and apply the same shift to the $z'$ magnitudes. For each of the color-color diagrams we fit and plot the relative Gaussian kernel density estimate (kde) and locate its peak identifying the maximum values in  $(g'-r')_{0}$ and $(i'-z')_{0}$ from the kde probability density function.

The M87 halo GCs are shown in Fig.~\ref{fig:CC_narrow} (green dots), where we also flag the position of the blue and red class of GCs as identified based on their $(g'-i')_{0}$ values: green dots in red circles refer to the metal-rich population, while green dots in blue circles represent metal-poor GCs.  
Within $\mathrm{R \le 40\, kpc}$ the SE and NW sides of M87 are characterized by GCs with similar colors, with the high-density peaks in the color-color space measured at $[(g'-r')_{0},(i'-z')_{0}]_{\mathrm{SE}} \sim [0.52, 0.15]\, [0.51,0.14]\,$, and  $[(g'-r')_{0},(i'-z')_{0}]_{\mathrm{NW}} \sim  [0.54, 0.14]\, [0.52,0.15]$. Similar are also the distributions of blue and red GCs in the color-color plane. At larger distances, i.e. $\mathrm{R > 40\, kpc}$ the color-color diagram extends to bluer $(i'-z')_{0}$ colors, with a significant fraction of object with $(i'-z')_{0} \le 0.1$. 
We note, however, that the average behavior differs between the SE and NW regions. NW of M87 the blue GCs define the high-density peak in the color-color diagram at $[(g'-r')_{0},(i'-z')_{0}] \sim [0.50,0.11]$. SE of M87 the blue GC population occupies a larger area and the high-density peak is measured at $[(g'-r')_{0},(i'-z')_{0}] \sim [0.48,0.14]$.

The same analysis is carried out for the ICGC sample, (we use the same binning as in Fig.\ref{fig:GC_RB}). The results are presented in Fig.~\ref{fig:CC_broad}, where this time black dots in red and blue circles identify the metal-rich and metal-poor populations of GCs, respectively. 
The ICGCs concentration in the color-color space differs from the one observed for the M87 halo GCs: independently from the distance a large fraction of ICGCs extend to $(i'-z')_{0} \leq 0.1$, the distribution is dominated by the blue ICGCs, and the few red GCs are bluer in the  $(g'-r')_{0}\, \mathrm{vs}\, (i'-z')_{0}$ plane. This implies blue high-density peaks measured at $[(g'-r')_{0},(i'-z')_{0}]_{\mathrm{SE}} \sim [0.44, 0.12]\, [0.48,0.11]$ and $[(g'-r')_{0},(i'-z')_{0}]_{\mathrm{NW}} \sim [0.47, 0.13]\, [0.49,0.07]$. These are bluer than the one measured for the M87 halo GCs of which we show a representative peak as green dashed line.We also compared the ICGC color-color distribution with the one measured for the 33 GCs that in Sec.~\ref{GC_Virgo_memb} we identified as bound to Virgo early-type dwarfs. Their peak (magenta dashed line in Fig.~\ref{fig:CC_broad}) is measured at redder colors. Besides the offset in the color peaks, the comparison between Fig.~\ref{fig:CC_narrow} and Fig.\ref{fig:CC_broad} also shows different slopes for the M87 halo and Virgo intra-cluster GCs color-color relation. A maximum-likelihood linear fit to the $(g'-r')_{0}$ and $(i'-z')_{0}$ results into a galaxy component that is usually steeper than the ICL. Also, by computing the median deviation of the GC $(i'-z')_{0}$ colors from the fitted linear relation, $\Delta$, the locus of the M87 halo GCs is tighter (i.e., stronger color-color relation) than the one of the ICGCs (fitted parameters and $\Delta$ values are given in the legend of Fig.~\ref{fig:CC_narrow} and Fig.~\ref{fig:CC_broad}). Such properties can be  interpreted in terms of different metallicity distributions \citep{powalka16b}.

Finally, \citet{peng06} find that the mean $(g'-z')_{0}$ colors of the GCs in ACSVCS galaxies define a color-magnitude sequence, such that more luminous/massive galaxies host, on average, redder GC systems. We than computed the mean $(g'-z')_{0}$ colors for the M87 halo and Virgo intra-cluster GCs and as function of the distance. The GC systems bound to the M87 halo show redder mean $(g'-z')_{0}$ values. Averaged across the distance, and with no distinction between NW and SE halves, we find $<(g'-z')_{0,\mathrm{M87}}> = 1.07 \pm 0.01$,  while for $\mathrm{R} > 40$ kpc we find $<(g'-z'){_0,\mathrm{M87}}> = 1.00 \pm 0.01$. The Virgo ICGCs are constant, within the uncertainties, to a mean color $<(g'-z')_{0,\mathrm{IC}}> ~ 0.94 \pm 0.02$. Such a result is consistent with the galaxy accreted component dominating the outer regions and reinforce our conclusion that the ICL is built up by less massive systems than the one contributing to the size growth of the M87 halo.

\subsection{Density profiles} 
\label{density_p}
We are now interested in studying the spatial distributions of the M87 halo and Virgo intra-cluster GCs to analyze their density profiles. This is because differences in their radial profiles may imply differences in the evolutionary paths of these two components as different progenitors are expected \citep{dolag10,cui14} and observed \citep{longobardi15a}
to contribute differently to the resulting density distribution.

\subsubsection{Spatial completeness}
\begin{figure}[!ht]
\centering

\includegraphics[width = 9cm]{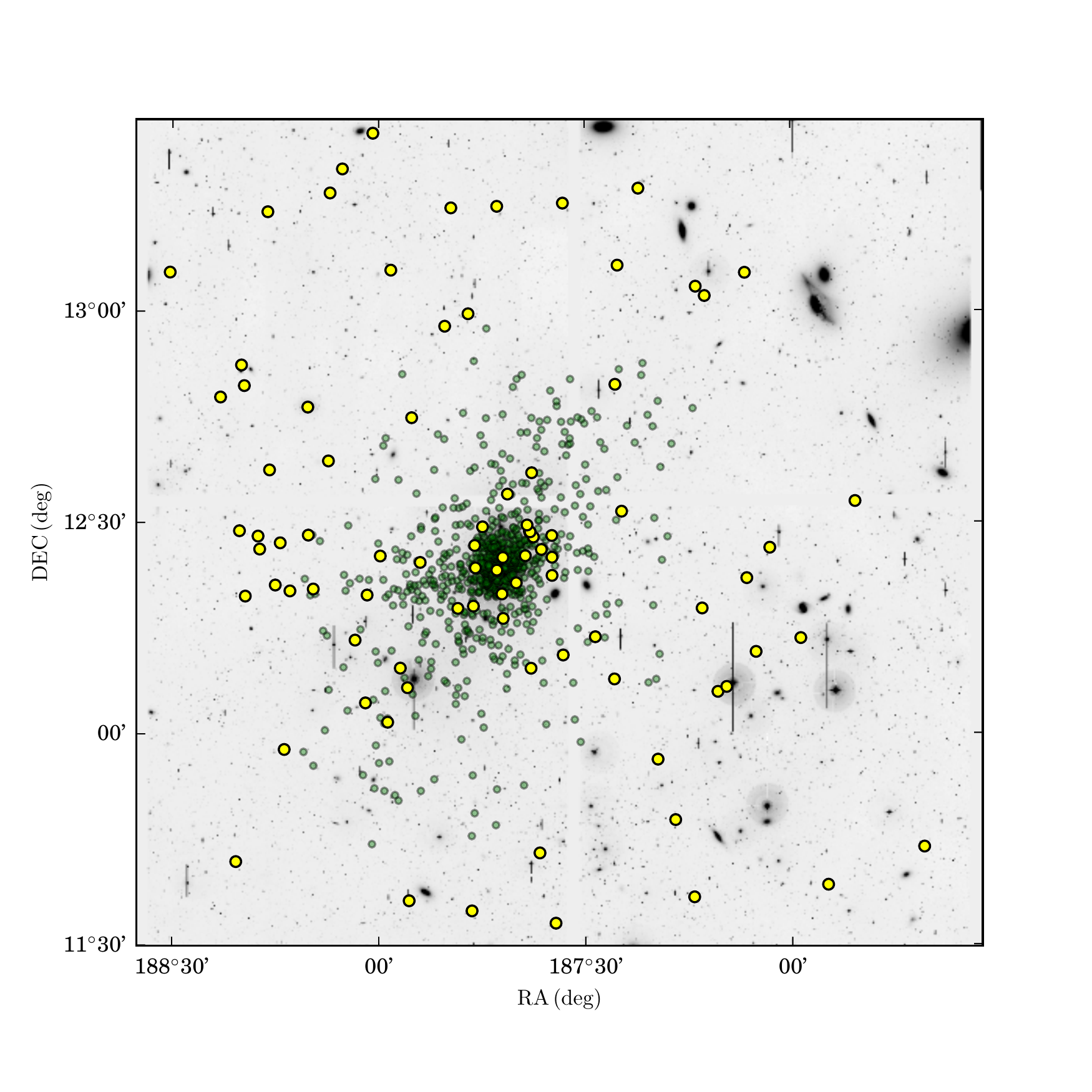}
\caption{Spatial distribution of the velocity-selected sample of GCs (green dots are M87 halo GCs and yellow dots are ICGCs)  superimposed on to the $g'$-band NGVS image of the central 2 by 2 square degree of Virgo. The NW field is more affected by spatial incompleteness. East is to the left, North is up.}
\label{fig:Spatial_Dist}
\end{figure}

In order to analyze the GC density profiles we need to compute a completeness factor as function of the distance, $C_{\mathrm{spec}}\, (\mathrm{R})$, that accounts for the spatial incompleteness of our spectroscopically selected sample of GCs. Such a completeness is estimated against the photometrically selected GCs, i.e. objects with high $\mathrm{p_{gc}}$ probability (see Sec.~\ref{back_contamination}), that we recall being 100\% complete down to our limiting magnitude, $g'=24.0\, \mathrm{mag}$. We then estimate the completeness function by computing the ratio between the number of photometric GCs with spectroscopic information, $\mathrm{N_{GC,spec}}$, and the total number of GCs in the photometric sample, $\mathrm{N_{GC,phot}}$, in each elliptical annulus\footnote{The completeness function $C_{\mathrm{spec}}\, (\mathrm{R})$ is computed using the spectroscopic sample of objects retaining those GCs that we flagged as bound to other Virgo members. This is because the photometric sample of GCs we use to complete our observations does contain such a contribution, and in this way the ratio is still conserved.}. Hence, the completeness factor as function of the distance from the M87's center, $C_{\mathrm{spec}}\, (\mathrm{R})$, can be written as:
\begin{displaymath}
C_{\mathrm{spec,j}}\, (\mathrm{R})=\mathrm{\frac{N_{GC,spec,j}}{N_{GC,phot,j}}},
\end{displaymath}
where $\mathrm{R}$ is the average major axis distance of all GCs falling within each elliptical annulus. The subscript $j$ indicates the two different GC components, red and blue.
Averaged over the elliptical bins the completeness factor amounts to $\langle C_{\mathrm{spec}}\rangle = 0.1$, however the NW field is more affected by incompleteness than the SE region. This effect is measured at larger distances so that in the last bin we miss 40\% more GCs NW to M87 as compared to the SE. This is due to observational strategies that observed the SE region more, but it is also related to the fact that the NW region has more photometric objects per unit area.

In Fig.~\ref{fig:Spatial_Dist} the sky positions of our sample of data (both M87 halo, green dots, and Virgo intra-cluster GCs, yellow dots) can be seen over-plotted on the NGVS $g'$-band image.  

\subsubsection{Density profile for the M87 halo and Virgo intra-cluster GCs}
The density profiles for the M87 halo and intra-cluster GCs are constructed by binning our GC samples in elliptical bins, that have as major axis distances the same we used for our GMM analysis int Sec.~\ref{kinematic_separation} (see also orange lines in Fig.~\ref{fig:GC_pspace}), and as position angle and ellipticity $\mathrm{P.A.=-25.0^{\circ}}$, and $e=0.4$, respectively \citep{ferrarese06,janowiecki10}. Hence, for each of the two dynamical components of GCs and as function of the distance, the density is computed as 
\begin{eqnarray}
\label{rho_M87}
\rho_{\mathrm{M87,j}}\mathrm{(R)}=\frac{\mathrm{\left(N_{obs,M87,j}(R)- N_{ov,j}(R)\right)}}{\mathrm{A(R)}}\times\frac{1}{C_{\mathrm{spec,j}}(\mathrm{R})}\\
\rho_{\mathrm{ICL,j}}\mathrm{(R)}=\frac{\mathrm{\left(N_{obs,ICL,j}(R)+ N_{ov,j}(R)\right)}}{\mathrm{A(R)}}\times\frac{1}{C_{\mathrm{spec,j}}(\mathrm{R})}
\label{rho_ICL}
\end{eqnarray} 
where $\mathrm{N_{obs,M87}(R)/N_{obs,ICL}(R)}$ is the observed number of M87 halo and Virgo intra-cluster GCs in the elliptical bin, respectively; $A(R)$ is the annulus area, estimated via Monte Carlo integration techniques if only a portion of the annulus intersects our FOV, and $C_{\mathrm{spec}}\, (\mathrm{R})$ is the spectroscopic completeness factor. The observed numbers of clusters together with the completeness factor differ for the two, red and blue, populations of GCs, that in Eq.~\ref{rho_M87} and Eq.~\ref{rho_ICL} are denoted with the subscript $j$. As can be seen in Fig.~\ref{fig:GMM_bin} the ICL and M87 halo velocity distributions overlap, and as result, the GMM algorithm assigns the ICGCs at low velocities relative to the galaxy systemic velocity to the galaxy halo component.
We statistically quantified this effect comparing the M87 halo and IC velocity distributions in each elliptical bin. The LOSVDs of the two components were approximated by Gaussian functions with mean values and dispersions as given in Table~\ref{GMM_parameter}. Hence, we calculated the fraction of ICGCs that lie inside the galaxy halo distribution as the area of overlap between the two curves. With this analysis we obtain a statistical estimate of the number of ICGCs, $\mathrm{N_{ov}(R)}$, contained in the M87 halo velocity distribution. It is a function of $\mathrm{R}$, however, averaged over the bins, 10\% of the GC sample counted as part of the M87 halo is estimated to be associated with the ICL. For each of the radial bins, this contribution was subtracted from the M87 halo and added to the IC component. We emphasize that to build the density profiles associated with the red and blue GCs we consider the different contributions these populations have within the ICGC component and account for it when computing $\mathrm{N_{ov}}$.

\begin{figure}[!ht]
\begin{center}
\includegraphics[width = 9cm]{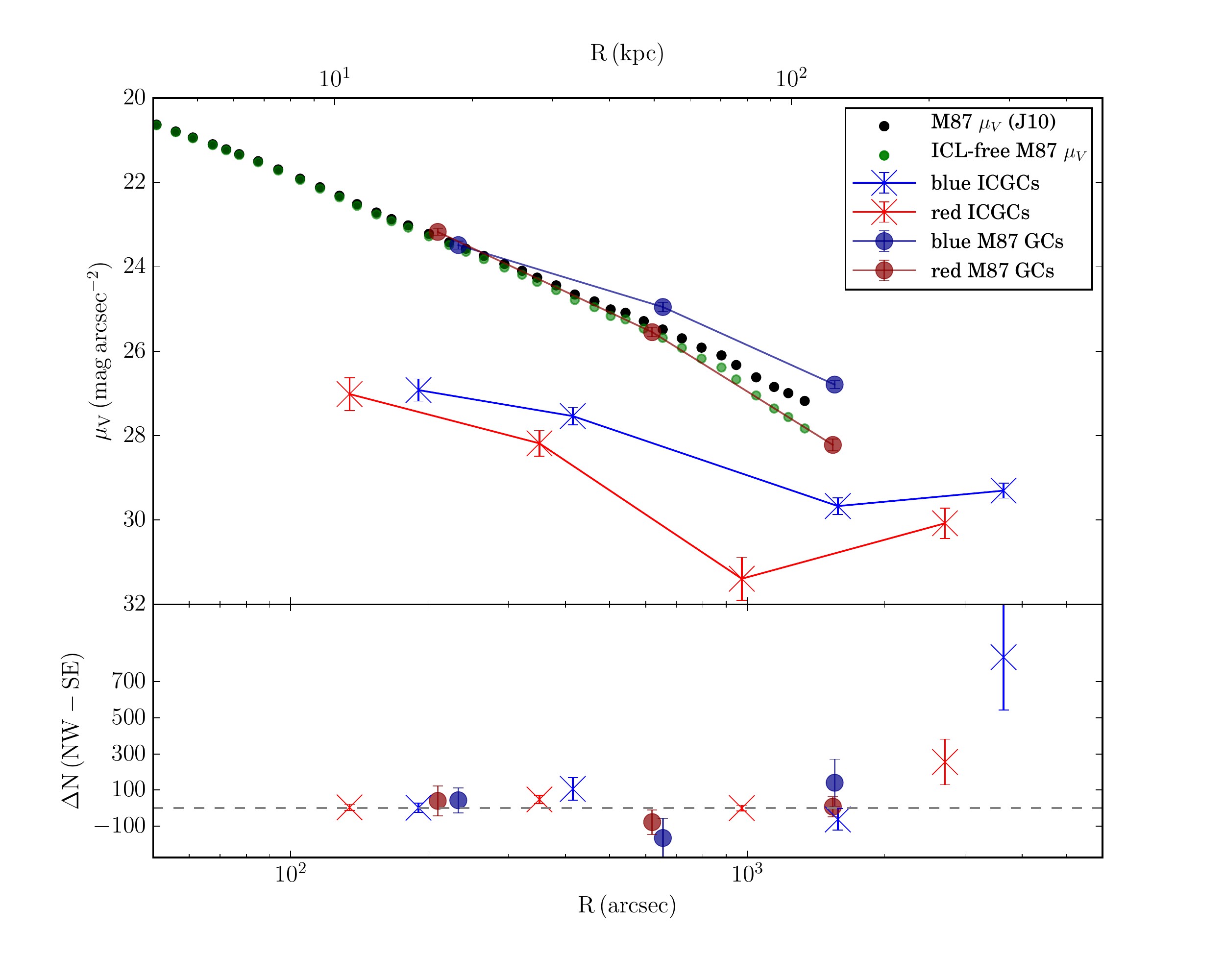}
\caption{{\bf Top Panel:} Azimuthally averaged density profiles as function of the major axis distance from M87 for the galaxy halo (circles) and IC (crosses) GCs, with the red and blue symbols identifying the red (metal-rich) and blue (metal-poor) populations of GCs. The M87 GCs show a distribution that is steeper than the IC counterpart, the latter being more extended, and shallower, however still centrally concentrated. The red GCs bound to the M87 halo (red circles) follow the ICL-free galaxy's surface brightness profile (green dots). The blue GCs bound to the M87 halo (blue circles) are more extended deviating from the M87 light profile for $\mathrm{R > 30\, kpc}$. {\bf Bottom Panel:} Difference in number of GCs between the NW and SE regions of M87 as function of the major axis distance from M87 (legend as in the top panel). A value of zero (dashed gray line) implies symmetry in number counts of GCs between the NW and SE fields. A significant deviation from such a symmetry is measured in the last bin for the ICGC population, implying a higher fraction of intra-cluster component in the NW region of M87.}
\label{fig:GC_density}
\end{center}
\end{figure}

In Fig.~\ref{fig:GC_density} we show the comparison between the $V$-band surface brightness measured in the M87 region \citep[black dots][]{janowiecki10}, the ICL-free M87 surface brightness (green dots), obtained subtracting from the measured $\mu_{\mathrm{V}}$ the ICL surface brightness (see Sec.~\ref{ICL_SN} for details about the adopted ICL surface brightness distribution),  and the density profile of the M87 halo GCs (full dots) and Virgo ICGCs (stars). To this end we plot the logarithmic GC profile with its uncertainties defined as:
\begin{eqnarray}
\mu_{\mathrm{GC_{l}}}(R) = -2.5 \log{[\rho_{\mathrm{l}}(\mathrm{R})]}+k_{0_{l}},\\
\sigma_{\mu_{\mathrm{GC_{l}}}}(R) = \left | \frac{\partial \mu (R)}{\partial _{\rho_{\mathrm{GC_{l}}}}} \right | \times \sigma_{\rho_{\mathrm{GC_{l}}}}(R),
\end{eqnarray} 
where the subscript $l$ denotes the two different GC components, and $k_{0_{l}}$ is an arbitrary constant that we added for a better comparison with the light. The uncertainty on the density is given as $\sigma_{\rho_{\mathrm{GC_{l}}}}(R) = 1/\sqrt{N_{\mathrm{GC_{l}}}(R)} \times \rho_{\mathrm{GC_{l}}}(R)$, with $N_{\mathrm{GC_{l}}}(R)$ being the final number of GC counted in each radial bin. The density profiles are shown separately for the blue and red populations of GCs (blue and red filled dots for the M87 halo and blue and red stars for the Virgo intra-cluster GCs, respectively). Note that the normalization factors differ for the red and blue populations associated with both the M87 halo and ICL.

We find the red component of the M87 halo GCs agrees well with the galaxy's light, following its surface brightness profile, while the blue component is more extended and flatter, in agreement with results from previous studies \citep[e.g.,][]{harris09,pota13,durrell14}. As discussed in the previous Section the total fraction of blue GCs in the M87 halo component amounts to 52\% of the entire sample. However Fig.~\ref{fig:GC_density} shows that its relative contribution increases as function of the distance:  close to the M87's center, $\mathrm{R \le 32\, kpc}$, 40\% of objects are blue, at intermediate distances, $\mathrm{32 < R \le 80\, kpc}$, this fraction increases to 50\%, and reaches 70\% at larger distances. The ICGC density profile, which is flatter than the galaxy halo distribution, is still centrally concentrated. These findings are consistent with predictions from hydro-dynamical simulations, where the radial
density profile of the component bound to the galaxy halo is much steeper than that of the diffuse IC component \citep{murante04,dolag10,cui14}.

Finally we compute the difference of GC counting for both the M87 halo and IC objects, and for the red and blue populations in the NW and SE fields with respect the M87's center (Fig.~\ref{fig:GC_density}), bottom panel). At all distances the galaxy halo component is consistent within the uncertainties with no asymmetry between the northern and southern fields ($\Delta\mathrm{N} = 0 $, dashed line). The cluster component, instead, shows asymmetry in the number of ICGCs when the NW and SE fields are studied separately, implying a higher fraction of IC component in the NW field in the outermost bin. This observation is consistent with what is measured by \citet{durrell14} from the photometric sample and is now validated on the basis of the spectroscopically confirmed GCs.

\subsubsection{The Virgo ICGCs $S_{N}$}
\label{ICL_SN}

The number of clusters per unit luminosity quantifies the GC specific frequency, $S_{N}$, that numerically is defined as 
\begin{equation}
S_{N} = \mathrm{2\times N_{GC,TO}} \times \mathrm{10^{0.4(M_{V} + 15)}},
\end{equation} 
for a system with absolute $V$-band magnitude $\mathrm{M_{V}}$. $\mathrm{N_{GC,TO}}$ is the number of GCs down to the GC luminosity function turnover magnitude, $\mu_{\mathrm{g,TO}}$, and, multiplied by two, it is used to represent the total cluster population using a symmetric parametrization of the GC luminosity function \citep{harris81}.
Assuming $\mu_{\mathrm{g,TO}}=24.0$ mag for the GC system at the center of Virgo \citep{jordan07}, $\mathrm{N_{ICGC,TO}}$ in our surveyed area can be computed by fitting a power law to the measured ICGC density distribution and extrapolating inward such a profile. Doing so, we find that the expected number of ICGCs in the central $2^{\circ}\times2^{\circ}$ of Virgo, and down to $g^\prime\le 24.0$ mag, is $\mathrm{N_{ICGC,TO}} = 2503 \pm 430$, where the error accounts for the uncertainty in the integrated density profile. The entire population of ICGCs can then be retrieved multiplying by two $\mathrm{N_{ICGC,TO}}$. However, such a choice is not free from caveats: we are assuming a constant GCLF turnover $g'$-band magnitude for the systems that contributed to the IC component that might result in a over-/underestimation of $\mathrm{N_{ICGC}}$ for brighter and fainter $\mu_{\mathrm{g,TO}}$, respectively. Thus, we computed the number of GCs we would add/lose if the 'true' GCLF would have magnitude cutoff in the range $\mu_{\mathrm{g,TO}} \in [23.8,24.2]$ with dispersion values $\sigma_{\mathrm{bright}} = 1.3$ and $\sigma_{\mathrm{faint}} = 0.8$. The latter represent the dispersion values found to be representative of the GCLF for bright and faint galaxies in Virgo, respectively. The range of $\mu_{\mathrm{g,TO}}$ traces, instead, the uncertainty on the $g'$-band turnover magnitude for the GCLF of Virgo galaxies, estimated to be constant to a value $\mu_{\mathrm{g,TO}} = 24.0 \pm 0.2$ \citep{jordan07}. Hence, we expect a total ICGC population of $\mathrm{N_{ICGC}} = 5006 \pm 1021$, where the error now accounts for the 11\% uncertainty due to the choice of the GCLF.

To estimate the total ICL luminosity we integrate the IC light profile traced by the ICPNs as given in Longobardi et al. (2018; in preparation) and in \citet{longobardi15a}. Analyzing a sample of ICPNs in the inner 0.5 square degree of Virgo, these authors computed the V-band surface brightness profile for the IC component in units of $L_{\odot,\mathrm{V}}$ pc$^{-2}$, as the scaled power-law fit to the IC PN surface density data \citep[more details can be found in,][] {longobardi15a}. They found that $I(\mathrm{R})_{\mathrm{ICL}} \propto \mathrm{R^{\gamma}}$, with $\gamma = -0.79$ (Longobardi et al. 2018, in prep) or $\gamma \in [-0.34,-0.04]$ \citep{longobardi15a}. Integrating over the NGVS pilot region, this profile gives a total V-band luminosity $\mathrm{L_{ICL} = 4.2\pm0.1 \times 10^{10}\, L_{\odot}}$, with the error on $\mathrm{L_{ICL}}$ computed as the variation in luminosity values when the different light profiles are used. This leads to our estimate for the ICGC specific frequency that is $S_{N,\mathrm{ICL}}=10.2\pm4.8$. \citet{peng08} presented specific frequencies for the GC systems of 100 early-type galaxies in the ACS Virgo Cluster Survey and found that luminous early-type galaxies have higher $S_{N}$ than intermediate-luminosity early-type galaxies by a factor of $\sim$2-3, with the latter class of objects ($-20.5 < M_{V} < -18$) generally having $S_{N}\sim1.5$. Early-type dwarf galaxies have a large spread in $S_{N}$, characterized by an average value $S_{N} = 3.1$, however reaching estimates as high as $S_{N} = 14$. They also observed that nearly all dwarf galaxies with high GC fractions are within 1 Mpc of M87, hence the globular cluster fraction that we associate with the Virgo ICGC component is as high as the one measured for dwarf ellipticals that have yet to be subjected to tidal stripping or disruption \citep{peng08}.

\section{Discussion} \label{sec:discussion} 

\subsection{The Virgo intra-cluster population of GCs}
\citet{durrell14} presented a large-scale study of the distribution of GCs in Virgo, statistically accounting for background/foreground contamination to the GC sample. By analyzing the two-dimensional maps of the GC distribution they showed that the blue GCs represent an extended population, still significant beyond several effective radii from the hosting galaxy's center. In the case of M87, they interpreted the GC population beyond 215 kpc as part of the intra-cluster component of Virgo, also suggested by a change in slope of the density profile of the blue GCs at major axis distances R $>$ 240 kpc,  and the evidence for a spatial asymmetry of GCs surrounding M87 for major axis distances larger than 20\arcmin, with  an  excess of  tracers  in  the NW  region (mostly the  blue  population).

On the spectroscopic front the contamination by foreground MW stars has not allowed a clear identification of Virgo ICGCs. \citet{strader11} compiled a large catalog of GCs to study the halo of M87 and reported no evidence of an IC population of GCs neither in terms of a flattening of the GC density distribution nor in terms of a steep rise in the velocity dispersion. They find a low value of $\sigma\sim 300\, \mathrm{km\, s^{-1}}$ in the galaxy's outer regions with no difference in kinematics between the red and blue GCs. However, despite their efforts to reduce the MW star contamination the information they gathered was insufficient to disentangle contaminants and genuine GCs in the most ambiguous velocity range $\mathrm{V_{LOS} < 350\, km\, s^{-1}}$, forcing them to classify all ambiguous objects as stars.

More recently, \citet{ko17} identified a sample of Virgo ICGCs that they associated with a main body of the cluster and with an infalling substructure. ICGCs with $\mathrm{V_{LOS} < \sim400\, kms^{-1}}$ were considered as representative of the infalling component, probably tracing systems that are dragged along by the M86 subgroup that is currently approaching the Virgo's center with negative LOS velocities. We describe a similar phenomenon in Sec.~\ref{sec:GC_PSpace}, where the comparison between the NW and SE LOSVDs shows a higher fraction of dE galaxies with negative $\mathrm{V_{LOS}}$ north of M87 (towards M86). However, our interpretation is that, in the process of falling, the M86 subgroup is contributing GCs to the IC component through the stripping of material coming from low mass galaxies that are being accreted into the Virgo core. These clusters hold the memory of the dynamics of their progenitors, however, they are not spatially associated with them, and we consider them as a genuine component of the ICL. We also note that the ICGC LOSVD in \citet{ko17} does not extend below $\mathrm{V_{LOS} \sim -200\, kms^{-1}}$, hence it is possible that they are missing some of the low velocity ICGCs due to their selection criteria, which were chosen to reduce foreground star contamination.

It is clear that the study of the Virgo ICGC component requires a proper accounting of contamination by foreground stars. In Sec.~\ref{sec:data} we showed that the photometric information provided by the NGVS in the pilot region resulted in a clean photometric-based selection of GCs, thus, allowing us to avoid hard cuts in velocity and explore the entire Virgo cluster velocity range. We showed that the IC component, overlapping onto the one of the central galaxy, results in a total LOSVD with strong and quite asymmetric tails, also making the GC kinematics gradually deviate from the one of M87 at larger distances until it merges with the Virgo cluster values as traced by the early-type dwarf galaxies. The Virgo IC component, then, has a distinct dynamical populations of GCs representative of a dynamically young system. This is shown by the ICGC velocity distribution that does not show signs of point-symmetry----independent of the distance, the ICGC velocity dispersion is higher in the SE region than in the NW region. Although more data is needed to make the latter statement statistically stronger, this result, together with the measured spatial asymmetry (see Sec.\ref{formatio_scenario}) are consistent with the growing observational evidence of the ongoing assembly of Virgo. However, we note that \citet{weil97} argued that SE of M87 there is a spray of material as the result of a past interaction that could be responsible of the twisted and flaring outer isophotes to the SE of M87 as measured in the recent work by \citet{mihos17}. Thus it is possible that there are multiple kinematic components in this region \citep[see also][]{powalka18}, which manifest as a high velocity dispersion due to the limited number of tracers.

Finally, the Virgo ICGC component is not responsible for the high specific frequency measured for M87 ($\sim\! 3 $ times higher than the average value observed for elliptical galaxies \citep{peng08}). \citet{west95}, suggested that $\sim$70\% of the total M87 GC population was associated with cluster GCs. Although the results presented in this work do identify a Virgo ICGC population that overlaps with the M87 halo GCs, we find that the cluster only accounts for roughly 13\% of the total number of GCs out to 216 kpc (see Table~\ref{GMM_parameter}), a distance beyond which we do not predominantly trace the galaxy component anymore. Hence the presence of the Virgo ICGC population cannot explain the high M87 $S_{N}$.

\subsection{Properties of the M87 halo and Virgo intra-cluster GCs and their connection to the systems formation}
\label{formatio_scenario}
The kinematic separation between M87 halo and Virgo intra-cluster GCs allowed us in Sec.~\ref{subsec:populations} to separately analyze their properties. We measured that both components have a red (metal-rich) and a blue (metal-poor) population of GCs, however while the majority of ICGCs are metal-poor, the red-to-blue GC ratio in M87 is close to 50\%. On the other hand, the blue GC component of M87 increases as function of the radial distance till it dominates in the outermost bin.
Different fractions of blue to red GCs also affect the $(g'-r')_{0}\, \mathrm{vs}\, (i'-z')_{0}$ color-color relation, that was recently shown to give insights in the evolution history of the host systems \citep{powalka16b}. We then find that the Virgo ICGCs peak on average at bluer colors than the M87 halo GCs. However the color-color regions occupied by the metal-poor GC component of the M87 halo population reach bluer colors moving towards large radii. In the NW region and for $\mathrm{R > 40\, kpc}$ the peak for the M87 halo GCs reaches similar values to those representative of the IC population at those distances. A similar trend is observed for the ICGCs, that on the northern side and for $\mathrm{R > 216\, kpc}$ are characterized by the bluest peak in the $(g'-r')_{0}\, \mathrm{vs}\, (i'-z')_{0}$ diagram.

We also have information on the density distribution of the M87 halo and intra-cluster GCs, with the latter measured to be shallower, although still centrally concentrated around M87. An excess of ICGCs in the NW region reveals the asymmetric nature of this component, while within the error the spatial distribution of the M87 halo does not deviate from point symmetry. Finally, we find a specific frequency for the Virgo ICGC population $S_{N,\mathrm{ICL}}=10.2\pm4.8$, assigning it to the category of high-$S_{N}$ systems.

The total number of GCs with respect to the parent stellar light, the relative fraction of blue-to-red objects, the GC color gradients as well as the different spatial distributions must relate to the dynamical evolution of the system that hosted these GCs. In what follows, we will review these properties and consider their implications for the formation and evolution of M87, the Virgo ICL and their GC populations.

\paragraph{The Virgo cluster core}
Galaxy interactions, as well as tidal interactions between galaxies and the cluster potential are the main mechanisms expected to build the IC component over time. The evidence in Virgo points to an IC component made up of GCs that are stripped from their host systems orbiting in the cluster potential. 1) It is well established that blue GCs have a shallower and more extended distribution than red GCs, and if so they are more likely to be stripped resulting in an IC component with a high fraction of blue to red GCs, and blue GC colors. Using N-body numerical simulations \citet{ramos15} analyzed the tidal removal of GCs in Virgo-like galaxies and found that on average halos lose $\sim$ 16\% and $\sim$29\% of their initial red and blue GC populations. On the other hand, observational works \citep[e.g.,][]{smith13}, also emphasize the impact of the orbital and spatial distributions on the stripping process. 2) The shallow and centrally concentrated density distribution that overlaps onto a steeper density is found to be representative of the simulated diffuse components at the center of clusters \citep{dolag10,cui14} and representative of the low-mass progenitors that went to build up the ICL at low-redshift \citep{cooper15}. 3) Numerical simulations \citep{murante04,dolag10} show that the stars that belong to the diffuse component did not lose memory of the kinematics of their progenitors, still reflecting the dynamics of the satellite. Thus, the kinematic result, revealing a GC system merging at large radii with the Virgo dwarf population, makes this class of galaxies the very likely progenitors of the IC component. This statement is also supported by the photometric properties of the Virgo ICGCs. The measured mean $(g'-z')_{0}$ colors are consistent with the one found to be representative of low mass systems \citep{peng06}. Furthermore, the $S_{N}$ value and the high blue-to-red ratio is consistent with the values measured for dwarf early-type galaxies found within 1 Mpc of the Virgo's center and mostly driven by the blue population of GCs \citep[e.g.,][]{peng08,mistani16}. The extreme environment these galaxies reside in favors a short star formation period (as indicated by their high-[$\alpha$/Fe] abundances) making them very efficient at forming massive star clusters \citep{liu16}. Closer to the center (within 40 kpc from M87), however, dwarf galaxies have few or no GCs, suggesting that GCs have been stripped and are contributing to what today we see as the IC population. This picture is consistent with what is observed in deep optical surveys, where the color and morphology of the ICL stellar streams suggests that the objects getting shredded are low luminosity systems \citep[e.g.,][]{rudick10,mihos17}. However, it is important to note that we do observe a red GC fraction kinematically associated with the IC component, adding the result that among the satellites that in in the past built up the diffuse light in Virgo there are also progenitors with $M_{\mathrm{prog}} > 10^{9} M_{\odot}$. This is because galaxies below such a mass have no measured red GC population \citep{peng08}.

In a recent work, \citet{ferrarese16} computed the expected total luminosity resulting from the disruption of galaxies in the Virgo core region, $\mathrm{L_{dis,V} = 7.7\pm2.0 \times 10^{10}\, L_{\odot}}$. Hence, our estimate of the Virgo ICL luminosity implies that the IC component can contribute up to $\sim$ 58\% of the total `accreted luminosity'. Such an estimate assumes that the ICL only originates from galaxies that have been disrupted (and that had GCs), while it is likely that a fraction of the IC component has been stripped from still existing galaxies, or from galaxies that had no GCs.

Finally, as previously discussed, dE galaxies are the most likely progenitors of the Virgo ICL as traced by GCs. Thus, we can convert the total IC luminosity associated to the cluster component into a mass estimate, assuming as a mass-to-light-ratio for the ICL the one measured for these systems. \citet{toloba11} analyzed a sample of dE galaxies in Virgo and estimated an average dynamical mass-to-light-ratio in the I band, $\Upsilon_{\mathrm{I,dyn}}=2.0 \pm 0.04 \Upsilon_{\odot}$. This implies a total mass for the IC component in the Virgo core $\mathrm{M_{ICC,tot}=10.8 \pm 0.1 \times 10^{11} M_{\odot}}$, that translates into a total stellar mass $\mathrm{M_{ICC,\star}=6.6\pm 5.0 \times 10^{11} M_{\odot}}$ if $\mathrm{M_{tot}/M_{\star} = 1.6\pm 1.2}$ \citep{toloba11} \footnote{to transform V-band luminosities into I-band luminosities we have used $\mathrm{V-I} = 1.03 \pm0.04$ as calculated by \citet{vanzee04} for dwarf galaxies.}. Needless to say, these results have important implications for the origin of the diffuse light in cluster cores,  essential for a fair comparison with hydrodynamical cosmological simulations. 

\paragraph{M87}
Galaxies like M87 are thought to form via a two-phase formation scenario following which, to a first growth via rapid star formation fueled by the infall of cold gas or through major merger events, follows a series of minor merger episodes which enrich the galaxy outer regions of stars and make them grow efficiently in size. Numerical simulations predict that only very massive systems, like M87, are going to experience mergers up to $\sim$5 times in their dynamical evolution history \citep{delucia06,delucia07}, and their halos are built up through the accretion of both massive and small systems \citep{cooper15}, so that massive galaxies ($\mathrm{\log{M_{\star} > 10.5\, M_{\odot}}}$) are observed to assemble stellar mass predominantly in their halos \citep{eigenthaler18}.

The properties of the GCs associated with M87 likely reflect the evolutionary path taken by the galaxy itself. 1) The origin of the GC color bimodality was explained as the natural consequence of hierarchical merging. Following this interpretation, the metal-poor population of GCs represents the accreted component as a consequence of the accretion of smaller satellites, that adds on to the preexisting (in-situ) metal-rich GC population, the latter that formed along with the body of the
galaxy itself \citep{cote98,kissler00,mackey04}. In this scenario, while the intrinsic (red) population of GCs dominates near the center of the galaxy, the ratio of metal-poor to metal-rich GCs increases with distance till the outer regions are characterized by an excess of blue clusters, as we measure for the M87 GCs. 2) On the other hand the higher fraction of red GCs measured moving closer to the center may also trace the accretion of more massive systems that due to dynamical friction would be dragged towards the central regions and there deposited their redder population of GCs.

Several questions remain unanswered. For example, can simulations of stripping of GCs from cluster {\bf dEs} reproduce the different dynamical components of GCs that also differ in kinematics and density distributions? Would such a set of simulations be able to reproduce an IC component with a metal-rich population of clusters, and if so would they measure the same ratio of blue-to-red GCs we observe? What would be the dynamical/stellar mass associated with the diffuse component in simulated galaxy cluster cores? On the observational front, what causes the different peak in the $(g'-r')_{0}\, \mathrm{vs}\, (i'-z')_{0}$ color-color relation for the M87 halo GCs? Is it tracing the presence of the M87 \textit{crown} NW of M87, that we know it is not a negligible perturbation of the galaxy halo properties, or is it tracing the accretion of different progenitors that differ in the SE and NW region? We could argue, indeed, that such a NW-SE division is the residual kinematical signature of progenitor galaxy groups having formed the center of Virgo, and we can speculate that if/when the M49 group will fall in, it will produce a similar structure for some Gyr in the N-S direction. Finally, it is possible that a fraction of the red GC component at the center of M87 can be traced back to the major merger event that caused the presence of the KDC \citep{emsellem14}. In this case, a different chemical composition could differentiate these objects from the red GC component that originated from other formation processes. Future projects focusing on GC stellar population properties derived through the analysis of the line-strengths of stellar absorption features in deep spectroscopy have the potential to untangle some of these issues.

\section{Summary} \label{sec:conclusion}

We have used the full photometric and spectroscopic information of the GCs in the NGVS Virgo core region (the $2^\circ\times2^\circ$ area around M87) to identify and separate the M87 halo GC population from the Virgo ICGCs, and study their properties separately. This study can be summarized as follows:

\begin{itemize}
\item We implemented the extreme deconvolution (XD) algorithm to model as a sum of Gaussians the multi-dimensional color-concentration-magnitude space occupied by the objects in the NGVS, and reduce the contamination by foreground stars. This allowed us to explore the entire Virgo velocity range, usually affected by hard cuts in velocity for $\mathrm{V_{LOS} < 500\, km\, s^{-1}}$ (Fig.~\ref{fig:GMM_total} and Fig.~\ref{fig:color_Star_GC}).

\item The GC kinematics reveal a system that, as function of the distance from the M87's center, deviates from the galaxy's kinematics. On both sides of M87, the GC system shows a decrease in the mean velocity and an increase in the velocity dispersion until it merges (for distances larger than 320 kpc) with the Virgo cluster potential, as traced by the dwarf early-type galaxies in Virgo (Fig.~\ref{fig:GC_pspace}).

\item Based on a Gaussian Mixture Model (GMM) analysis, the mixture that best describes the total LOSVD is a two component model, that we then interpret as arising from the Virgo's IC and M87's halo components. M87 halo and ICL overlap out to 216 kpc. Further out, the GMM identifies only one component with mean velocity  and velocity dispersion consistent with tracing the ICL (Fig.~\ref{fig:GMM_bin}). This analysis, run separately for the sample of data in the SE and NW regions of M87, shows an IC component that is not characterized by point symmetry: the SE region shows higher values of velocity dispersion with respect to the NW counterpart. This finding can be interpreted as the result of an IC component that is not relaxed. However, due to the limited number of GC velocity measurements at these extreme distances, the level of significance is within 1 $\sigma$. 

\item{The analysis of the $(g'-i')_{0}$ vs. $g'$ color-magnitude diagram shows that both M87 galaxy and Virgo ICGC populations have a blue (metal-poor, $(g'-i')_{0} < 0.77$) and a red (metal-rich, $(g'-i')_{0} \ge 0.77$) component of GCs. However the IC population is richer in blue GCs, the latter representing $\sim$70\% of the IC sample. The fraction of blue GCs in the M87 component is $\sim$50\% of the sample (Fig.\ref{fig:GC_RB}). On the other hand, the blue GC component of the M87 halo increases as function of the distances reaching 70\% in the outermost bin.}

\item{The ICGCs are characterized by bluer colors than the galaxy halo GC population as shown by the study of the $(g'-r')_{0}\, \mathrm{vs.}\, (i'-z')_{0}$ color-color diagrams. Within the M87 halo population the GCs in the NW region and for major axis distances $\mathrm{40\, kpc < R \le 216\, kpc}$ are bluer, and they may be driven by the presence of a recent accretion event that in that region we know is contributing 60\% of the M87 halo light \citep{longobardi15b}. Within the IC population, the GCs in the NW region and for major axis distances $\mathrm{R > 216\, kpc}$ are the bluest (Fig.~\ref{fig:CC_narrow} and Fig.~\ref{fig:CC_broad}).}

\item{The intra-cluster GC population has a shallower density profile than the M87 galaxy GC population, consistent with results from theoretical simulations of diffuse stellar component (ICL) around central cluster galaxies \citep{dolag10}. The different slopes can also be interpreted as due to the different origins of the two components: steeper profiles come from the accretion of more massive systems at higher redshifts, while the shallower and more extended ICL component would be the result of less massive systems whose accretion happened at lower redshifts \citep[e.g.,][]{cooper15}. This is consistent with the measured mean $(g'-z')_{0}$ colors that are consistent with the one found to be representative of low luminosity/mass systems \citep{peng06} }. We find evidence for an asymmetry in the number of ICGCs when the NW and SE fields are studied separately, implying a higher fraction of IC component in the NW field for distances larger than $\mathrm{R > 216\, kpc}$ (Fig.~\ref{fig:GC_density}).

\item{ The population of ICGCs falls under the category of high-$S_{N}$ systems with a measured specific frequency $S_{N,\mathrm{ICL}} =  10.2\pm4.8$. This value is consistent with the one observed for the dwarf early-type galaxies that reside within 1 Mpc from the cluster center, supporting a stripping / disruption scenario for the ICGCs.} 

\item{The total V-band luminosity relative to the ICL at the center of Virgo is $\mathrm{L_{ICL} = 4.2\pm0.1 \times 10^{10}\, L_{\odot}}$. This value translates into a total (dark plus luminous) and stellar masses
$\mathrm{M_{ICC,tot}=10.8 \pm 0.1}$ $\times 10^{11} \mathrm{M_{\odot}}$, and $\mathrm{M_{ICC,\star}=6.6\pm5.0\times 10^{11} M_{\odot}}$, respectively.}
\end{itemize}

These results show that the GCs successfully trace the galaxy halo-ICL interface, and their properties allow us to understand the formation and evolution of these two components.
This work depicts a picture in which the M87 halo and Virgo intra-cluster GCs have different progenitors, with the latter most likely representing the Virgo accreted component from low and intermediate mass dwarf-ellipticals. 

\section*{Acknowledgments}
AL is thankful to E. Emsellem and L. Sales for constructive and useful discussion on the analysis and interpretation of the data.
EWP and SL acknowledge support from the National Natural Science Foundation of China through Grant No. 11573002. EWP also acknowledges travel support from the Chinese Academy of Sciences South America Center for Astronomy (CASSACA) that was essential for developing the photometric classification techniques used in this paper. SE and AJ acknowledge support from project IC120009 ``Millennium Institute of Astrophysics (MAS)'' of the Millennium Science Initiative, Chilean Ministry of Economy. AJ acknowledges additional support from project BASAL CATA PFB-06. THP. acknowledges support through the FONDECYT Regular Project No. 1161817 and the BASAL Center for Astrophysics and Associated Technologies (PFB-06). THP and AL acknowledge ECOS-Sud/CONICYT project C15U02.  H.-X.Z. acknowledges a support from the CAS Pioneer Hundred Talents Program. C.L. acknowledges the NSFC grants 11673017, 11203017, 11433002.  C.L. is supported by Key Laboratory for Particle Physics, Astrophysics and Cosmology, Ministry of Education. This paper is based on observations obtained with MegaPrime/MegaCam, a joint project of CFHT and CEA/IRFU, at the Canada- France-Hawaii Telescope (CFHT) which is operated by the National Research Council (NRC) of Canada, the Institut National des Sciences de l'Univers of the Centre National de la Recherche Scientifique (CNRS) of France, and the University of Hawaii. This research used the facilities of the Canadian Astronomy Data Centre operated by the National Research Council of Canada with the support of the Canadian Space Agency. Observations reported here were obtained at the MMT Observatory, a joint facility of the University of Arizona and the Smithsonian Institution. MMT telescope time was granted, in part, by NOAO, through the Telescope System Instrumentation Program (TSIP). TSIP is funded by NSF. Data presented in this paper were obtained at the Anglo-Australian Telescope, which is operated by the Australian Astronomical Observatory. Data presented herein were obtained at the W.M. Keck Observatory, which is operated as a scientific partnership among the California Institute of Technology, the University of California and the National Aeronautics and Space Administration. The Observatory was made possible by the generous financial support of the W.M. Keck Foundation. This research made use of Astropy (http://www.astropy.org), a community-developed core Python package for Astronomy (Astropy Collaboration, 2018).




\begin{thebibliography}{}

\bibitem[Agnello et al.(2014)]{agnello14} Agnello, A., Evans, N.~W., Romanowsky, A.~J., \& Brodie, J.~P.\ 2014, \mnras, 442, 3299 


\bibitem[{{Aguerri} {et~al.}(2005){Aguerri}, {Gerhard}, {Arnaboldi},
  {Napolitano}, {Castro-Rodriguez}, \& {Freeman}}]{aguerri05}
{Aguerri}, J.~A.~L., {Gerhard}, O.~E., {Arnaboldi}, M., {et~al.} 2005, \aj,
  129, 2585

\bibitem[Alam et al.(2015)]{alam15} Alam, S., Albareti, F.~D., Allende Prieto, C., et al.\ 2015, \apjs, 219, 12 

\bibitem[Alamo-Mart{\'{\i}}nez \& Blakeslee(2017)]{alamo17} Alamo-Mart{\'{\i}}nez, K.~A., \& Blakeslee, J.~P.\ 2017, \apj, 849, 6 



\bibitem[{{Arnaboldi} {et~al.}(1996){Arnaboldi}, {Freeman}, {Mendez},
  {Capaccioli}, {Ciardullo}, {Ford}, {Gerhard}, {Hui}, {Jacoby}, {Kudritzki},
  \& {Quinn}}]{arnaboldi96}
{Arnaboldi}, M., {Freeman}, K.~C., {Mendez}, R.~H., {et~al.} 1996, \apj, 472,
  145

\bibitem[{{Arnaboldi} {et~al.}(2004){Arnaboldi}, {Gerhard}, {Aguerri},
  {Freeman}, {Napolitano}, {Okamura}, \& {Yasuda}}]{arnaboldi04}
{Arnaboldi}, M., {Gerhard}, O., {Aguerri}, J.~A.~L., {et~al.} 2004, \apjl, 614,
  L33

\bibitem[Arp(1967)]{arp67} Arp, H.~C.\ 1967, \aplett, 1, 1 

\bibitem[Ashman \& Zepf(1992)]{az92} Ashman, K.~M., \& Zepf, S.~E.\ 1992, \apj, 384, 50 


\bibitem[Ashman \& Bird(1993)]{ashman93} Ashman, K.~M., \& Bird, C.~M.\ 1993, \aj, 106, 2281 


\bibitem[Barbary et al.(2012)]{barbary12} Barbary, K., Aldering, G., Amanullah, R., et al.\ 2012, \apj, 745, 31 

\bibitem[Barbosa et al.(2018)]{barbosa18} Barbosa, C.~E., Arnaboldi, M., Coccato, L., et al.\ 2018, \aap, 609, A78 


\bibitem[Battaglia et al.(2005)]{battaglia05} Battaglia, G., Helmi, A., Morrison, H., et al.\ 2005, \mnras, 364, 433 

\bibitem[Baum(1955)]{baum55} Baum, W.~A.\ 1955, \pasp, 67, 328

\bibitem[Beasley et al.(2002)]{beasley02} Beasley, M.~A., Baugh, C.~M., Forbes, D.~A., Sharples, R.~M., \& Frenk, C.~S.\ 2002, \mnras, 333, 383 

\bibitem[Bender et al.(1994)]{bender94} Bender, R., Saglia, R.~P., \& Gerhard, O.~E.\ 1994, \mnras, 269, 785 

\bibitem[Bender et al.(2015)]{bender15} Bender, R., Kormendy, J., Cornell, M.~E., \& Fisher, D.~B.\ 2015, \apj, 807, 56 

\bibitem[Bekki \& Yahagi(2006)]{bekki06} Bekki, K., \& Yahagi, H.\ 2006, \mnras, 372, 1019 

\bibitem[Bertin \& Arnouts(1996)]{bertin96} Bertin, E., \& Arnouts, S.\ 1996, \aaps, 117, 393 

\bibitem[Bertin et al.(2002)]{bertin02} Bertin, E., Mellier, Y., Radovich, M., et al.\ 2002, Astronomical Data Analysis Software and Systems XI, 281, 228 

\bibitem[Binggeli et al.(1985)]{binggeli85} Binggeli, B., Sandage, A., \& Tammann, G.~A.\ 1985, \aj, 90, 1681 

\bibitem[{{Binggeli} {et~al.}(1987){Binggeli}, {Tammann}, \&
  {Sandage}}]{binggeli87}
{Binggeli}, B., {Tammann}, G.~A., \& {Sandage}, A. 1987, \aj, 94, 251

\bibitem[{{Binggeli} {et~al.}(1993){Binggeli}, {Popescu}, \&
  {Tammann}}]{binggeli93}
{Binggeli}, B., {Popescu}, C.~C., \& {Tammann}, G.~A. 1993, \aaps, 98, 275
\bibitem[Blakeslee et al.(2009)]{blakeslee09} Blakeslee, J.~P., Jord{\'a}n, A., Mei, S., et al.\ 2009, \apj, 694, 556 

\bibitem[Blakeslee et al.(2012)]{blakeslee12} Blakeslee, J.~P., Cho, H., Peng, E.~W., et al.\ 2012, \apj, 746, 88 

\bibitem[Boselli et al.(2008)]{boselli08} Boselli, A., Boissier, S., Cortese, L., \& Gavazzi, G.\ 2008, \apj, 674, 742-767 

\bibitem[Boselli et al.(2014)]{boselli14} Boselli, A., Voyer, E., Boissier, S., et al.\ 2014, \aap, 570, A69 

\bibitem[Bovy et al.(2011)]{bovy11} Bovy, J., Hennawi, J.~F., Hogg, D.~W., et al.\ 2011, \apj, 729, 141 


\bibitem[Brodie \& Huchra(1991)]{brodie91} Brodie, J.~P., \& Huchra, J.~P.\ 1991, \apj, 379, 157 

\bibitem[Brodie \& Strader(2006)]{brodie06} Brodie, J.~P., \& Strader, J.\ 2006, \araa, 44, 193 

\bibitem[Brodie et al.(2014)]{brodie14} Brodie, J.~P., Romanowsky, A.~J., Strader, J., et al.\ 2014, \apj, 796, 52 

\bibitem[Cantiello \& Blakeslee(2007)]{cantiello07} Cantiello, M., \& Blakeslee, J.~P.\ 2007, \apj, 669, 982 

\bibitem[Carter \& Dixon(1978)]{carter78} Carter, D., \& Dixon, K.~L.\ 1978, \aj, 83, 574 

\bibitem[Coccato et al.(2013)]{coccato13} Coccato, L., Arnaboldi, M., \& Gerhard, O.\ 2013, \mnras, 436, 1322 

\bibitem[{{Conselice} {et~al.}(2001){Conselice}, {Gallagher}, \&
  {Wyse}}]{conselice01}
{Conselice}, C.~J., {Gallagher}, III, J.~S., \& {Wyse}, R.~F.~G. 2001, \apj,
  559, 791


\bibitem[{{Contini} {et~al.}(2014){Contini}, {De Lucia}, {Villalobos}, \&
  {Borgani}}]{contini14}
{Contini}, E., {De Lucia}, G., {Villalobos}, {\'A}., \& {Borgani}, S. 2014,
  \mnras, 437, 3787

\bibitem[Cooper et al.(2015)]{cooper15} Cooper, A.~P., Gao, L., Guo, Q., et al.\ 2015, \mnras, 451, 2703 


\bibitem[C{\^o}t{\'e} et al.(1998)]{cote98} C{\^o}t{\'e}, P., Marzke, R.~O., \& West, M.~J.\ 1998, \apj, 501, 554 

\bibitem[C{\^o}t{\'e} et al.(2000)]{cote00} C{\^o}t{\'e}, P., Marzke, R.~O., West, M.~J., \& Minniti, D.\ 2000, \apj, 533, 869 


\bibitem[C{\^o}t{\'e} et al.(2001)]{cote01} C{\^o}t{\'e}, P., McLaughlin, D.~E., Hanes, D.~A., et al.\ 2001, \apj, 559, 828 

\bibitem[C{\^o}t{\'e} et al.(2002)]{cote02} C{\^o}t{\'e}, P., West, M.~J., \& Marzke, R.~O.\ 2002, \apj, 567, 853 
\bibitem[C{\^o}t{\'e} et al.(2003)]{cote03} C{\^o}t{\'e}, P., McLaughlin, D.~E., Cohen, J.~G., \& Blakeslee, J.~P.\ 2003, \apj, 591, 850 

\bibitem[C{\^o}t{\'e} et al.(2004)]{cote04} C{\^o}t{\'e}, P., Blakeslee, J.~P., Ferrarese, L., et al.\ 2004, \apjs, 153, 223 


\bibitem[Cui et al.(2014)]{cui14} Cui, W., Murante, G., Monaco, P., et al.\ 2014, \mnras, 437, 816 

\bibitem[{{De Lucia} {et~al.}(2006){De Lucia}, {Springel}, {White}, {Croton},
  \& {Kauffmann}}]{delucia06}
{De Lucia}, G., {Springel}, V., {White}, S.~D.~M., {Croton}, D., \&
  {Kauffmann}, G. 2006, \mnras, 366, 499

\bibitem[{{De Lucia} \& {Blaizot}(2007)}]{delucia07}
{De Lucia}, G. \& {Blaizot}, J. 2007, \mnras, 375, 2

\bibitem[Dilday et al.(2010)]{dilday10} Dilday, B., Bassett, B., Becker, A., et al.\ 2010, \apj, 715, 1021 

\bibitem[{{Dolag} {et~al.}(2010){Dolag}, {Murante}, \& {Borgani}}]{dolag10}
{Dolag}, K., {Murante}, G., \& {Borgani}, S. 2010, \mnras, 405, 1544


\bibitem[Drinkwater et al.(2000)]{drinkwater00} Drinkwater, M.~J., Jones, J.~B., Gregg, M.~D., \& Phillipps, S.\ 2000, \pasa, 17, 227 


\bibitem[{{Durrell} {et~al.}(2002){Durrell}, {Ciardullo}, {Feldmeier},
  {Jacoby}, \& {Sigurdsson}}]{durrell02}
{Durrell}, P.~R., {Ciardullo}, R., {Feldmeier}, J.~J., {Jacoby}, G.~H., \&
  {Sigurdsson}, S. 2002, \apj, 570, 119

\bibitem[Durrell et al.(2014)]{durrell14} Durrell, P.~R., C{\^o}t{\'e}, P., Peng, E.~W., et al.\ 2014, \apj, 794, 103 


\bibitem[{{Emsellem} {et~al.}(2014){Emsellem}, {Krajnovi{\'c}}, \&
  {Sarzi}}]{emsellem14}
{Emsellem}, E., {Krajnovi{\'c}}, D., \& {Sarzi}, M. 2014, \mnras, 445, L79

\bibitem[Eigenthaler et al.(2018)]{eigenthaler18} Eigenthaler, P., Puzia, T.~H., Taylor, M.~A., et al.\ 2018, \apj, 855, 142 

\bibitem[{{Feldmeier} {et~al.}(2004{\natexlab{a}}){Feldmeier}, {Ciardullo},
  {Jacoby}, \& {Durrell}}]{feldmeier04}
{Feldmeier}, J.~J., {Ciardullo}, R., {Jacoby}, G.~H., \& {Durrell}, P.~R.
  2004{\natexlab{a}}, \apj, 615, 196

\bibitem[{{Ferguson} {et~al.}(1998){Ferguson}, {Gallagher}, \&
  {Wyse}}]{ferguson98}
{Ferguson}, A.~M.~N., {Gallagher}, J.~S., \& {Wyse}, R.~F.~G. 1998, \aj, 116,
  673

\bibitem[Ferrarese et al.(2006)]{ferrarese06} Ferrarese, L., C{\^o}t{\'e}, P., Jord{\'a}n, A., et al.\ 2006, \apjs, 164, 334 


\bibitem[Ferrarese et al.(2012)]{ferrarese12} Ferrarese, L., C{\^o}t{\'e}, P., Cuillandre, J.-C., et al.\ 2012, \apjs, 200, 4 

\bibitem[Ferrarese et al.(2016)]{ferrarese16} Ferrarese, L., C{\^o}t{\'e}, P., S{\'a}nchez-Janssen, R., et al.\ 2016, \apj, 824, 10 

\bibitem[Firth et al.(2008)]{firth08} Firth, P., Drinkwater, M.~J., \& Karick, A.~M.\ 2008, \mnras, 389, 1539 

\bibitem[Forbes et al.(1997)]{forbes97} Forbes, D.~A., Brodie, J.~P., \& Grillmair, C.~J.\ 1997, \aj, 113, 1652 




\bibitem[Forbes \& Forte(2001)]{forbes01} Forbes, D.~A., \& Forte, J.~C.\ 2001, \mnras, 322, 257 

\bibitem[Forbes(2017)]{forbes17} Forbes, D.~A.\ 2017, \mnras, 472, L104 

\bibitem[Ford \& Butcher(1979)]{ford79} Ford, H.~C., \& Butcher, H.\ 1979, \apjs, 41, 147 

\bibitem[Forte et al.(1982)]{forte82} Forte, J.~C., Martinez, R.~E., \& Muzzio, J.~C.\ 1982, \aj, 87, 1465 


\bibitem[{{Gal-Yam} {et~al.}(2003){Gal-Yam}, {Maoz}, {Guhathakurta}, \&
  {Filippenko}}]{galyam03}
{Gal-Yam}, A., {Maoz}, D., {Guhathakurta}, P., \& {Filippenko}, A.~V. 2003,
  \aj, 125, 1087


\bibitem[Gavazzi et al.(2000)]{gavazzi00} Gavazzi, G., Boselli, A., V{\'{\i}}lchez, J.~M., Iglesias-Paramo, J., \& Bonfanti, C.\ 2000, \aap, 361, 1 

\bibitem[Gavazzi et al.(2003)]{gavazzi03} Gavazzi, G., Boselli, A., Donati, A., Franzetti, P., \& Scodeggio, M.\ 2003, \aap, 400, 451 

\bibitem[Gebhardt \& Kissler-Patig(1999)]{gebhardt99} Gebhardt, K., \& Kissler-Patig, M.\ 1999, \aj, 118, 1526 

\bibitem[Georgiev et al.(2010)]{georgiev10} Georgiev, I.~Y., Puzia, T.~H., Goudfrooij, P., \& Hilker, M.\ 2010, \mnras, 406, 1967 


\bibitem[Gerhard(1993)]{gerhard93} Gerhard, O.~E.\ 1993, \mnras, 265, 213 


\bibitem[{{Gerhard} {et~al.}(2005){Gerhard}, {Arnaboldi}, {Freeman},
  {Kashikawa}, {Okamura}, \& {Yasuda}}]{gerhard05}
{Gerhard}, O., {Arnaboldi}, M., {Freeman}, K.~C., {et~al.} 2005, \apjl, 621,
  L93

\bibitem[{{Gnedin}(2003)}]{gnedin03}
{Gnedin}, O.~Y. 2003, \apj, 589, 752



\bibitem[{{Gonzalez} {et~al.}(2005){Gonzalez}, {Zabludoff}, \&
  {Zaritsky}}]{gonzales05}
{Gonzalez}, A.~H., {Zabludoff}, A.~I., \& {Zaritsky}, D. 2005, \apj, 618, 195

\bibitem[{{Gonzalez} {et~al.}(2007){Gonzalez}, {Zaritsky}, \&
  {Zabludoff}}]{gonzales07}
{Gonzalez}, A.~H., {Zaritsky}, D., \& {Zabludoff}, A.~I. 2007, \apj, 666, 147

\bibitem[Grossauer et al.(2015)]{grossauer15} Grossauer, J., Taylor, J.~E., Ferrarese, L., et al.\ 2015, \apj, 807, 88 

\bibitem[Hanes et al.(2001)]{hanes01} Hanes, D.~A., C{\^o}t{\'e}, P., Bridges, T.~J., et al.\ 2001, \apj, 559, 812 


\bibitem[Harris \& van den Bergh(1981)]{harris81} Harris, W.~E., \& van den Bergh, S.\ 1981, \aj, 86, 1627 

\bibitem[Harris(1991)]{harris91} Harris, W.~E.\ 1991, \araa, 29, 543 

\bibitem[Harris et al.(2009)]{harris09} Harris, W.~E., Kavelaars, J.~J., Hanes, D.~A., Pritchet, C.~J., \& Baum, W.~A.\ 2009, \aj, 137, 3314 

\bibitem[Harris et al.(2013)]{harris13} Harris, W.~E., Harris, G.~L.~H., \& Alessi, M.\ 2013, \apj, 772, 82 


\bibitem[Harris et al.(2017)]{harris17} Harris, W.~E., Ciccone, S.~M., Eadie, G.~M., et al.\ 2017, \apj, 835, 101 

\bibitem[Ha{\c s}egan et al.(2005)]{hasegan05} Ha{\c s}egan, M., Jord{\'a}n, A., C{\^o}t{\'e}, P., et al.\ 2005, \apj, 627, 203 

\bibitem[Hartke et al.(2017)]{hartke17} Hartke, J., Arnaboldi, M., Longobardi, A., et al.\ 2017, \aap, 603, A104 

\bibitem[Hartke et al.(2018)]{hartke18} Hartke, J., Arnaboldi, M., Gerhard, O., et al.\ 2018, arXiv:1805.03092 


\bibitem[Hilker et al.(1999)]{hilker99} Hilker, M., Infante, L., Vieira, G., Kissler-Patig, M., \& Richtler, T.\ 1999, \aaps, 134, 75 

\bibitem[Hudson \& Robison(2018)]{hudson18} Hudson, M.~J., \& Robison, B.\ 2018, \mnras 


\bibitem[Janowiecki et al.(2010)]{janowiecki10} Janowiecki, S., Mihos, J.~C., Harding, P., et al.\ 2010, \apj, 715, 972 

\bibitem[Jord{\'a}n et al.(2002)]{jordan02} Jord{\'a}n, A., C{\^o}t{\'e}, P., West, M.~J., \& Marzke, R.~O.\ 2002, \apjl, 576, L113 
\bibitem[Jord{\'a}n et al.(2003)]{jordan03} Jord{\'a}n, A., West, M.~J., C{\^o}t{\'e}, P., \& Marzke, R.~O.\ 2003, \aj, 125, 1642 
\bibitem[Jord{\'a}n et al.(2004)]{jordan04} Jord{\'a}n, A., C{\^o}t{\'e}, P., West, M.~J., et al.\ 2004, \aj, 127, 24 
\bibitem[Jord{\'a}n et al.(2007)]{jordan07} Jord{\'a}n, A., McLaughlin, D.~E., C{\^o}t{\'e}, P., et al.\ 2007, \apjs, 171, 101 

\bibitem[Kartha et al.(2014)]{kartha14} Kartha, S.~S., Forbes, D.~A., Spitler, L.~R., et al.\ 2014, \mnras, 437, 273

\bibitem[Kelson et al.(2002)]{kelson02} Kelson, D.~D., Zabludoff, A.~I., Williams, K.~A., et al.\ 2002, \apj, 576, 720

\bibitem[Kim et al.(2014)]{kim14} Kim, S., Rey, S.-C., Jerjen, H., et al.\ 2014, \apjs, 215, 22 

\bibitem[King(1966)]{king66} King, I.~R.\ 1966, \aj, 71, 64 

\bibitem[Kissler-Patig(2000)]{kissler00} Kissler-Patig, M.\ 2000, Reviews in Modern Astronomy, 13, 13 

\bibitem[Ko et al.(2017)]{ko17} Ko, Y., Hwang, H.~S., Lee, M.~G., et al.\ 2017, \apj, 835, 212 

\bibitem[{{Kormendy} {et~al.}(2009){Kormendy}, {Fisher}, {Cornell}, \&
  {Bender}}]{kormendy09}
{Kormendy}, J., {Fisher}, D.~B., {Cornell}, M.~E., \& {Bender}, R. 2009, \apjs,
  182, 216

\bibitem[Kruijssen(2014)]{kruijssen14} Kruijssen, J.~M.~D.\ 2014, Classical and Quantum Gravity, 31, 244006 

\bibitem[Kundu \& Whitmore(2001)]{kundu01} Kundu, A., \& Whitmore, B.~C.\ 2001, \aj, 121, 2950 

\bibitem[{{Laporte} {et~al.}(2013){Laporte}, {White}, {Naab}, \&
  {Gao}}]{laporte13}
{Laporte}, C.~F.~P., {White}, S.~D.~M., {Naab}, T., \& {Gao}, L. 2013, \mnras,
  435, 901



\bibitem[Larsen et al.(2001)]{larsen01} Larsen, S.~S., Brodie, J.~P., Huchra, J.~P., Forbes, D.~A., \& Grillmair, C.~J.\ 2001, \aj, 121, 2974 

\bibitem[Leaman et al.(2013)]{leaman13} Leaman, R., VandenBerg, D.~A., \& Mendel, J.~T.\ 2013, \mnras, 436, 122 

\bibitem[Lee et al.(2010)]{lee10} Lee, M.~G., Park, H.~S., \& Hwang, H.~S.\ 2010, Science, 328, 334 

\bibitem[Lim et al.(2017)]{lim17} Lim, S., Peng, E.~W., Duc, P.-A., et al.\ 2017, \apj, 835, 123 

\bibitem[{{Liu} {et~al.}(2005){Liu}, {Zhou}, {Ma}, {Wu}, {Yang}, {Li}, \&
  {Chen}}]{liu05}
{Liu}, Y., {Zhou}, X., {Ma}, J., {et~al.} 2005, \aj, 129, 2628


\bibitem[Liu et al.(2015a)]{liu15a} Liu, C., Peng, E.~W., C{\^o}t{\'e}, P., et al.\ 2015, \apj, 812, 34 
\bibitem[Liu et al.(2015b)]{liu15b} Liu, C., Peng, E.~W., Toloba, E., et al.\ 2015, \apjl, 812, L2 
\bibitem[Liu et al.(2016)]{liu16} Liu, Y., Peng, E.~W., Blakeslee, J., et al.\ 2016, \apj, 818, 179 


\bibitem[Longobardi et al.(2013)]{longobardi13} Longobardi, A., Arnaboldi, M., Gerhard, O., et al.\ 2013, \aap, 558, A42 

\bibitem[Longobardi et al.(2015a)]{longobardi15a} Longobardi, A., Arnaboldi, M., Gerhard, O., \& Hanuschik, R.\ 2015, \aap, 579, A135 

\bibitem[Longobardi et al.(2015b)]{longobardi15b} Longobardi, A., Arnaboldi, M., Gerhard, O., \& Mihos, J.~C.\ 2015, \aap, 579, L3 


\bibitem[Mackey \& Gilmore(2004)]{mackey04} Mackey, A.~D., \& Gilmore, G.~F.\ 2004, \mnras, 355, 504 

\bibitem[McDonald et al.(2011)]{mcdonald11} McDonald, M., Courteau, S., Tully, R.~B., \& Roediger, J.\ 2011, \mnras, 414, 2055 
\bibitem[McLaughlin(1999)]{mclaughlin99} McLaughlin, D.~E.\ 1999, \apjl, 512, L9 

\bibitem[{{Mei} {et~al.}(2007){Mei}, {Blakeslee}, {C{\^o}t{\'e}}, {Tonry},
  {West}, {Ferrarese}, {Jord{\'a}n}, {Peng}, {Anthony}, \& {Merritt}}]{mei07}
{Mei}, S., {Blakeslee}, J.~P., {C{\^o}t{\'e}}, P., {et~al.} 2007, \apj, 655,
  144

\bibitem[McLaughlin et al.(1994)]{mclaughlin94} McLaughlin, D.~E., Harris, W.~E., \& Hanes, D.~A.\ 1994, \apj, 422, 486 


\bibitem[{{Merritt}(1984)}]{merritt84}
{Merritt}, D. 1984, \apj, 276, 26

\bibitem[{{Mihos} {et~al.}(2005){Mihos}, {Harding}, {Feldmeier}, \&
  {Morrison}}]{mihos05}
{Mihos}, J.~C., {Harding}, P., {Feldmeier}, J., \& {Morrison}, H. 2005, \apjl,
  631, L41

\bibitem[Mihos et al.(2017)]{mihos17} Mihos, J.~C., Harding, P., Feldmeier, J.~J., et al.\ 2017, \apj, 834, 16 

\bibitem[Misgeld \& Hilker(2011)]{misgeld11} Misgeld, I., \& Hilker, M.\ 2011, \mnras, 414, 3699 

\bibitem[Mistani et al.(2016)]{mistani16} Mistani, P.~A., Sales, L.~V., Pillepich, A., et al.\ 2016, \mnras, 455, 2323 


\bibitem[{{Montes} {et~al.}(2014){Montes}, {Trujillo}, {Prieto}, \&
  {Acosta-Pulido}}]{montes14}
{Montes}, M., {Trujillo}, I., {Prieto}, M.~A., \& {Acosta-Pulido}, J.~A. 2014,
  \mnras, 439, 990

\bibitem[{{Moore} {et~al.}(1998){Moore}, {Lake}, \& {Katz}}]{moore98}
{Moore}, B., {Lake}, G., \& {Katz}, N. 1998, \apj, 495, 139

\bibitem[Mould et al.(1987)]{mould87} Mould, J.~R., Oke, J.~B., \& Nemec, J.~M.\ 1987, \aj, 93, 53

\bibitem[Mu{\~n}oz et al.(2014)]{munoz14} Mu{\~n}oz, R.~P., Puzia, T.~H., Lan{\c c}on, A., et al.\ 2014, \apjs, 210, 4 

\bibitem[{{Murante} {et~al.}(2004){Murante}, {Arnaboldi}, {Gerhard}, {Borgani},
  {Cheng}, {Diaferio}, {Dolag}, {Moscardini}, {Tormen}, {Tornatore}, \&
  {Tozzi}}]{murante04}
{Murante}, G., {Arnaboldi}, M., {Gerhard}, O., {et~al.} 2004, \apjl, 607, L83

\bibitem[{{Naab} {et~al.}(2009){Naab}, {Johansson}, \& {Ostriker}}]{naab09}
{Naab}, T., {Johansson}, P.~H., \& {Ostriker}, J.~P. 2009, \apjl, 699, L178

\bibitem[{{Nulsen} \& {Bohringer}(1995)}]{nulsen95}
{Nulsen}, P.~E.~J. \& {Bohringer}, H. 1995, \mnras, 274, 1093

\bibitem[Oldham \& Evans(2016)]{oldham16} Oldham, L.~J., \& Evans, N.~W.\ 2016, \mnras, 462, 298 


\bibitem[{{Oser} {et~al.}(2010){Oser}, {Ostriker}, {Naab}, {Johansson}, \&
  {Burkert}}]{oser10}
{Oser}, L., {Ostriker}, J.~P., {Naab}, T., {Johansson}, P.~H., \& {Burkert}, A.
  2010, \apj, 725, 2312


\bibitem[Ohyama \& Hota(2013)]{oyama13} Ohyama, Y., \& Hota, A.\ 2013, \apjl, 767, L29 


\bibitem[{Pedregosa {et~al.}(2011)Pedregosa, Varoquaux, Gramfort, Michel,
  Thirion, Grisel, Blondel, Prettenhofer, Weiss, Dubourg, Vanderplas, Passos,
  Cournapeau, Brucher, Perrot, \& Duchesnay}]{pedregosa11}
Pedregosa, F., Varoquaux, G., Gramfort, A., {et~al.} 2011, Journal of Machine
  Learning Research, 12, 2825

\bibitem[Peng et al.(2004)]{peng04} Peng, E.~W., Ford, H.~C., \& Freeman, K.~C.\ 2004, \apj, 602, 705 
\bibitem[Peng et al.(2006)]{peng06} Peng, E.~W., Jord{\'a}n, A., C{\^o}t{\'e}, P., et al.\ 2006, \apj, 639, 95 
\bibitem[Peng et al.(2008)]{peng08} Peng, E.~W., Jord{\'a}n, A., C{\^o}t{\'e}, P., et al.\ 2008, \apj, 681, 197-224 
\bibitem[Peng et al.(2009)]{peng09} Peng, E.~W., Jord{\'a}n, A., Blakeslee, J.~P., et al.\ 2009, \apj, 703, 42 
\bibitem[Peng et al.(2011)]{peng11} Peng, E.~W., Ferguson, H.~C., Goudfrooij, P., et al.\ 2011, \apj, 730, 23 
\bibitem[Peng \& Lim(2016)]{peng16} Peng, E.~W., \& Lim, S.\ 2016, \apjl, 822, L31 


\bibitem[Perrett et al.(2003)]{perrett03} Perrett, K.~M., Stiff, D.~A., Hanes, D.~A., \& Bridges, T.~J.\ 2003, \apj, 589, 790 



\bibitem[Powalka et al.(2016a)]{powalka16a} Powalka, M., Lan{\c c}on, A., Puzia, T.~H., et al.\ 2016, \apjs, 227, 12 
\bibitem[Powalka et al.(2016b)]{powalka16b} Powalka, M., Puzia, T.~H., Lan{\c c}on, A., et al.\ 2016, \apjl, 829, L5 
\bibitem[Powalka et al.(2017)]{powalka17} Powalka, M., Lan{\c c}on, A., Puzia, T.~H., et al.\ 2017, \apj, 844, 104 
\bibitem[Powalka et al.(2018)]{powalka18} Powalka, M., Puzia, T.~H., Lan{\c c}on, A., et al.\ 2018, \apj, 856, 84 


\bibitem[Pota et al.(2013)]{pota13} Pota, V., Graham, A.~W., Forbes, D.~A., et al.\ 2013, \mnras, 433, 235 

\bibitem[{{Puchwein} {et~al.}(2010){Puchwein}, {Springel}, {Sijacki}, \&
  {Dolag}}]{puchwein10}
{Puchwein}, E., {Springel}, V., {Sijacki}, D., \& {Dolag}, K. 2010, \mnras,
  406, 936
  
\bibitem[Puzia et al.(1999)]{puzia99} Puzia, T.~H., Kissler-Patig, M., Brodie, J.~P., \& Huchra, J.~P.\ 1999, \aj, 118, 2734 

\bibitem[Puzia et al.(2004)]{puzia04} Puzia, T.~H., Kissler-Patig, M., Thomas, D., et al.\ 2004, \aap, 415, 123 


\bibitem[Puzia et al.(2005a)]{puzia05a} Puzia, T.~H., Perrett, K.~M., \& Bridges, T.~J.\ 2005, \aap, 434, 909 

\bibitem[Puzia et al.(2005b)]{puzia05b} Puzia, T.~H., Kissler-Patig, M., Thomas, D., et al.\ 2005, \aap, 439, 997 


\bibitem[Roediger et al.(2017)]{roediger17} Roediger, J.~C., Ferrarese, L., C{\^o}t{\'e}, P., et al.\ 2017, \apj, 836, 120 

\bibitem[Sand et al.(2011)]{sand11} Sand, D.~J., Graham, M.~L., Bildfell, C., et al.\ 2011, \apj, 729, 142 
\bibitem[Schuberth et al.(2010)]{schuberth10} Schuberth, Y., Richtler, T., Hilker, M., et al.\ 2010, \aap, 513, A52 

\bibitem[Smith et al.(2013)]{smith13} Smith, R., S{\'a}nchez-Janssen, R., Fellhauer, M., et al.\ 2013, \mnras, 429, 1066 

\bibitem[Sparks et al.(1993)]{sparks93} Sparks, W.~B., Ford, H.~C., \& Kinney, A.~L.\ 1993, \apj, 413, 531 
\bibitem[Stetson et al.(1989)]{stetson89} Stetson, P.~B., Vandenberg, D.~A., Bolte, M., Hesser, J.~E., \& Smith, G.~H.\ 1989, \aj, 97, 1360 

\bibitem[Strader et al.(2005)]{strader05} Strader, J., Brodie, J.~P., Cenarro, A.~J., Beasley, M.~A., \& Forbes, D.~A.\ 2005, \aj, 130, 1315 


\bibitem[Ramos et al.(2015)]{ramos15} Ramos, F., Coenda, V., Muriel, H., \& Abadi, M.\ 2015, \apj, 806, 242 

\bibitem[Richtler(2013)]{richtler13} Richtler, T.\ 2013, 370 Years of Astronomy in Utrecht, 470, 327 



\bibitem[{{Romanowsky} {et~al.}(2012){Romanowsky}, {Strader}, {Brodie},
  {Mihos}, {Spitler}, {Forbes}, {Foster}, \& {Arnold}}]{romanowsky12}
{Romanowsky}, A.~J., {Strader}, J., {Brodie}, J.~P., {et~al.} 2012, \apj, 748,
  29




\bibitem[{{Rudick} {et~al.}(2006){Rudick}, {Mihos}, \& {McBride}}]{rudick06}
{Rudick}, C.~S., {Mihos}, J.~C., \& {McBride}, C. 2006, \apj, 648, 936

\bibitem[{{Rudick} {et~al.}(2009){Rudick}, {Mihos}, {Frey}, \&
  {McBride}}]{rudick09}
{Rudick}, C.~S., {Mihos}, J.~C., {Frey}, L.~H., \& {McBride}, C.~K. 2009, \apj,
  699, 1518

\bibitem[{{Rudick} {et~al.}(2010){Rudick}, {Mihos}, {Harding}, {Feldmeier},
  {Janowiecki}, \& {Morrison}}]{rudick10}
{Rudick}, C.~S., {Mihos}, J.~C., {Harding}, P., {et~al.} 2010, \apj, 720, 569

\bibitem[S{\'a}nchez-Janssen et al.(2016)]{sanchez16} S{\'a}nchez-Janssen, R., Ferrarese, L., MacArthur, L.~A., et al.\ 2016, \apj, 820, 69 

\bibitem[Santos(2003)]{santos03} Santos, M.~R.\ 2003, Extragalactic Globular Cluster Systems, 348 


\bibitem[Schroder et al.(2002)]{schroder02} Schroder, L.~L., Brodie, J.~P., Kissler-Patig, M., Huchra, J.~P., \& Phillips, A.~C.\ 2002, \aj, 123, 2473 

\bibitem[Schuberth et al.(2008)]{shubert08} Schuberth, Y., Richtler, T., Bassino, L., \& Hilker, M.\ 2008, \aap, 477, L9 

\bibitem[{{Seigar} {et~al.}(2007){Seigar}, {Graham}, \& {Jerjen}}]{seigar07}
{Seigar}, M.~S., {Graham}, A.~W., \& {Jerjen}, H. 2007, \mnras, 378, 1575

\bibitem[Spengler et al.(2017)]{spengler17} Spengler, C., C{\^o}t{\'e}, P., Roediger, J., et al.\ 2017, \apj, 849, 55 

\bibitem[{{Strader} {et~al.}(2011){Strader}, {Romanowsky}, {Brodie}, {Spitler},
  {Beasley}, {Arnold}, {Tamura}, {Sharples}, \& {Arimoto}}]{strader11}
{Strader}, J., {Romanowsky}, A.~J., {Brodie}, J.~P., {et~al.} 2011, \apjs, 197,
  33
  
 \bibitem[Taylor et al.(2010)]{taylor10} Taylor, M.~A., Puzia, T.~H., Harris, G.~L., et al.\ 2010, \apj, 712, 1191 
 
\bibitem[Toloba et al.(2014)]{toloba14} Toloba, E., Guhathakurta, P., Peletier, R.~F., et al.\ 2014, \apjs, 215, 17 

\bibitem[Toloba et al.(2011)]{toloba11} Toloba, E., Boselli, A., Cenarro, A.~J., et al.\ 2011, \aap, 526, A114 

\bibitem[Toloba et al.(2016)]{toloba16} Toloba, E., Li, B., Guhathakurta, P., et al.\ 2016, \apj, 822, 51 



\bibitem[{{Thomas} {et~al.}(2005){Thomas}, {Maraston}, {Bender}, \& {Mendes de
  Oliveira}}]{thomas05}
{Thomas}, D., {Maraston}, C., {Bender}, R., \& {Mendes de Oliveira}, C. 2005,
  \apj, 621, 673
  
  
 \bibitem[Toledo et al.(2011)]{toledo11} Toledo, I., Melnick, J., Selman, F., et al.\ 2011, \mnras, 414, 602  
 
\bibitem[Tonini(2013)]{tonini13} Tonini, C.\ 2013, \apj, 762, 39 

\bibitem[Tully \& Shaya(1984)]{tully84} Tully, R.~B., \& Shaya, E.~J.\ 1984, \apj, 281, 31

\bibitem[Usher et al.(2015)]{usher15} Usher, C., Forbes, D.~A., Brodie, J.~P., et al.\ 2015, \mnras, 446, 369 
\bibitem[van Zee et al.(2004)]{vanzee04} van Zee, L., Barton, E.~J., \& Skillman, E.~D.\ 2004, \aj, 128, 2797 

\bibitem[Veale et al.(2017)]{veale17} Veale, M., Ma, C.-P., Greene, J.~E., et al.\ 2017, \mnras, 471, 1428 

\bibitem[{{Ventimiglia} {et~al.}(2011){Ventimiglia}, {Arnaboldi}, \&
  {Gerhard}}]{ventimiglia11}
{Ventimiglia}, G., {Arnaboldi}, M., \& {Gerhard}, O. 2011, \aap, 528, A24+

\bibitem[{{V{\'{\i}}lchez-G{\'o}mez}(1999)}]{vilchez99}
{V{\'{\i}}lchez-G{\'o}mez}, R. 1999, in Astronomical Society of the Pacific
  Conference Series, Vol. 170, The Low Surface Brightness Universe, ed. J.~I.
  {Davies}, C.~{Impey}, \& S.~{Phillips}, 349


\bibitem[{{Watson} \& {Conroy}(2013)}]{watson13}
{Watson}, D.~F. \& {Conroy}, C. 2013, \apj, 772, 139

\bibitem[{{Weil} {et~al.}(1997){Weil}, {Bland-Hawthorn}, \& {Malin}}]{weil97}
{Weil}, M.~L., {Bland-Hawthorn}, J., \& {Malin}, D.~F. 1997, \apj, 490, 664


\bibitem[West et al.(1995)]{west95} West, M.~J., Cote, P., Jones, C., Forman, W., \& Marzke, R.~O.\ 1995, \apjl, 453, L77 
\bibitem[West et al.(2011)]{west11} West, M.~J., Jord{\'a}n, A., Blakeslee, J.~P., et al.\ 2011, \aap, 528, A115 



\bibitem[Williams et al.(2007)]{williams07} Williams, B.~F., Ciardullo, R., Durrell, P.~R., et al.\ 2007, \apj, 654, 835 

\bibitem[Yoshida et al.(2002)]{yoshida02} Yoshida, M., Yagi, M., Okamura, S., et al.\ 2002, \apj, 567, 118 


\bibitem[Zepf et al.(2000)]{zepf00} Zepf, S.~E., Beasley, M.~A., Bridges, T.~J., et al.\ 2000, \aj, 120, 2928 


\bibitem[Zhang et al.(2015)]{zhang15} Zhang, H.-X., Peng, E.~W., C{\^o}t{\'e}, P., et al.\ 2015, \apj, 802, 30 
\bibitem[Zhu et al.(2014)]{zhu14} Zhu, L., Long, R.~J., Mao, S., et al.\ 2014, \apj, 792, 59 

\bibitem[Zhang et al.(2018)]{zhang18} Zhang, H.-X., Puzia, T.~H., Peng, E.~W., et al.\ 2018, \apj, 858, 37


\bibitem[{{Zibetti} {et~al.}(2005){Zibetti}, {White}, {Schneider}, \&
  {Brinkmann}}]{zibetti05}
{Zibetti}, S., {White}, S.~D.~M., {Schneider}, D.~P., \& {Brinkmann}, J. 2005,
  \mnras, 358, 949


\bibitem[Zwicky(1951)]{zwicky51} Zwicky, F.\ 1951, \pasp, 63, 61 

\bibitem[{{Zwicky}(1952)}]{zwicky52}
{Zwicky}, F. 1952, \pasp, 64, 242
\end{thebibliography}
\end{document}